\documentclass[]{aa} 

\usepackage{graphicx}
\usepackage{txfonts}
\usepackage{natbib}
\usepackage{longtable}
\usepackage{lscape}
\usepackage{color}

\providecommand{\sorthelp}[1]{}

\newcommand{\Planckn}{{\it Planck}}
\newcommand{\Herscheln}{{\it Herschel}}
\newcommand{\Planck}{{\it Planck }}
\newcommand{\Herschel}{{\it Herschel }}
\newcommand{\target}{{G035.39-00.33 }}
\newcommand{\targetn}{{G035.39-00.33}}

\begin{document}
 
\title{
{Dust spectrum and polarisation at 850\,$\mu$m in the massive
IRDC \target}
%
%
}

\author{Mika  Juvela\inst{1},
Vincent       Guillet\inst{2,3},
Tie           Liu\inst{4,5},
Isabelle      Ristorcelli\inst{6,7},
Veli-Matti    Pelkonen\inst{1},
Dana          Alina\inst{8},
Leonardo      Bronfman\inst{9},
David J.      Eden\inst{10},
Kee Tae       Kim\inst{4},
Patrick M.    Koch\inst{11},
Woojin        Kwon\inst{4,12},
Chang Won     Lee\inst{4,12},
Johanna       Malinen\inst{13},
Elisabetta    Micelotta\inst{1},
Julien        Montillaud\inst{14},
Mark G.       Rawlings\inst{5},
Patricio      Sanhueza\inst{15},
Archana       Soam\inst{4,16},
Alessio       Traficante\inst{17},
Nathalie      Ysard\inst{2},
Chuan-Peng    Zhang\inst{18,19}
}

\institute{
Department of Physics, P.O.Box 64, FI-00014, University of Helsinki,
Finland, {\em mika.juvela@helsinki.fi}
\and 
Institut d'Astrophysique Spatiale, CNRS, Univ. Paris-Sud,
Universit\'{e} Paris-Saclay, B\^{a}t. 121, 91405 Orsay cedex, France
\and 
Laboratoire Univers et Particules de Montpellier, Universit\'{e}
de Montpellier, CNRS/IN2P3, CC 72, Place Eug\`{e}ne Bataillon, 34095
Montpellier Cedex 5, France
\and 
Korea Astronomy and Space Science Institute, 776
Daedeokdaero, Yuseong-gu, Daejeon 34055, Republic of Korea
\and 
East Asian Observatory, 660 N. A'oh\={o}k\={u} Place, 
Hilo, Hawaii 96720-2700, USA
\and 
Universit\'e de Toulouse, UPS-OMP, IRAP, F-31028 Toulouse cedex 4, France   
\and 
CNRS, IRAP, 9 Av. colonel Roche, BP 44346, F-31028 Toulouse cedex 4, France  
\and 
Department  of  Physics,  School  of  Science  and  Technology,
Nazarbayev University, Astana 010000, Kazakhstan
\and 
Departamento de Astronom\'ia, Universidad de Chile, Casilla 36-D,
Santiago, Chile
\and 
Astrophysics Research Institute, Liverpool John Moores University,
Ic2, Liverpool Science Park, 146 Brownlow Hill, Liverpool, L3 5RF, UK
\and 
Academia Sinica, Institute of Astronomy and Astrophysics, Taipei, Taiwan
\and 
Korea University of Science and Technology, 217 Gajang-ro,
Yuseong-gu, Daejeon 34113, Republic of Korea 
\and 
Institute of Physics I, University of Cologne, Germany
\and 
Institut UTINAM, CNRS UMR 6213, OSU THETA, Universit\'e de
Franche-Comt\'e, 41bis avenue de l’Observatoire,
25000 Besan\c{c}on, France
\and 
National Astronomical Observatory of Japan, National Institutes of
Natural Sciences, 2-21-1 Osawa, Mitaka, Tokyo 181-8588, Japan
\and 
SOFIA Science Center, Universities Space Research Association, NASA
Ames Research Center, Moffett Field, California 94035
\and 
IAPS - INAF, via Fosso del Cavaliere, 100, I-00133 Roma, Italy
\and 
National Astronomical Observatories, Chinese Academy of Sciences,
100012 Beijing, PR China
\and 
Max-Planck-Institut f\"ur Astronomie, K\"onigstuhl 17, 69117
Heidelberg, Germany 
}

\authorrunning{M. Juvela et al.}

\date{Received September 15, 1996; accepted March 16, 1997}

\abstract { 
The sub-millimetre polarisation of dust emission from star-forming clouds
carries information on grain properties and on the effects that
magnetic fields have on cloud evolution.
} 
{
Using observations of a dense filamentary cloud \targetn, we aim to
characterise the dust emission properties and the variations of the
polarisation fraction.
}
{
JCMT SCUBA-2/POL-2 observations at 850\,$\mu$m are combined with
\Planck 850\,$\mu$m (353\,GHz) data to map polarisation
fraction at small and large scales. With previous total intensity
SCUBA-2 observations (450\,$\mu$m and 850\,$\mu$m) and \Herschel data,
the column densities are determined via modified blackbody fits and
via radiative transfer modelling.
Models are constructed to examine how the observed polarisation angles
and fractions depend on potential magnetic field geometries and grain
alignment processes.
}
{
POL-2 data show clear changes in the magnetic field orientation.  
%
%
These are not in contradiction with the uniform orientation and
almost constant polarisation fraction seen by Planck, because of the
difference in the beam sizes and the POL-2 data being affected by
spatial filtering.
The filament has a peak column density
of $N({\rm H}_2)\sim 7\times 10^{22}$\,cm$^{-2}$, a minimum dust
temperature of $T\sim$12\,K, and a mass of $\sim$4300\,M$_{\sun}$ for
the area $N({\rm H}_2)> 5\times 10^{21}$\,cm$^{-2}$.
The estimated average value of the dust opacity spectral index is
$\beta \sim$ 1.9. The ratio of sub-millimetre and J band optical
depths is $\tau(250\mu{\rm m})/\tau({\rm J}) \sim 2.5\times 10^{-3}$,
more than four times the typical values for diffuse medium.
The polarisation fraction decreases as a function of column density to
$p\sim$1\% in the central filament. Because of noise, the
observed decrease of $p(N)$ is significant only at $N({\rm H}_2)> 2
\times 10^{22}$\,cm$^{-2}$. The observations suggest that the grain
alignment is not constant. Although the data can be explained with a
complete loss of alignment at densities above $\sim 10^4$\,cm$^{-3}$ or
using the predictions of radiative torques alignment, the uncertainty
of the field geometry and the spatial filtering of the SCUBA-2 data
prevent strong conclusions.
}
{ 
The \target filament shows strong signs of dust evolution and the low
polarisation fraction is suggestive of a loss of polarised emission
from its densest parts.
}


\maketitle

\section{Introduction} \label{sect:intro}

Filamentary structures play an important role in star formation, from
cloud formation to the birth of clumps and gravitationally bound
pre-stellar cores. Filaments range from infrared dark clouds (IRDCs),
with lengths up to tens of parsecs \citep{Elmegreen1979,Egan1998,
Goodman2014, Wang2015}, to the parsec-scale star-forming filaments of
nearby molecular clouds
\citep{Bally1987,Andre2010,Menshchikov2010,Arzoumanian2011,Hill2011,Schneider2012,
Hennemann2012,GCC-III,Andre2014,GCC-VII}, and further down in linear
scale to thin fibres as sub-structures of dense filaments
\citep{Hacar2013, FernandezLopez2014,Hacar2018} and to 
low-column-density striations
\citep{Palmeirim2013,Cox2016,Heyer2016,Miettinen2018}.

Most likely all filaments do not have a common origin. The formation
of an individual structure can be the result of random turbulent
motions \citep{BallesterosParedes1999,Padoan2001,Klassen2017,
LiKleinMcKee2018}, cloud-cloud collisions, triggering by external
forces \citep{Hennebelle2008, Federrath2010, WuTanNakamura2017,
Anathpindika2018,SCOPE,Liu_G35_pol}, or a combination of several
factors. The effects on star formation are closely connected to the
role that magnetic fields have in the formation of filaments and later
in the fragmentation and the support of gravitationally bound
structures.

Our knowledge of the magnetic fields in filamentary clouds is largely
based on polarisation, the optical and near-infrared (NIR)
polarisation observations of the light from background stars and the
polarised dust emission at far-infrared (FIR), sub-millimetre, and
radio wavelengths. 
The methods are partly complementary, extinction studies probing
diffuse regions and clouds up to visual extinctions of $A_{\rm V}\sim
20^{\rm m}$ \citep{Goodman1995,Neha2018,Kandori2018}, while emission
studies cover the range of $A_{\rm V} \sim 1-100^{\rm m}$
\citep{WardThompson2000,Liu_G35_pol, Pattle2017_BISTRO, Kwon2018}.
The magnetic field appears to be mainly (but not
perfectly) orthogonal to the main axis of some nearby filamentary
clouds such as the Musca \citep{Pereyra2004,Cox2016}, Taurus
\citep{Heyer1987, Goodman1990, planck2015-XXXV}, Pipe
\citep{Alves2008}, and Lupus I \citep{Matthews2014} molecular clouds.
This also means that the fainter striations, which tend to be
perpendicular to high-column-density filaments, are aligned with the
magnetic field orientation. It has been suggested that the striations
represent accretion onto the potentially star-forming filaments, the
inflow thus being funnelled by the magnetic fields
\citep{Palmeirim2013}. Studies with \Planck data have found that the
column density structures tend to be aligned with the magnetic field
in diffuse clouds while in the molecular clouds and at higher
densities the orthogonal configuration is more typical
\citep{planck2014-XXXII, planck2015-XXXV, Malinen2016, Alina2018}. The
orthogonal configuration was typical also for the dense clouds that
were observed with ground-based telescopes in \citet{Koch2014}. 
A similar trend in the relative orientations at low and high column
densities has been reported for numerical simulations
\citep{Soler2013,Klassen2017,LiKleinMcKee2018}.
The orthogonal geometry seems dominant even in the most massive
filaments and in regions of active star formation. However, the
situation can be complicated by the effects of local gravitational
collapse, stellar feedback, and the typically higher levels of
background and foreground emission
\citep{Santos2016,Pattle2017_BISTRO,Hoq2017}.

The polarisation fraction $p$ appears to be negatively correlated with
the column density \citep{Vrba1976, Gerakines1995, WardThompson2000,
Alves2014_pol, planck2014-XX} although sometimes the relation is
difficult to separate from the noise-induced bias that affects
observations at low signal-to-noise ratios (SNR). The column-density
dependence of $p$ has also been studied statistically in connection
with clumps and filaments \citep[][Ristorcelli et al., in
prep.]{planck2014-XXXIII}.
This raises the question whether the decrease is caused by a specific
magnetic field geometry (such as small-scale line tangling or changes
in the large-scale magnetic field orientation) or by factors related
to the grain alignment. The radiative torques (RAT) are a strong
candidate for a mechanism behind the grain alignment
\citep{Lazarian1997,HoangLazarian2014}. Because RAT require radiation
to spin up the dust grains, they naturally predicts a loss of
polarisation at high $A_{\rm V}$. The effect depends on the grain
properties and is thus affected by the grain growth that is known to
take place in dense environments \citep{Stepnik2003, Ysard2013,
Whittet2001, Voshchinnikov2013}. If RAT are the main cause of grain
alignment, it is difficult to produce any significant polarised
emission from very dense clumps and filaments \citep{Pelkonen2009}. On
the other hand, numerical simulations also have shown the significance
of geometrical depolarisation, which would still probe the magnetic
field configurations at lower column densities
\citep{planck2014-XX, Chen2016_B}.

We have studied the filamentary IRDC \targetn, which has a mass of some
17000\,$M_{\sun}$ \citep{KainulainenTan2013} and is located at a
distance of 2.9\,kpc \citep{Simon2006b}. The source corresponds to
PGCC~G35.49-0.31 in the \Planck catalogue of Galactic Cold Clumps
\citep{PGCC}. The field has been targeted by several recent studies in
both molecular lines and in continuum \citep[e.g.][]{Zhang2017,
Liu_G35_pol}. Although the single-dish infrared and sub-millimetre
images of \target are dominated by a single $\sim$5\,pc long
structure, high resolution line observations have revealed the
presence of velocity-coherent, $\sim$0.03\,pc wide sub-filaments or
fibres \citep{Henshaw2017}. The filament is associated with a number
of dense cores that, while being cold ($T\la$16\,K) and IR-quiet, may
have potential for future high-mass star formation
\citep{NguyenLuong2011,Liu_G35_pol}. There are a number of low
luminosity (Class 0) protostars but \target appears to be in an early
stage of evolution where the cloud structure is not yet strongly
affected by the stellar feedback. This makes \target a good target for
studies of dust polarisation. \citet{Liu_G35_pol} already discussed
the magnetic field morphology in \target based on POL-2 observations
made with the JCMT SCUBA-2 instrument.
\citet{Liu_G35_pol} estimated that the average plane-of-the-sky (POS)
magnetic field strength is $\sim 50\,\mu$G and the field might provide
significant support for the clumps in the filament against
gravitational collapse. The pinched magnetic field morphology in its
southern part was suggested to be related to accretion flows along the
filament.

In this paper we will use \Planckn, \Herscheln, and JCMT/POL-2
observations to study the structure, dust emission spectrum, and
polarisation properties of \target. In particular, we investigate the
polarisation fraction variations, its column-density dependence, and the
interpretations in terms of magnetic field geometry and grain
alignment efficiency. After describing the observations in
Sect.~\ref{sect:obs} and the methods in Sect.~\ref{sect:methods}, the
main results are presented in Sect.~\ref{sect:results}. These include
estimates of dust opacity (Sect.~\ref{sect:opacity}) and polarisation
fraction (Sect.~\ref{sect:polfrac}). The radiative transfer models for
the total intensity and for the polarised emission are presented in
Sect.~\ref{sect:RT}. We discuss the results in
Sect.~\ref{sect:discussion} before presenting the conclusions in
Sect.~\ref{sect:conclusions}.

\section{Observational data}  \label{sect:obs}

\subsection{JCMT observations} \label{sect:obs_SCUBA}

The observations with the JCMT SCUBA-2 instrument \citep{Holland2013} 
are described in detail in \citet{Liu_G35_pol}. We use the 850\,$\mu$m
(total intensity and polarisation) and 450\,$\mu$m (total intensity)
data. First total intensity observations were carried out in April
2016 as part of the SCOPE program \citep{SCOPE}.

The POL-2 polarisation measurements were made between June and
November 2017 using the POL-2 DAISY mapping mode (project code:
M17BP050; PI: Tie Liu).
The field was covered by two mappings, each covering a circular region
with a diameter of 12$\arcmin$. The maps were created with the pol2map
routine of the Starlink SMURF package. The final co-added maps have an
rms noise of $\sim$1.5 mJy/beam.
The map making employed a filtering scale of $\theta_{\rm
F}$=200$\arcsec$, which removes extended emission but results in good
fidelity to structures smaller than $\theta_{\rm F}$
\citep{Mairs2015}. For further detail of the observations, see 
\citet{Liu_G35_pol}.

We assume for SCUBA-2 a 10\% uncertainty, which covers the calibration
uncertainty as an absolute error relative to the other data sets.
The contamination of the 850\,$\mu$m band by CO($J=$3-2) could be a
source of systematic positive error. Although the CO contribution in
850\,$\mu$m measurements can sometimes reach tens of percent
\citep{Drabek2012}, it is usually below 10\% \citep[e.g.][]{Moore2015,
Mairs2016, Juvela_2018_pilot}. 
Parts of the \target field have been mapped with the JCMT/HARP
instrument (observation ID JCMT\_1307713342\_798901). The
$^{12}$CO(3-2) line area (in main beam temperature $T_{\rm MB}$)
towards the northern clump reaches 66\,K\,km\,s$^{-1}$. This
corresponds to a 8.3\,MJy\,sr$^{-1}$ (46\,mJy\,beam$^{-1}$)
contamination in the 850\,$\mu$m continuum value, which is some 8\% of
the measured surface brightness. However, this does not take into
account that observations filter out all large-scale emission. The
average $^{12}$CO signal at 2$\arcmin$ distance of this position is
still some $7$\,MJy\,sr$^{-1}$. When the large-scale emission is
filtered out, the residual effect on the 850\,$\mu$m surface
brightness should be $\sim$2\% or less and small compared to the
assumed total uncertainty of 10\%. Therefore, we do not apply any
corrections to the 850\,$\mu$m values.

The FWHM of the SCUBA-2 main beam is $\sim 14\arcsec$ at 850\,$\mu$m
and $\sim 8\arcsec$ at 450\,$\mu$m. Because the beam patterns include
a wider secondary component \citep{Dempsey2013}, we used Uranus
measurements (see Table~\ref{table:Uranus}) to derive spherically
symmetric beam patterns. The planet size, which was $\sim 3.7\arcsec$
at the time of the observations, has little effect on the estimated
beams and is not explicitly taken into account \citep[see
also][]{Pattle2015}.


\begin{table}
\caption[]{Observation IDs of the data used}
\begin{center}
\begin{tabular}{ll}
Observation &  Observation ID  \\
\hline
\hline
\targetn/SCUBA-2  &  scuba2\_00063\_20160413T170550  \\
\targetn/POL-2    &  scuba2\_00011\_20170814T073201  \\
Uranus/SCUBA-2    &  scuba2\_00021\_20171109T074149  \\
                  &  scuba2\_00027\_20171110T094533  \\
                  &  scuba2\_00042\_20171110T125528  \\
                  &  scuba2\_00032\_20171120T092415  \\
\hline            
\end{tabular}
\end{center}
\label{table:Uranus}
\end{table}

\subsection{\Herschel observations} \label{sect:obs_Herschel}

The \Herschel SPIRE data at 250\,$\mu$m, 350\,$\mu$m, and 500\,$\mu$m
were taken from the \Herschel Science Archive
(HSA)\footnote{http://archives.esac.esa.int/hsa/whsa/}.  We use the
level 2.5 maps produced by the standard data reduction pipelines and
calibrated for extended emission (the so-called Photometer Extended
Map Product). The observations ID numbers are 1342204856 and
1342204857 and the data were originally observed in the HOBYS
programme \citep{Motte2010}. 

The resolutions of the SPIRE observations are 18.4$\arcsec$,
25.2$\arcsec$, and 36.7$\arcsec$ in the 250\,$\mu$m, 350\,$\mu$m, and
500\,$\mu$m bands, respectively\footnote{The Spectral and Photometric
Imaging Receiver (SPIRE) Handbook,
http://herschel.esac.esa.int/Docs/SPIRE/spire\_handbook.pdf}. The beam
sizes and shapes depend on the source spectrum
\footnote{http://herschel.esac.esa.int/twiki/bin/view/Public/
SpirePhotometerBeamProfileAnalysis}. We use beams that are calculated
for a modified blackbody spectrum with a colour temperature of
$T$=15\,K and a dust emission spectral index of $\beta$=1.8.  The beam
shapes are not sensitive to small variations in $T$ and $\beta$
\citep{Griffin2013, GCC-VI} but could be less accurate for hot point
sources. We adopt for the SPIRE bands a relative uncertainty of 4\%.

The surface brightness scale of the archived \Herschel SPIRE maps have
an absolute zero point that is based on a comparison with \Planck
measurements (e.g. Fig.~\ref{fig:overview}). We convolved the maps
to 40$\arcsec$ resolution, fitted the data with modified blackbody (MBB)
curves with $\beta=1.8$, and used these spectral energy distributions
(SEDs) to colour correct the SPIRE and SCUBA-2 data. In the
temperature range of $T=10-20$\,K, the corrections are less than 2\%.
For example, the SPIRE colour corrections remain essentially identical
irrespective on whether the colour temperatures are estimated using the
total intensity or the background-subtracted surface brightness data
(see Sect.~\ref{sect:MBB_Herschel}).

We show some \Herschel maps from the PACS instrument
\citep{Poglitsch2010} but these data are not used in the analysis of
dust emission. At 70\,$\mu$m the filament is seen in absorption
(except for a number of point sources) and at 160\,$\mu$m the filament
is seen neither in absorption nor as an excess over the background
(see Fig.~\ref{fig:overview_2}). Even without this significant
contribution of the extincted background component, the inclusion of
shorter wavelengths would bias the estimates of the dust SED
parameters \citep[e.g.][]{Shetty2009a, Malinen2011, Juvela2012_Tmix}.

\subsection{Other data on infrared and radio dust emission} \label{sect:obs_Planck}

\Planck 850\,$\mu$m (353\,GHz) data are used to examine the dust
emission and the dust polarisation at scales larger than the $\sim
5\arcmin$ \Planck beam. The data were taken from the Planck Legacy
Archive\footnote{https://www.cosmos.esa.int/web/planck/pla} and
correspond to the 2015 maps \citep{planck2014-a01} where the CMB emission
has been subtracted.
We make no corrections for the cosmic infrared background (CIB)
because its effect ($\sim$0.13\,MJy\,sr$^{-1}$ at 353\,GHz
\citep{planck2014-XXIX}) is insignificant compared to the strong cloud
emission. The \Planck 850\,$\mu$m data has some contamination from CO
$J=3-2$ line emission. We do not correct for this, because the effect
is small and these data are not used for SED analysis \citep[see
also][]{GCC-VI}. The estimated effect of the (unpolarised) CO emission
on the polarisation fraction $p$ is not significant, $\sim$1\% or less
of the $p$ values.

\begin{figure*}
\includegraphics[width=18.0cm]{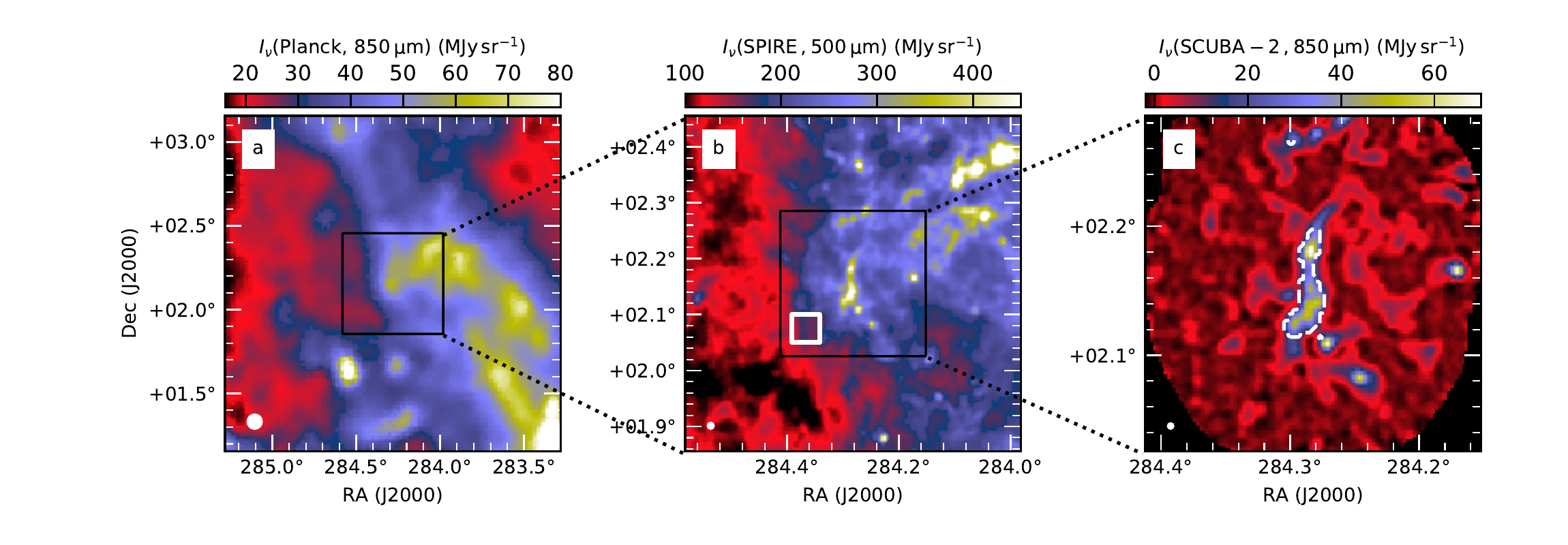}
\caption{
\Planck 850\,$\mu$m (353\,GHz), \Herschel 500\,$\mu$m, and
SCUBA-2/POL-2 850\,$\mu$m surface brightness maps of the \target
filament and its surroundings. The beam sizes are indicated in the
lower left corner of each frame (5$\arcmin$, 37$\arcsec$, and
14$\arcsec$, respectively). In frame b the white box indicates a
reference region for background subtraction. The dashed contour in
frame c is drawn at the level of 180\,MJy\,sr$^{-1}$ of the
background-subtracted 500\,$\mu$m surface brightness. 
}
\label{fig:overview}
\end{figure*}

Figure~\ref{fig:overview} shows \Planckn, \Herscheln, and SCUBA-2
surface brightness maps of the \target region.

\begin{figure}
\includegraphics[width=8.8cm]{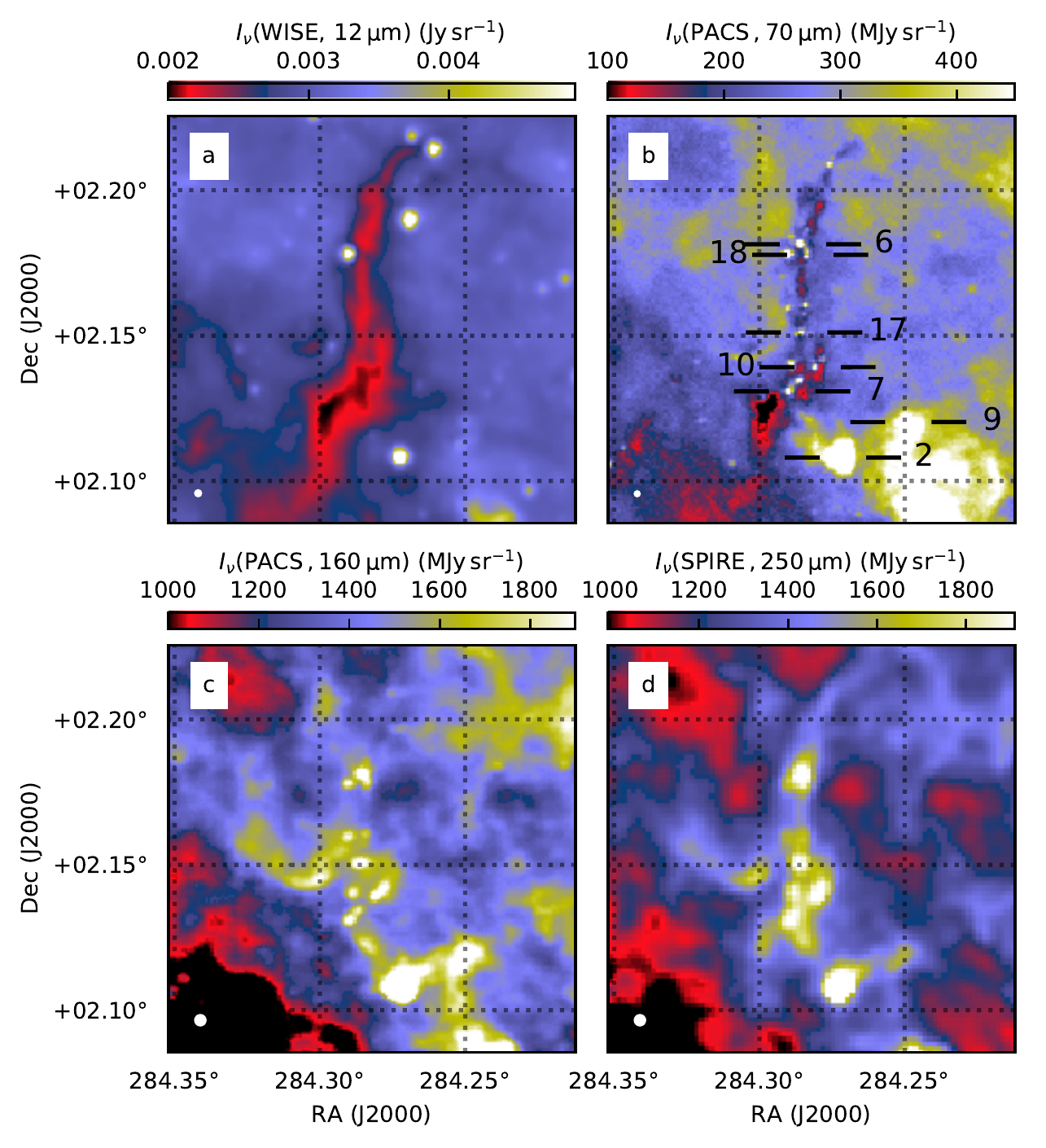}
\caption{
WISE 12\,$\mu$m, and \Herschel PACS instrument 70\,$\mu$m and
160\,$\mu$m and SPIRE instrument 250\,$\mu$m surface brightness images
of \target. In the 70\,$\mu$m image some of the highest-luminosity
sources from \citet{NguyenLuong2011} are indicated. 
}
\label{fig:overview_2}
\end{figure}

Figure~\ref{fig:overview_2} shows surface brightness images from
mid-infrared (MIR) to sub-millimetre wavelengths. In addition to
\Herschel data, the figure shows the 12\,$\mu$m surface brightness
from the WISE survey \citep{Wright2010}. The filament is seen in
absorption up to the 70\,$\mu$m band. The 160\,$\mu$m image is
dominated by warm dust and the main high-column-density structure is
not visible, before again appearing in emission at 250\,$\mu$m. The
ratio of the 100\,$\mu$m and 250\,$\mu$m dust opacities is $\sim$5,
which suggests (although does not directly prove) that the filament is
optically thin at 250\,$\mu$m. This is later corroborated by the
derived $\tau(250\,\mu{\rm m})$ estimates and by independent column
density estimates.

The 70\,$\mu$m image shows more than ten point-like sources that
appear to be associated to the main filament. Only one of them is
visible at 12\,$\mu$m, showing that they are either in an early stage
or otherwise heavily obscured by high column densities. The sources
can be identified also in the PACS 160\,$\mu$m image but not at
250\,$\mu$m, because of the lower resolution and lower sensitivity to
high temperatures, many of the sources are blended together or not
visible above the extended cold dust emission. The sources were
studied by \citet{NguyenLuong2011}, who also estimated their
bolometric luminosities. The sources with luminosity (or with an
estimated upper limit) above 100\,$L_{\sun}$ are marked in
Fig.~\ref{fig:overview_2}b and are listed in Table~\ref{table:PS_L}.
The numbering refers to that in \citet{NguyenLuong2011} Table~1. The
most luminous source \#2 is outside the main filament. The others have
bolometric luminosities of the order of $100\,L_{\sun}$. The low dust
temperatures indicate that the internal heating caused by these
(probably) embedded sources is not very significant.

\begin{table}
\caption{List of the most luminous sources in the \target field.}
\begin{center}
\begin{tabular}{rllccc}
\hline \hline 
\#\tablefootmark{a} & RA      & DEC        &  $T_{\rm dust}$     &   $M$          &  $L$         \\ 
                    & (J2000) &  (J2000)   &  (K)     &  ($M_{\sun}$)  & (L$_{\sun}$) \\ 
\hline                    
2  &  18:57:05.1 &   2:06:29  &   27$\pm$6  &     24$\pm$16  &   4700       \\ 
6  &  18:57:08.4 &   2:10:53  &   16$\pm$3  &     20$\pm$12  &   70-200     \\ 
7  &  18:57:09.3 &   2:07:51  &   12$\pm$1  &     49$\pm$17  &   50-130     \\ 
9  &  18:56:59.7 &   2:07:13  &   22$\pm$4  &     3 $\pm$1   &    70-120    \\ 
17 &  18:57:08.3 &   2:09:04  &   13$\pm$2  &     50$\pm$19  &   50-140     \\ 
18 &  18:57:07.8 &   2:10:40  &   14$\pm$2  &     20$\pm$9   &   40-120     \\ 
\hline
\end{tabular}
\tablefoot{
\tablefoottext{a}{Source numbering from the Table~1 of \cite{NguyenLuong2011}.}
}
\end{center}
\label{table:PS_L}
\end{table}

\subsection{Extinction data} \label{sect:obs_AV}

\citet{KainulainenTan2013} calculated for the \target region 
high-dynamical-range extinction maps using a combination of NIR
observations of reddened background stars and the MIR extinction of
extended emission. A NIR extinction map was made at 30$\arcsec$
resolution using UKIDSS data \citep{Lawrence2007} and an adaptation of
the NICER method in \citet{Kainulainen2011_NICER} \citep[see
also][]{Lombardi2001}. The assumed extinction curve has
$\tau(V)=3.54\,\tau(J)$ \citep{Cardelli1989}. The MIR extinction was
measured using Spitzer 8\,$\mu$m images from the GLIMPSE survey
\citep{Butler2012}. This enabled the extension of the estimates to
higher column densities and down to a nominal resolution of
2$\arcsec$. The MIR data suffer from spatial filtering (low
sensitivity to extended structures) and exhibit some differences
relative to the NIR data that could be caused by fluctuations in the
brightness of the background (or foreground).
\citet{KainulainenTan2013} compensated for these effects by combining
the two data sets into a single $A_{\rm V}$ map. The correlation
between the NIR and MIR data was best in the $A_{\rm V}=10-15$\,mag
range while at higher column densities the NIR estimates are, as
expected, smaller because the background stars do not provide a good
sampling of the highest column densities. 
The morphology and relative extinction values in the combined
extinction map are not dependent
on a priori assumptions of the absolute dust opacities but do depend
on the assumed opacity ratio of $\kappa(8\,\mu{\rm m})/\kappa({\rm
K})=0.29$. 

We converted the \citet{KainulainenTan2013} $A_{\rm V}$ estimates
(data provided by J. Kainulainen) to J band optical depth $\tau({\rm
J})$ using the opacity ratio quoted above. In Sect.~\ref{sect:opacity}
the $\tau({\rm J})$ map will be compared to observations of dust
emission.

\section{Methods} \label{sect:methods}

\subsection{Column density estimates} \label{methods:MBB}

Basic column density estimates can be derived via modified blackbody
(MBB) fits that model the observed intensities as
\begin{equation}
I_{\nu}(\nu) = I_{\nu}(\nu_0) \frac{B_{\nu}(\nu, T)}{B_{\nu}(\nu_0, T)}
       \left( \frac{\nu}{\nu_0} \right)^{\beta},
\label{eq:MBB}
\end{equation}
where $B_{\nu}$ is the Planck law, $T$ the colour temperature, $\beta$
the dust opacity spectral index, $I_{\nu}(\nu)$ the observed
intensities, and $I_{\nu}(\nu_0)$ the intensity at the reference
frequency $\nu_0$. In the MBB fit the free parameters are
$I_{\nu}(\nu_0)$, $T$, and $\beta$, although in many cases the
spectral index $\beta$ is kept fixed. With an assumption of the value
of $\kappa(\nu_0)$, the dust opacity relative to the total gas mass,
the MBB result can be converted to estimates of the column density,
\begin{equation}
N({\rm H}_2) = \frac{I_{\nu}(\nu_0)}{B_{\nu}(T) \kappa(\nu_0) \, \mu
m_{\rm H}}.
\label{eq:colden}
\end{equation}
Here $\mu$ is the total mass per Hydrogen molecule and $\mu=2.8$
atomic mass units the total gas mass per hydrogen molecule. The mass
surface density (g/cm$^2$) is $\Sigma = N({\rm H}_2) \, \mu$. We adopt
dust opacities $\kappa(\nu) =
0.1\,(\nu/1000\,GHz)^{\beta}$\,cm$^2$\,g$^{-1}$ \citep{Beckwith1990,
GCC-III}.
The above assumes that the observed intensities can be represented by
a single MBB formula like in Eq.~(\ref{eq:MBB}). This is not generally
true and, in particular, leads to an underestimation of the column
densities of non-isothermal sources \citep{Shetty2009a,
Malinen2011,Juvela2012_Tmix,Juvela2013_colden}.
Equation~(\ref{eq:MBB}) also explicitly assumes that the emission is
optically thin, which is probably the case for \target observations at
wavelengths $\lambda \ge 250$\,$\mu$m. For optically thick emission
the column density estimates would always be highly unreliable and the
use of the full formula instead of the optically thin approximation of
Eq.~(\ref{eq:MBB}) is not likely to improve the accuracy
\citep{Malinen2011,Menshchikov2016}.

If all the maps used in the fits are first convolved to a common
low resolution, the previous formulas provide column density maps at
this resolution. We also made column density maps at a higher
resolution by making a model that consisted of high-resolution
$I_{\nu}(\nu_0)$ and $T$ maps, keeping the spectral index $\beta$
constant. This model provides predictions at the observed frequencies
according to Eq.~(\ref{eq:MBB}). Each model-predicted map was
convolved to the resolution of the corresponding observed map using
the convolution kernels described in Sects.~\ref{sect:obs_SCUBA} and
\ref{sect:obs_Herschel}. The minimisation of the weighted least
squares residuals provided the final model maps for $I_{\nu}(\nu_0)$
and $T$. The free parameters thus consisted of the intensity values
$I_{\nu}(\nu_0)$ and the temperature values $T$ of each pixel of the
model maps. The chosen pixel size was 6$\arcsec$, more than two times
smaller than the resolution of the observed surface brightness maps.
Because the solution at a given position depends on the solution at
nearby positions, the $I_{\nu}(\nu_0)$ and $T$ maps need to be
estimated through a single optimisation problem rather than for each
pixel separately.
The optimised model maps were convolved to FWHM$_{\rm MOD}$ and were
then used to calculate column density maps at that same resolution. We
used FWHM$_{\rm MOD}$=20$\arcsec$, when fitting \Herschel data, and
FWHM$_{\rm MOD}$=15$\arcsec$, when fitting combined \Herschel and
SCUBA-2 observations. The procedure is discussed further in
Appendix~\ref{app:MBB}. Because the method is simply fitting the
observed surface brightness values, it is still subject to all the
caveats regarding the line-of-sight (LOS) temperature variations.

\subsection{Polarisation quantities} \label{methods:pol}

The polarisation fraction could be calculated as
\begin{equation}
p = \frac{\sqrt{ Q^2+U^2 }}{I},
\end{equation}
but this estimate is biased because of observational noise and because
$p$ depends on the squared sum of $Q$ and $U$. Therefore, we use the
modified asymptotic estimator of \citet{P14},
\begin{equation}
p_{\rm mas} = p - b^2 \frac{ 1-\exp(-p^2/b^2) }{2p},
\end{equation}
where $b^2$ is 
\begin{equation}
b^2 = \frac{
\sigma_U^{\prime 2} \cos^2(2\psi_0-\theta) + \sigma_Q^{\prime 2} (2\psi_0-\theta)
}{I_0^2} \, ,
\label{eq:b2}
\end{equation}
with 
\begin{equation}
\theta = \frac{1}{2} \mathrm{atan} \left(
\frac{2 \sigma_{QU}}{\sigma_Q^2-\sigma_U^2}
\right)\, ,
\end{equation}
\begin{equation}
\sigma_Q^{\prime 2} = 
\sigma_Q^2 \cos^2\theta + \sigma_U^2 \sin^2\theta
+ \sigma_{QU} \sin 2\theta,
\end{equation}
\begin{equation}
\sigma_U^{\prime 2} = 
\sigma_Q^2 \sin^2\theta + \sigma_U^2 \cos^2\theta
- \sigma_{QU} \sin 2\theta.
\end{equation}
In Eq.~(\ref{eq:b2}) $\psi_0$ stands for the true polarisation angle
and is in practice replaced by its estimate (see below). The error
estimates of $p$ are calculated from
\begin{equation}
\sigma_{p, \rm \,mas} = 
\sqrt{ \sigma_Q^{\prime 2}  \cos^2(2\psi-\theta) + 
       \sigma_U^{\prime 2}  \sin^2(2\psi-\theta)  }/I.
\end{equation}
\citep{P14, Montier2015b}. 
The $p_{\rm mas}$ estimator is reliable at $p_{\rm mas}/\sigma_{p, \rm
mas}>2$ \citep{Montier2015b}.
In this paper, polarisation fractions are calculated using the $p_{\rm
mas}$ estimator, both in the case of real observations and in the
simulations of Appendix~\ref{app:sim}. The only exception is the
analysis of radiative transfer models (Sect.~\ref{sect:RT}), because
these are free of noise that could affect the $p$ estimates.

The polarisation angle depends on Stokes $Q$ and $U$ as
\begin{equation}
\psi = 0.5 \arctan(U, Q).
\end{equation}
We use the IAU convention where the angle increases from north towards
east. The estimated POS magnetic field orientation
is obtained by adding $\pi/2$ radians to $\psi$. The uncertainties of
$\psi$ are estimated as
\begin{equation}
\sigma_{\psi} = 
\sqrt{
\frac{Q^2\sigma_U^2 + U^2\sigma_Q^2 - 2QU\sigma_{QU}}
{Q^2\sigma_Q^2+U^2\sigma_U^2+2QU\sigma_{QU}}
}
\, \, \frac{\sigma_p}{2p} \,{\rm rad},
\end{equation}
based on the error estimates of the Stokes parameters $\sigma_Q$ and
$\sigma_U$ and the covariance between Stokes $Q$ and $U$,
$\sigma_{QU}$ \citep{P14, Montier2015b}. All the quantities in the
above formulas are available from the data reduction except for
the SCUBA-2 covariances $\sigma_{QU}$, which are set to
zero. \citet{Montier2015a} note that the $\psi$ error estimates are
reliable for SNR>4 but can be strongly underestimated for lower SNR
because of the bias of the $p$ parameter.

The uniformity of the polarisation vector orientations and thus the
regularity of the underlying magnetic field can be characterised with
the polarisation angle dispersion function $S$
\citep{planck2014-XIX}. It is calculated as a function of position
$\bar r$ as
\begin{equation}
S(\bar r, \bar \delta) = 
\sqrt{  
\frac{1}{N}  \sum_{i=1}^N \left( \psi(\bar r) - \psi(\bar r + \bar \delta_i) \right)^2
}.
\label{eq:S}
\end{equation}
Here $\bar \delta_i$ is an offset for $N$ map pixels at distances
[$\delta/2$, $3\delta/2$] from the central position $\bar r$. The
scalar $\delta$ thus defines the spatial scale at which the dispersion
is estimated.  We set the $\delta$ values according to the present
data resolution as $\delta=$FWHM/2.
The angle difference is calculated directly from the
Stokes parameters as
\begin{equation}
\psi(\bar r) - \psi(\bar r + \bar \delta_i) = \arctan(Q_r
U_{\delta}-Q_{\delta} U_r, Q_r
Q_{\delta}+U_r U_{\delta})/2,
\end{equation}
where the indices $r$ and $\delta$ refer to the positions $\bar r$ and
$\bar r + \bar \delta_i$, respectively.
In the convolution of the Stokes vector images and in the calculation
of the polarisation angle dispersion function, we take into account
the rotation of the polarisation reference frame as described in
Appendix~A of \citet{planck2014-XIX}. However, these corrections are
not very significant at the angular scales discussed in this paper.
All $S$ values presented in this paper are bias-corrected as
$\sqrt{S^2 - \sigma(S)^2}$, where $\sigma(S)$ is the estimated
uncertainty for $S$ in Eq.(\ref{eq:S}) \citep{planck2014-XIX}.

\subsection{Radiative transfer models} \label{methods:RT}

We complemented the analysis described in Sects.~\ref{methods:MBB} and
\ref{methods:pol} with radiative transfer (RT) calculations. These
have the advantage of providing a more realistic description of the
temperature variations and, in the case of polarisation, allow the
explicit testing of the effects of imperfect grain alignment and
different magnetic field geometries.

The models cover an area of $13\arcmin \times 13\arcmin$ on the sky
with a regular grid where the size of the volume elements corresponds
to 6$\arcsec$. The LOS density profile was assumed to have a
functional form of $n(z) \propto (1+(z/R)^2)^{p/2}$, where $z$ is the
LOS coordinate. With parameters $R=0.16$\,pc and $p=2$ this gives for
the filament similar extent in the LOS direction as observed in the
POS. Such a short LOS extent is appropriate only for the densest
regions. Therefore, we used a scaled LOS coordinate $z/z_{N}$ where
$z_{N}$ is linear with respect to the logarithm of the column density
and increases from 1 for $N>5\times 10^{22}$\,cm$^{-2}$ to 5 for a
factor of ten smaller column densities.

The RT model initially corresponded to the column densities estimated
from MBB fits at 40$\arcsec$ resolution. The cloud was illuminated by
the normal interstellar radiation field (ISRF) according to the
\citet{Mathis1983} model. The dust properties were taken from
\citet{Compiegne2011} but the dust opacity at wavelengths
$\lambda>100\mu$m were increased to give $\tau(250\,\mu{\rm
m})/\tau({\rm J})$ ratios of $10^{-3}$ or $1.6\times 10^{-3}$. The
extinction curve was rescaled to give the same $\kappa(250\,\mu{\rm
m})$ value as quoted in Sect.~\ref{methods:MBB}. The latter scaling
has no real effect on the RT modelling itself but simplifies the
comparison with values derived from observations.

The models were optimised to match a set of surface brightness
observations. The free parameters included the scaling of the column
densities, one parameter per a 6$\arcsec$ map pixel, and the scaling
of the external radiation field, $k_{ISRF}$. 
The \target region includes a number of radiation sources with
luminosities $\sim 10\,{\rm L}_{\sun}$ or less. Because their location
along the line of sight is not known, the qualitative effects of
internal heating were tested by including in the model an optional
diffuse emission component. The diffuse emission has the same spectrum
as the external radiation field and it was scaled with a parameter
$k_{\rm diff}$, the value of 1 corresponding to a bolometric
luminosity of 1\,$L_{\sun}\,{\rm pc}^{-3}$. 

The radiative transfer problem was solved with the Monte Carlo program
SOC \citep[Juvela et al. in prep.;][]{TRUST-I}. Because the fitted observations are at
long wavelengths $\lambda \ge 250\,\mu$m, the dust grains were assumed
to be in equilibrium with the radiation field and the emission from
stochastically heated grains was omitted. SOC calculates the dust
temperatures based on the radiative transfer simulation and writes out
surface brightness maps at the requested wavelengths.

SOC can be used to calculate estimates of the polarised dust emission.
This was done using grain alignment that was either constant, had an ad 
hoc density-dependence, or was predicted by RAT calculations
\citep{Lazarian_Hoang_2007}. For the RAT case, the radiative transfer
modelling provided the intensity and anisotropy of the radiation
field, which were then used to estimate the minimum size of aligned
grains and thus a reduction factor $R$ for the polarised emission
originating in each model cell. The calculations were done as
described in \citet{Pelkonen2009}. The polarisation signal is
dependent on the minimum size of the grains that remain aligned in a
magnetic field. This is dependent on the ratio between the angular
velocity produced by the radiation field and the thermal rotation
rate,
\begin{equation}
\left(  \frac{\omega_{\rm rad}}{\omega_{\rm T}} \right)^2
\propto  
\frac{a}{(n_{\rm H} T )^2} 
\left[
\int  (Q_{\Gamma} \cdot \hat a) \lambda J_{\lambda} d \lambda 
\right]^2,
\label{eq:RAT}
\end{equation}
where $n_{\rm H}$ is the volume density, $T$ the temperature, $a$ the
grain size, $Q_{\Gamma}$ the wavelength-dependent efficiency of RAT
(dependent on the grain properties), $\hat a$ the unit vector of the
rotational axis, and $J_{\lambda}$ the radiation field intensity.
Thus, grain alignment is promoted by larger grain sizes and larger
intensity and anisotropy of the radiation field. Conversely, higher
density and temperature tend to reduce the grain alignment and
subsequently the polarised intensity. 

Given a model of the 3D magnetic field within the model volume, SOC
gives synthetic maps for $I$, $Q$, and $U$. We used these to examine the
effect that imperfect grain alignment can have on the observed
polarisation fraction distributions. For comparison with the full
calculations with grain alignment, synthetic maps were also 
produced assuming a constant value of $R$ or an ad hoc density
dependence of $R$.

\section{Results} \label{sect:results}

\begin{figure}
\includegraphics[width=8.8cm]{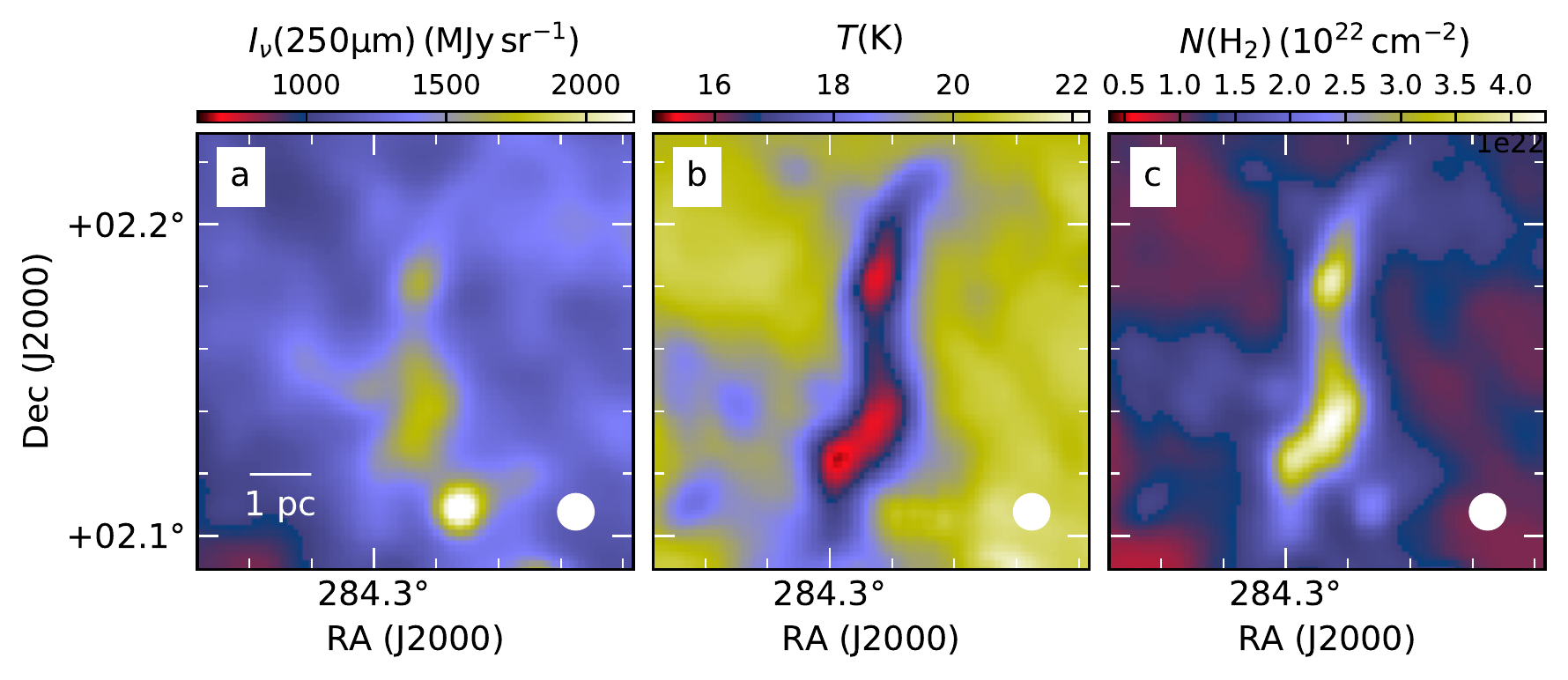}
\caption{
Results of modified blackbody fits to SPIRE data at 40$\arcsec$
resolution and without background subtraction: 250\,$\mu$m intensity
(frame a), colour temperature (frame b), and column density (frame c).
}
\label{fig:SPIRE_MBB}
\end{figure}

\subsection{\Herschel data} \label{sect:MBB_Herschel}

Figure~\ref{fig:SPIRE_MBB} shows the results of MBB fits using SPIRE
surface brightness maps at 40$\arcsec$ resolution. The fits were done
to data before background subtraction and thus correspond to emission
from the full LOS. The extended cloud component has a significant
contribution of almost $N({\rm H}_2)= 10^{22}$\,cm$^{-2}$ to the total
column density. The peak column densities of both the northern and
the southern parts are $N({\rm H}_2) \sim 4\times 10^{22}$\,cm$^{-2}$.
The colour temperatures are 20-21\,K in the background, below 18\,K
within the dense filament ($N({\rm H}_2) > 2\times
10^{22}$\,cm$^{-2}$), and reach minimum values of 15.5\,K and 15.2\,K
in the northern and southern clumps, respectively.
PACS data were not used (see Sect.~\ref{sect:obs_Herschel}), but at
the temperatures of the main filament ($T\sim$15\,K), the \Herschel
250\,$\mu$m, 350\,$\mu$m, and 500\,$\mu$m SPIRE bands give reliable
measurements of the dust colour temperature \citep[see][]{GCC-III}. On
the other hand, they do not give strong simultaneous constraints for
both the colour temperature and the spectral index. Therefore, the
SPIRE data were fitted using a constant value of $\beta$=1.8.

\begin{figure}
\includegraphics[width=8.8cm]{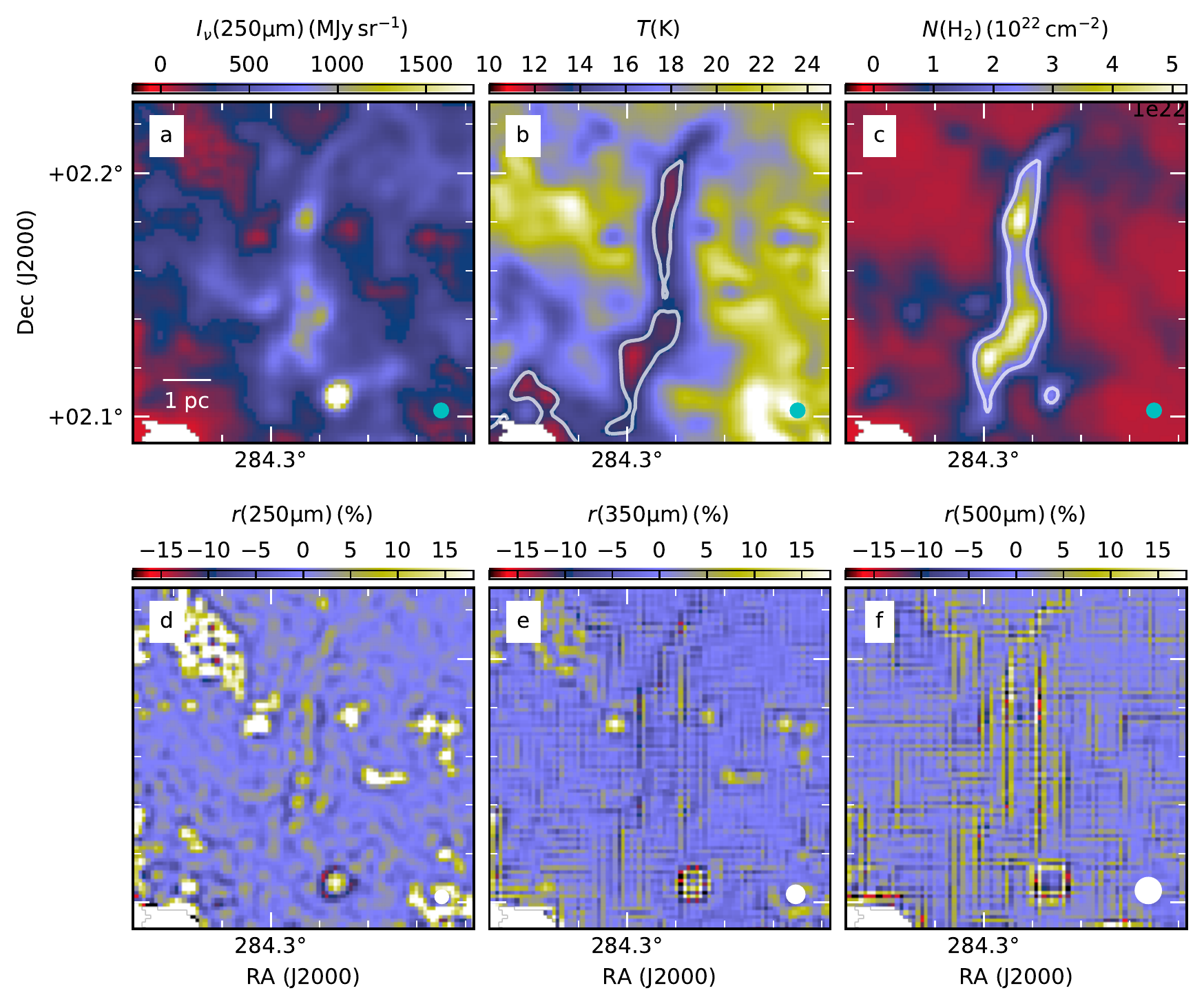}
\caption{
Results of modified blackbody fits to SPIRE data at a resolution of
20$\arcsec$. The upper frames show the fitted 250\,$\mu$m intensity,
dust colour temperature, and column density. The lower frames show the
relative error of the fits for the individual SPIRE bands. We have
masked a region in the SE corner where the background-subtracted
250\,$\mu$m intensities are below 10\,MJy\,sr$^{-1}$. The beam sizes
of the observations are shown in the bottom right-hand corner of each
frame. The contours are drawn at $T$=13.5\,K and $N({\rm H}_2)=2\times
10^{22}\,{\rm cm}^{-2}$.
}
\label{fig:OPT2D_SPIRE}
\end{figure}

We created column density maps at a resolution of 20$\arcsec$, as
described in Sect.~\ref{methods:MBB} using background-subtracted data.
The background was determined as the average signal in a $3\arcmin
\times 3\arcmin$ area centred at RA=18$^{\rm h}$57$^{\rm m}$28$^{\rm
s}$, DEC=2$\degr4\arcmin 30\arcsec$ (see Fig.~\ref{fig:overview}b).
Compared to Fig.~\ref{fig:SPIRE_MBB}, the filament is colder, mainly
because of the background subtraction (see
Fig.~\ref{fig:OPT2D_SPIRE}). The minimum temperatures are 12.4\,K in
the northern and 11.7\,K in the southern part (13.7\,K and 12.7\,K,
respectively, if this map is convolved down to 40$\arcsec$
resolution). At the 20$\arcsec$ resolution the fitted
$I_{\nu}(250\,\mu{\rm m})$ map shows local maxima at the positions of
the MIR sources (Fig.~\ref{fig:overview_2}b) but are not similarly
visible in column density. In spite of the background subtraction, the
peak column densities are higher, slightly above $5\times
10^{22}$\,cm$^{-2}$ in both the northern and the southern parts. This
is a consequence of the lower colour temperatures. The column
densities are probably still underestimated because of LOS temperature
variations. We will refer to this version of the column density map as
$N_3({\rm H}_2)$, the sub-index referring to the number of bands
fitted.

Unlike in the standard MBB fits that are done for each pixel
separately, Fig.~\ref{fig:OPT2D_SPIRE} corresponds to a global fit
over the map. The fit residuals (Fig.~\ref{fig:OPT2D_SPIRE}d-f) are
dominated by small-scale artefacts (below the beam size) that are
connected with the finite pixel size and possibly with imperfections
in the beam model. If these residual maps are convolved to the
resolution of the observations, they are smooth with peak-to-peak
errors below $\sim$4\%.

\subsection{Combined \Herschel and SCUBA-2 data} \label{sect:MBB_all}

\begin{figure}
\includegraphics[width=8.8cm]{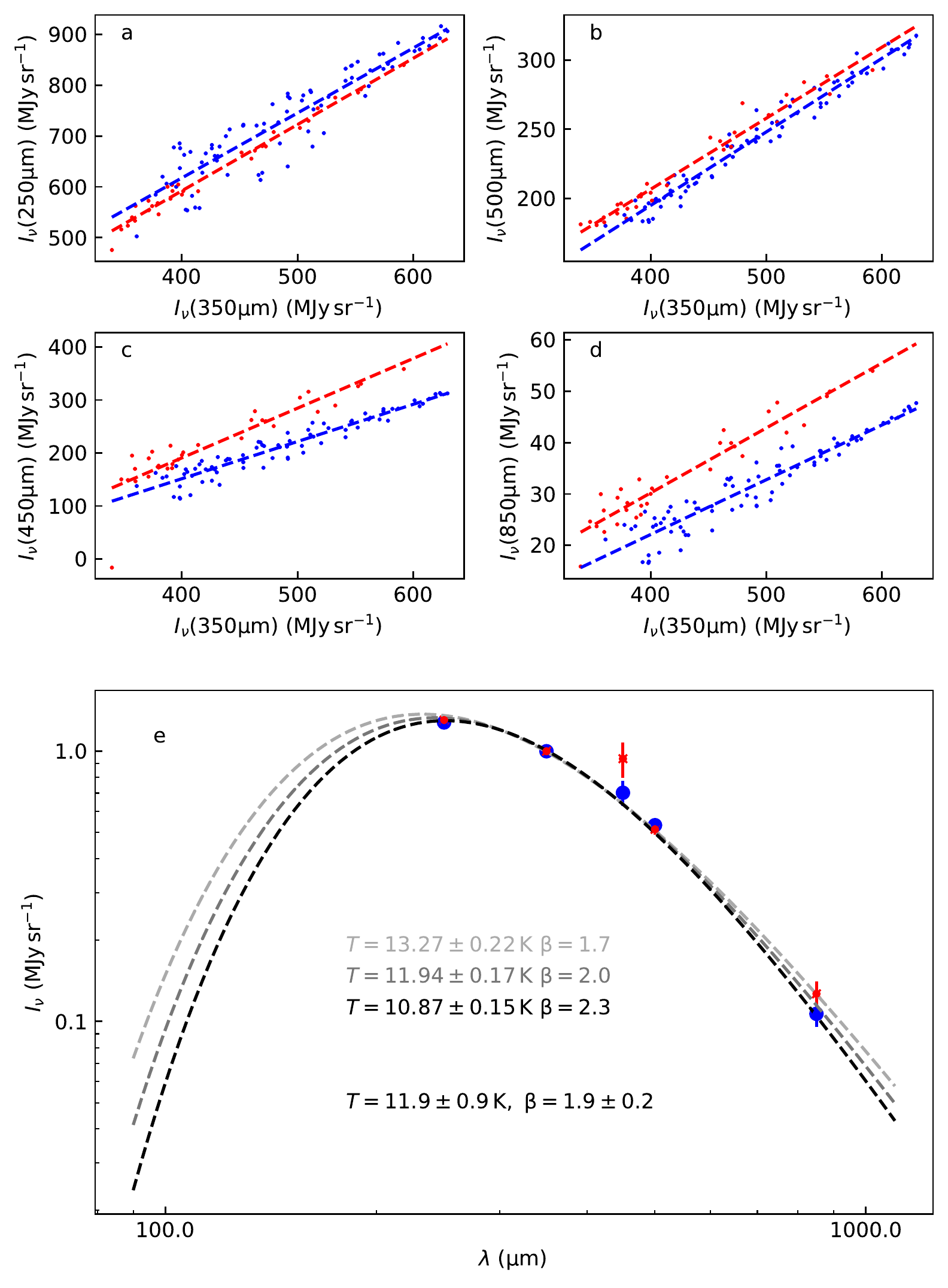}
\caption{
SEDs from the correlations of surface brightness values over the main
filament. Frames a-d show the band-to-band correlations for the
northern (red) and southern (blue) parts of the filaments. The points
correspond to an oversampling by a factor of two relative to the data
resolution. Frame e shows the resulting SEDs and modified blackbody
fits to the combined data for three fixed values of the spectral index
$\beta$. The result for a fit with free $\beta$ are shown below the
results of the fixed-$\beta$ fits.
}
\label{fig:SED_corr}
\end{figure}

We estimated the average SED of the main filament using band-to-band
correlations. We selected data where the background-subtracted SPIRE
500\,$\mu$m values were above of 180\,MJy\,sr$^{-1}$ (see
Fig.~\ref{fig:overview}c), further dividing the filament into a northern
and a southern part along $\delta=2\degr10\arcmin$. The data were
convolved to the resolution of the 500\,$\mu$m band, each band was
correlated with the 350\,$\mu$m data, and the uncertainties of the
linear fits were estimated with bootstrapping. The correlations in the
northern and the southern regions and the SED fit to the combined data
are shown in Fig.~\ref{fig:SED_corr}.

The data were fitted with MBB functions using the Markov chain Monte
Carlo method and flat priors with temperatures in the range 7-30\,K
and spectral indices in the range 0.5-3.5. Fits to all five bands gave
$T=11.29 \pm 0.83$\,K, $\beta=2.06 \pm 0.22$
for the southern part,
$T=12.46 \pm 0.95$\,K, $\beta=1.82 \pm 0.24$
for the northern part, and
$T=11.92 \pm 0.87$\,K, $\beta=1.94 \pm 0.22$
for the combined data. In this last case, the fitted SED consisted of
the weighted average of the SEDs points of the southern and northern
parts.
The effects from the spatial filtering of the SCUBA-2 data should be
small because the selected data only cover a $\sim 1.5 \arcmin$ wide
part of the filament. The 450\,$\mu$m SCUBA-2 point of the
northern region is significantly above the fitted SED. However, if
this point is omitted, the spectral index estimate remains almost
unchanged, $\beta=1.95$. The fit to the three SPIRE channels without
SCUBA-2 data gave $T=12.66 \pm 1.42$\,K, $\beta=1.76 \pm 0.33$.


We fitted the SPIRE and the 850\,$\mu$m data also with a model that
had one free parameter for $\beta$, one free parameter for the
relative offset of the 850\,$\mu$m surface brightness map, and one
free parameter for the colour temperature in each pixel.  We used the
same relative uncertainties as above but further assumed a correlation
$\rho=0.5$ between the errors of the SPIRE channels. Unlike in the
previous surface brightness correlations, the fit relies on the
consistency of the intensity zero points of the background-subtracted
SPIRE maps. The results from Markov chain Monte Carlo calculations
were 
$\beta=1.84 \pm 0.02$ for the southern part and
$\beta=1.69 \pm 0.02$ for the northern part. 
These are close to the SPIRE-only fits, partly because the SCUBA-2
data have less leverage on the $\beta$ values once the 850\,$\mu$m
surface brightness offset is included as a separate free parameter.
All error estimates above correspond to the 4\% (SPIRE) and 10\%
(SCUBA-2) uncertainties of the surface brightness measurements. The
true uncertainties can be larger because of the systematic errors.

We fitted the three SPIRE and two SCUBA-2 bands together to derive
maps of dust colour temperature and of optical depth at 15$\arcsec$
resolution, under the assumption of $\beta=1.8$. The optimisation
procedure is the same as in Sect.~\ref{sect:MBB_Herschel} (see
Sect.~ref{methods:MBB}). We used background-subtracted SPIRE data but
also had to correct the zero points of the SCUBA-2 data. This was done
by taking the predictions of SPIRE fits with $\beta=1.8$ at the
wavelengths 450\,$\mu$m and 850\,$\mu$m and comparing these to the
SCUBA-2 maps at the same resolution. The 450\,$\mu$m offset was
calculated using the average surface brightness values of the pixels
where the original SCUBA-2 450\,$\mu$m value was above
160\,MJy\,sr$^{-1}$. For the 850\,$\mu$m map the corresponding
threshold was 30\,MJy\,sr$^{-1}$. These offset-corrected maps were
used as additional constraints in the area where the signal was above
the quoted surface brightness thresholds. This means that SCUBA-2 data
were used over a narrow region around the main filament where the loss
of low spatial frequencies should be small. Because the offsets were
based on the SPIRE SEDs, these data cannot be used to draw any
conclusions on the SED shape at wavelengths beyond 500\,$\mu$m. The
SCUBA-2 data only provide additional constraints on the small-scale
column density structure. The resulting 250\,$\mu$m optical depth
estimates are referred to as $\tau_5(250\,\mu{\rm m})$ and the column
density estimates as $N_5({\rm H}_2)$.

\begin{figure*}
\sidecaption
\includegraphics[width=12cm]{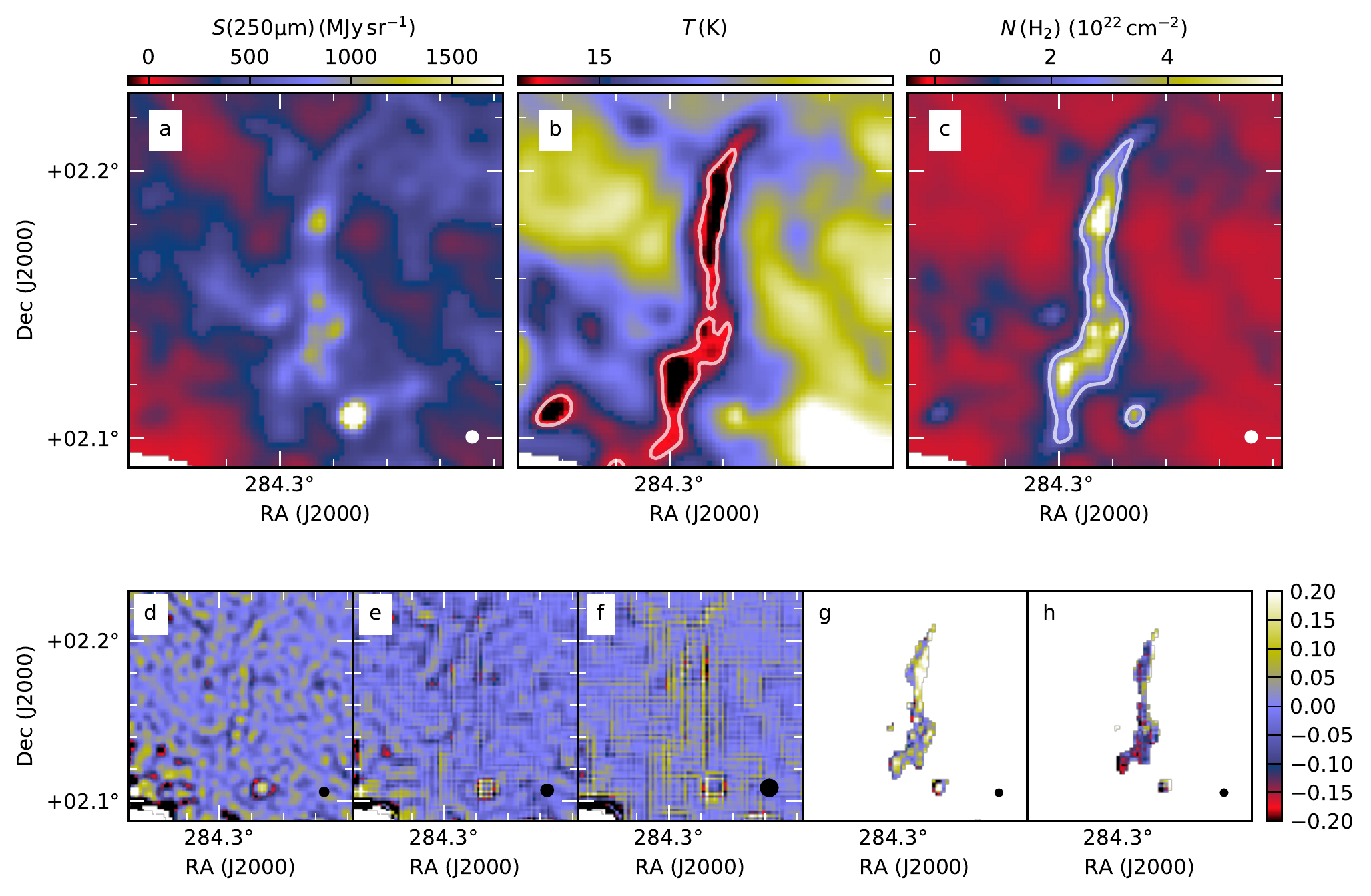}
\caption{
Modified blackbody fit to the combined SPIRE and SCUBA-2 data at
15$\arcsec$ resolution, using SCUBA-2 data as a constraint only over
the main filament. Frames a-c show the fitted 250\,$\mu$m surface
brightness, the colour temperature, and the estimated column density.
The frames d-h show the residuals relative to the observed,
background-subtracted surface brightness values at 250, 350, 500, 450,
and 850\,$\mu$m, respectively. The fits used $\beta=1.8$.
The contours are drawn at $T$=13.5\,K and $N({\rm H}_2)=2\times
10^{22}\,{\rm cm}^{-2}$.
}
\label{fig:OPT2D_ALLX}
\end{figure*}

The results are shown in Fig.~\ref{fig:OPT2D_ALLX} at the resolution
of FWHM=15$\arcsec$ (Gaussian beam). In principle, this is the
resolution also outside the main filament, where SCUBA-2 data were not
used. However, there FWHM=15\,$\arcsec$ corresponds to a deconvolution
below the SPIRE resolution and the small-scale structure is not
reliable. The peak column densities are $7.2\times 10^{22}$\,cm$^{-2}$
and $7.0\times 10^{22}$\,cm$^{-2}$ for the northern and the southern
part, respectively. Unlike in Fig.~\ref{fig:OPT2D_SPIRE}, there are
several local column density maxima NW of the southern clump that are
related to the 70--250\,$\mu$m sources of Fig.~\ref{fig:overview_2}.
They are more visible because of the higher resolution (15$\arcsec$
vs. 20$\arcsec$). However, if the effective resolution of the fitted
temperature map (which is dependent on longer-wavelength SPIRE
channels) is lower than the effective resolution of the fitted surface
brightness map, the column density estimates could be biased upwards
at the location of warm point-like sources.

\subsection{Dust opacity} \label{sect:opacity}

\begin{figure*}
\includegraphics[width=17cm]{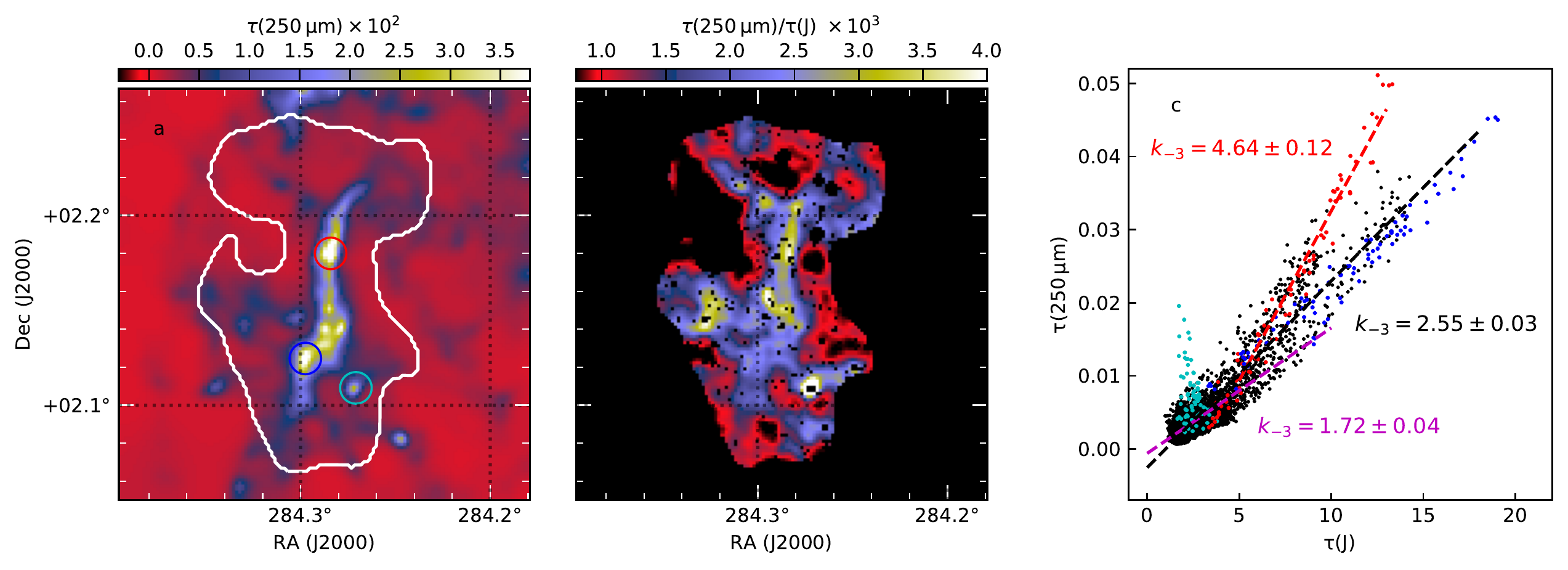}
\caption{
Comparison of $\tau(250\mu {\rm m})$ and $\tau({\rm J})$ estimates: 
the $\tau(250\mu {\rm m})$ map (frame a), the ratio $k_{-3} = 10^{-3}
\times \tau(250\mu {\rm m})/\tau({\rm J})$ (frame b), and the same
ratio as a correlation plot (frame c). The last two frames are limited
to data in the area indicated in the first frame, also excluding
pixels for which $\tau({\rm J})$ is not defined because of masked
stars. Selected $30\arcsec$ radius regions (see frame a) are marked in
frame c with different colours: a point source SE of the southern
clump (cyan) and the northern and southern $\tau(250\mu {\rm m})$
peaks (red and blue circles, respectively). The black line in frame c
is a least squares fit to the data, only excluding the SE point
source. Fits to the northern clump and to data $\tau({\rm J})<6$ are
shown in red and magenta, respectively.
\label{fig:tau250_vs_tauJ}
}
\end{figure*}

The extinction map of \citet{KainulainenTan2013}
(Sect.~\ref{sect:obs_AV}) enables us to compare dust opacities between
the NIR/MIR and sub-millimetre regimes. The correlations of these 
$\tau({\rm J})$ values with the $\tau_5(250\,\mu{\rm m})$ optical
depth estimates are shown in Fig.~\ref{fig:tau250_vs_tauJ}.

The least squares fit gave an average ratio of 
$\tau(250\,\mu{\rm m})/\tau({\rm J}) = (2.55 \pm 0.03) \times 10^{-3}$.
The error estimate only refers to the uncertainty of the fit itself,
which was estimated by bootstrapping. The relation is found to be
steeper in the northern clump and shallower $\tau({\rm J})<6$ (see
Fig.~\ref{fig:tau250_vs_tauJ}).

In addition to the correlation plot of Fig.~\ref{fig:tau250_vs_tauJ},
we estimated the $\tau(250\,\mu{\rm m})/\tau({\rm J})$ ratio based on
the absolute values. We subtracted from the $\tau({\rm J})$ and
$\tau(250\,\mu{\rm m})$ maps a background that was estimated as the
average along a 1$\arcmin$-wide boundary that follows the contour in
Fig.~\ref{fig:tau250_vs_tauJ}a. After the subtraction of the local
background, the average values inside the contour gave
$\tau(250\,\mu{\rm m}))/\tau({\rm J})= (2.1 \pm 0.5) \times 10^{-3}$.
The error estimate is based on the total signal fluctuations over the
area used for background subtraction.

\subsection{Polarisation data} \label{sect:poldata}

\begin{figure*}
\includegraphics[width=17.7cm]{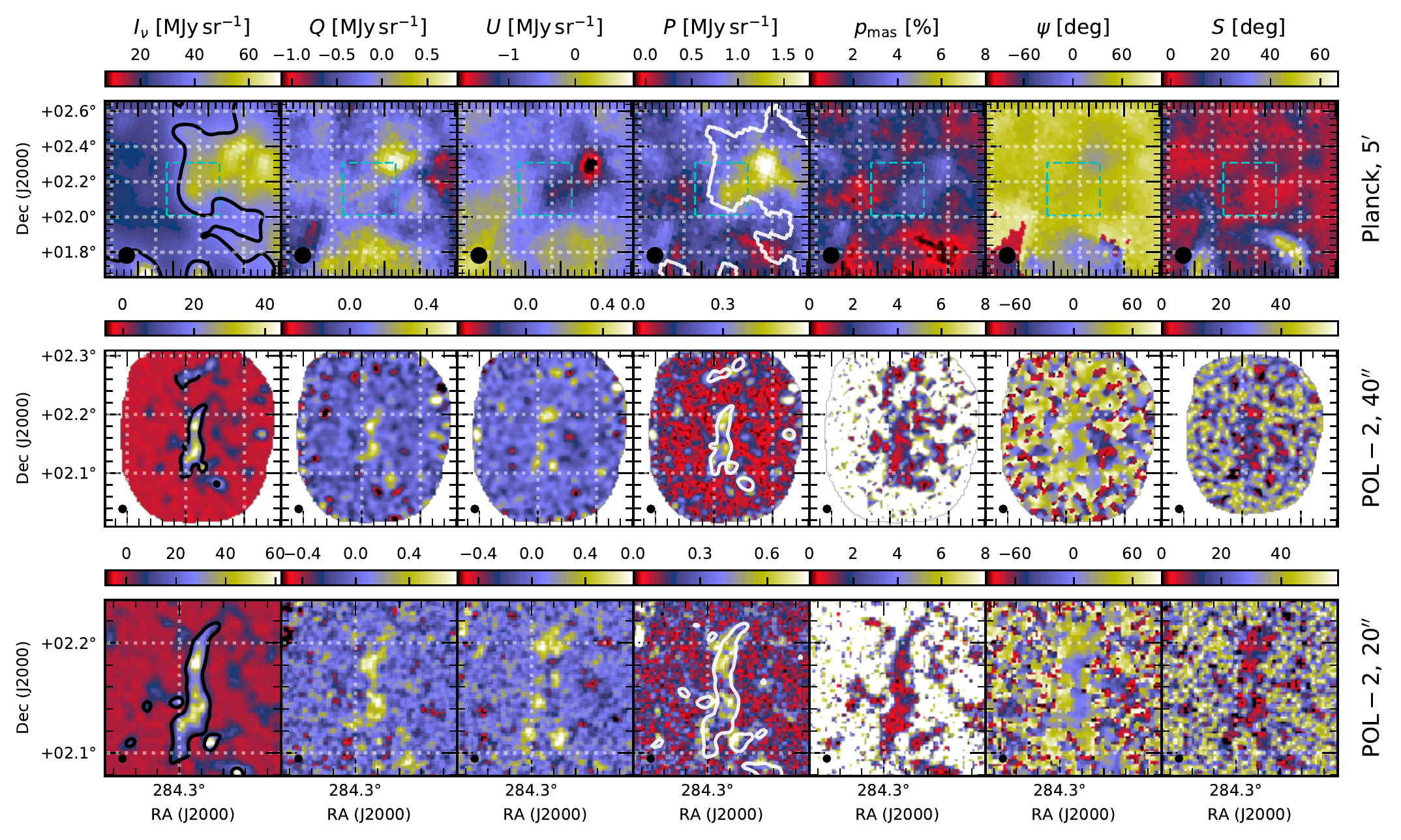}
\caption{
Overview of polarisation data of the \target region. The three rows
show, respectively, \Planck data at 5$\arcmin$ resolution, POL-2 
data at 40$\arcsec$ resolution, and POL-2 data at 20$\arcsec$
resolution. The seven frames on each row are: $I$, $Q$, $U$, polarised
intensity $P$, polarisation fraction $p_{\rm mas}$, polarisation angle
$\psi$, and polarisation angle dispersion function $S$ for a lag of
$\delta=$FWHM/2. The dashed boxes on \Planck maps correspond to the
size of the second row images. The column densities $N({\rm
H}_2)=10^{22}\,{\rm cm}^{-2}$ (at the corresponding resolutions) are
indicated with contours on the $I$ maps. On the second and third row
these are calculated from background-subtracted \Herschel and SCUBA-2
data. The white contours on the $P$ maps correspond to an intensity
$I(850\,\mu{\rm m})$ of 40\,MJy\,sr$^{-1}$ on the \Planck map and
10\,MJy\,sr$^{-1}$ on the POL-2 maps. 
\label{fig:VI_maps}
}
\end{figure*}

Figure~\ref{fig:VI_maps} shows an overview of the \Planck and POL-2
polarisation data. \Planck maps have very little noise. When
POL-2 data are convolved to a 40$\arcsec$ resolution, the polarisation
angle dispersion $S$ is clearly affected by noise outside the $N({\rm
H}_2)=10^{22}\,{\rm cm}^{-2}$ contour and $p$ becomes dominated by
noise closer to the map edges. At the higher 20$\arcsec$ resolution,
polarisation fraction values become uncertain as soon as column
density drops below $N({\rm H}_2)=10^{22}\,{\rm cm}^{-2}$.

\begin{figure}
\includegraphics[width=8.8cm]{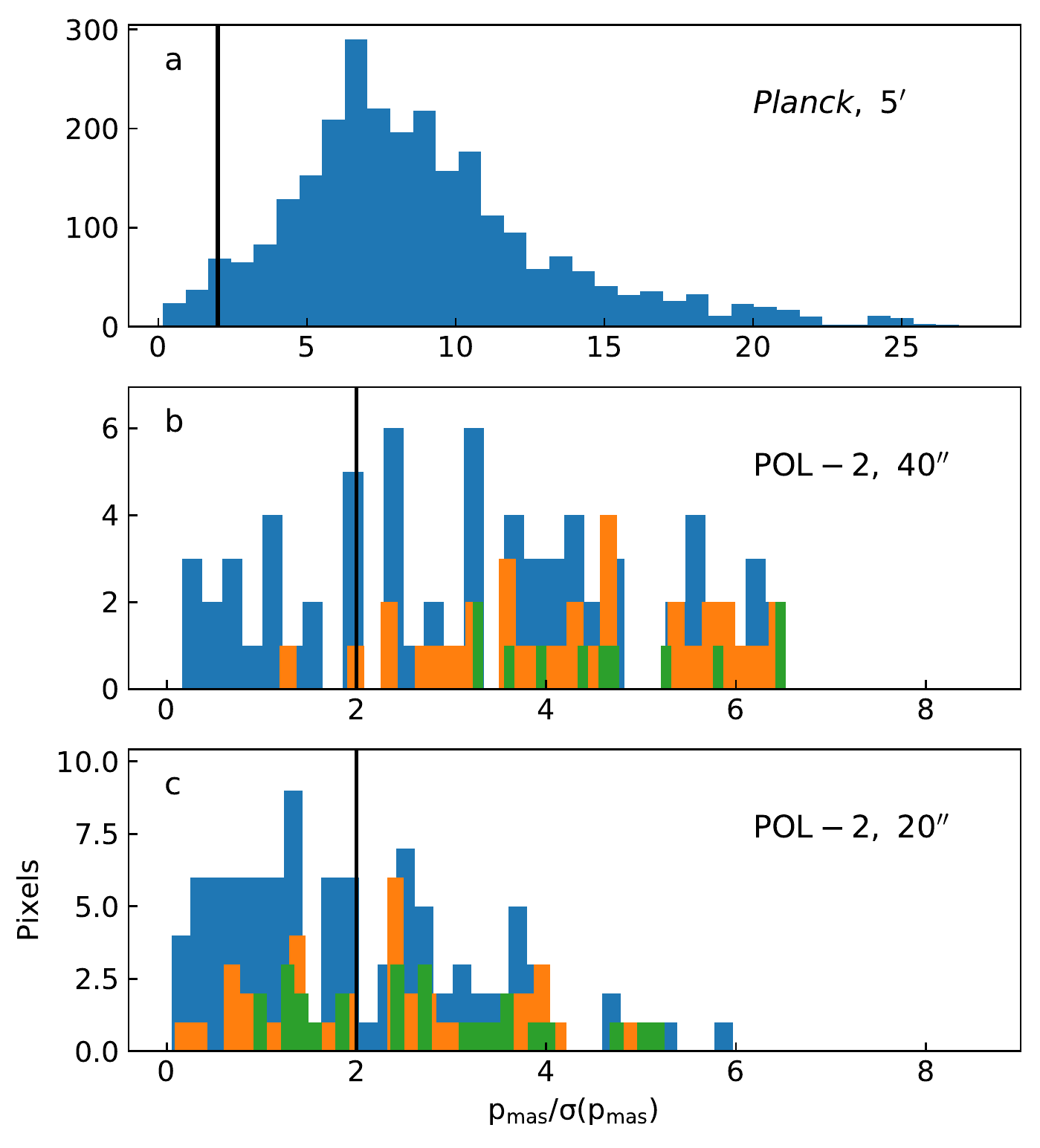}
\caption{
SNR of polarisation fraction. The frames correspond to \Planck data at
5$\arcmin$ resolution (frame a) and to POL-2 data at 40$\arcsec$
(frame b) and 20$\arcsec$ (frame c) resolution. Frame a includes all
data over a $2\degr \times 2\degr$ area. In frames b-c, histograms
show POL-2 data above column density thresholds of 1, 2, and 3 times
$N({\rm H_2})=10^{22}$\,cm$^{-2}$ (blue, orange, and green histograms,
respectively). Vertical lines at 2 indicate an approximate limit
above which the $p_{\rm mas}$ estimates are reliable. All data are
sampled at steps FWHM/2.
\label{fig:SNR}
}
\end{figure}

Figure~\ref{fig:SNR} shows histograms for the SNR of polarisation
fraction, $p_{\rm mas}/\sigma_{p, \rm mas}$. The plot includes
histograms for \Planck data at 5$\arcmin$ resolution and for the POL-2
data at 20$\arcsec$ and 40$\arcsec$ resolutions. According to
\citet{Montier2015b}, $p_{\rm mas}$ is unbiased for $p_{\rm
mas}/\sigma_{p, \rm mas}>2$. The SNR is sufficient for almost all
\Planck data at the full resolution and most of the POL-2 data at
40$\arcsec$ resolution, when selected at $N({\rm
H_2})>10^{22}$\,cm$^{-2}$. Data cannot be thresholded directly
using the SNR because that would lead to a biased selection of $p$
values \citep{planck2016-l11B}. Figure~\ref{fig:SNR}c shows that at
20$\arcsec$ resolution a significant part of POL-2 $p_{\rm mas}$
estimates may be biased (at SNR$<$2 the modified asymptotic estimator
may not remove all the bias in $p$) and a higher column density
threshold does not fully remove the problem.

The polarisation angle estimates are mainly unbiased but since they
are affected by noise, at low SNR the polarisation angle dispersion
function $S$ will have systematic positive errors that are not fully
removed by the bias correction. The appearance of the
Fig.~\ref{fig:VI_maps} maps is in qualitative agreement with this.

\subsection{Magnetic field geometry} \label{sect:geometry}

The magnetic field geometry of the cloud \target has been discussed in
detail in \citet{Liu_G35_pol} based on the POL-2 observations.
However, we present some plots on the magnetic field morphology before
concentrating on the polarisation fraction in the following sections.

\begin{figure}
\includegraphics[width=8.8cm]{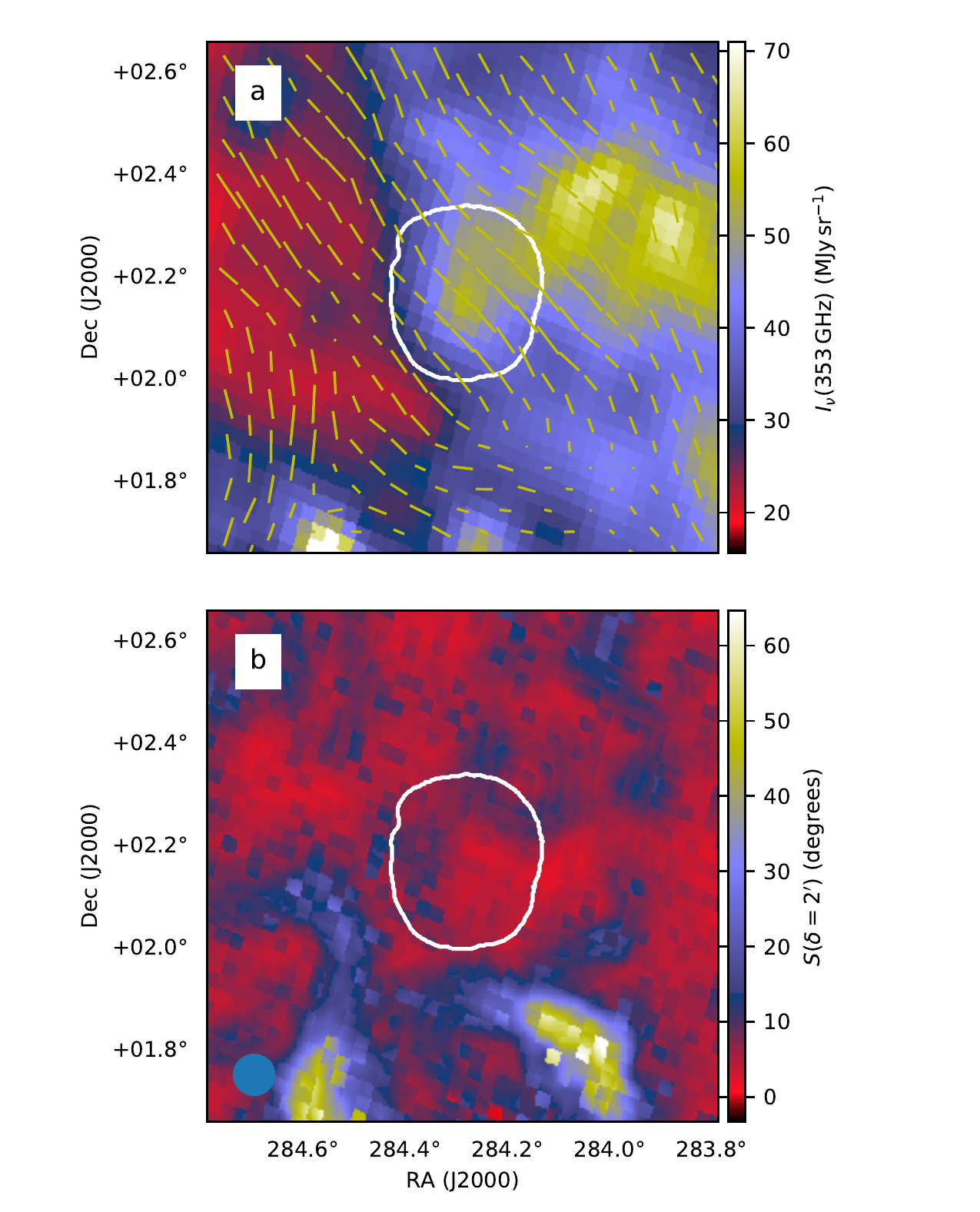}
\caption{
\Planck 850\,$\mu$m (353\,GHz) surface brightness with vectors showing
the POS magnetic field orientation (frame a) and polarisation angle
dispersion function $S(\delta=2.5\arcmin)$ calculated from \Planck
data (frame b). The maps are at the original $5\arcmin$ resolution.
The white contours indicate the area covered by SCUBA-2 observations.
The length of the polarisation vectors is proportional to the
polarisation fraction (90$\arcsec$ for 1\% of polarisation).
}
\label{fig:S_Planck}
\end{figure}

\begin{figure*}
\includegraphics[width=17cm]{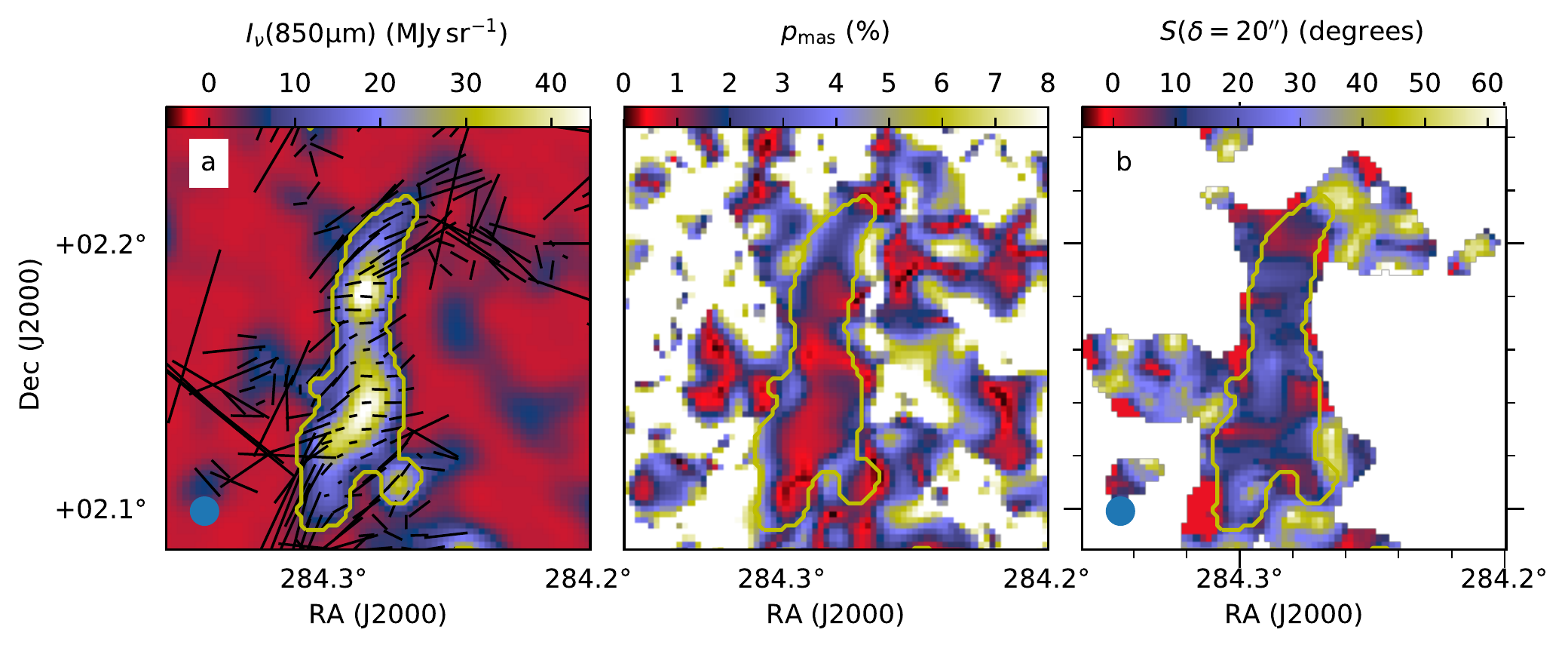}
\caption{
Field orientation on the POL-2 850\,$\mu$m intensity map (frame a),
polarisation fraction $p_{\rm mas}$ (frame b), and polarisation
angle dispersion function $S(\delta=20\arcsec)$ (frame c) calculated
from data at 40$\arcsec$ resolution. Polarisation vectors (at
20$\arcsec$ steps, rotated to show the magnetic field orientation) and
$S$ are shown for pixels with $N({\rm H_2})>5\times
10^{21}$\,cm$^{-2}$. The length of the polarisation vectors per cm is
the same as in Fig.~\ref{fig:S_Planck} (14.4$\arcsec$ for a
polarisation fraction of 1\%). The yellow contours correspond to
$N({\rm H_2})=10^{22}$\,cm$^{-2}$, outside of which polarisation angles
are uncertain.
\label{fig:S_SCUBA}
}
\end{figure*}

Figure~\ref{fig:S_Planck}a shows a large-scale polarisation map based
on \Planck 850\,$\mu$m. This is dominated by a regular field that in
equatorial coordinates runs from NE to SW. At the 5$\arcmin$
resolution the \target filament is not prominent because of the strong
background emission (see Fig.~\ref{fig:overview}). The SCUBA-2
850\,$\mu$m surface brightness map in Fig.~\ref{fig:S_SCUBA}a shows
the main ridge and some other filamentary features that were discussed
in \citet{Liu_G35_pol}. At this $\sim 14 \arcsec$ resolution the
polarisation vectors show a less ordered field. In the central part,
the field is partly perpendicular to the filament. In the north, the
field turns parallel to the filament and is thus almost perpendicular
to the large-scale field observed by \Planckn. The SE-NW orientation
observed in the northern end is actually common to filament boundary
regions and is particularly clear on the eastern side.

The second frames in Figs.~\ref{fig:S_Planck}-\ref{fig:S_SCUBA} show
maps of the bias-corrected polarisation angle dispersion
function $S$. For \Planck these are calculated at the scale of
$\delta=2.5\arcmin$ using the \Planck observations at their native
resolution of FWHM=5$\arcmin$. In the case of SCUBA-2, to increase the
SNR, the data were smoothed to a resolution of 40$\arcsec$ and $S$ was
calculated with $\delta=20\arcsec$.
Figure~\ref{fig:S_SCUBA} shows that in POL-2 observations
$S(20\arcsec)$ goes in some areas below $\sim$10$\degr$. Higher values
are found for example in the northern clump. There the change in the
magnetic field orientation coincides with the intensity maximum and
large $S$ avalues are not produced by noise alone. Similarly, at the
eastern filament edge, the polarisation angles are uniform along the
boundary but change systematically between the high and low column
densities, contributing to the variation seen inside the $N({\rm
H}_2)=10^{22}$\,cm$^{-2}$ contour.

The \Planck polarisation vectors are quite uniform over the \target
filament while the field geometry in SCUBA-2 850\,$\mu$m data is
different and partly orthogonal. One may ask whether the
observations are consistent or whether the locally changing magnetic
field orientation should be visible in \Planck data as a drop in the 
polarisation fraction. We tested this by making simultaneous fits to
the $I$, $Q$, and $U$ data of both \Planck and SCUBA-2. The results in
Appendix~\ref{sect:IQU_fit} show that the observations are not
contradictory. This is possible because of the large difference in the
beam sizes and because the SCUBA-2 data are not sensitive to emission
at scales larger than 200$\arcsec$. Thus, most information about the
large-scale field is filtered out in the SCUBA-2 data.  


\subsection{Polarisation fraction} \label{sect:polfrac}

\subsubsection{Polarisation fraction from \Planck observations}
\label{sect:polfrac_Planck}

\begin{figure}
\includegraphics[width=8.8cm]{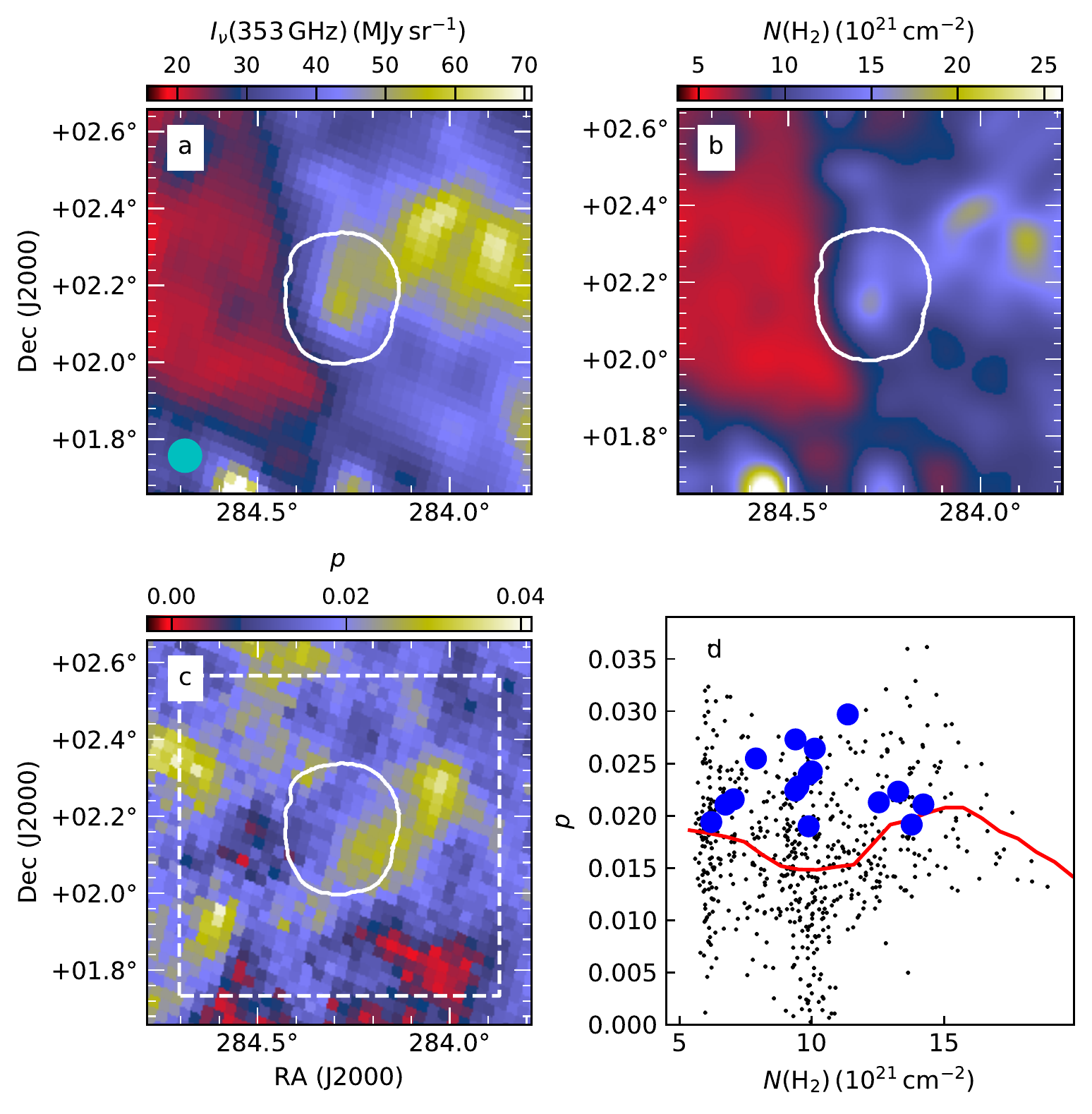}
\caption{
Polarisation fraction of the \target region based on \Planck data.
Frame a shows the 850\,$\mu$m surface brightness, frame b the
\Herschel column density map convolved to 5$\arcmin$ resolution, and
frame c a map of the polarisation fraction $p_{\rm mas}$ from \Planck
data.  The white contour indicates the coverage of the SCUBA-2 map.
Frame d shows the polarisation fraction as a function of column
density, sampled with 2$\arcmin$ steps from maps at 5$\arcmin$
resolution, excluding the map boundaries (indicated by dashed lines in
Frame c), without a SNR cut-off. The red line is the running mean and
the blue circles show values from the area covered by SCUBA-2
observations. 
}
\label{fig:plot_Planck_INP}
\end{figure}

Figure~\ref{fig:plot_Planck_INP} shows the bias-corrected
polarisation fraction estimate $p_{\rm mas}$ from \Planck
observations over a $1\degr \times 1\degr$ region and at a resolution
of 5$\arcmin$. The average value is $p\sim$2\%. For comparison, the
\Herschel column density map was convolved to the same resolution but
the polarisation fraction does not show clear dependence on the column
density. At this resolution, the \target filament shows up in the
column density map only as a minor local maximum and the polarised
signal appears to be dominated by more extended emission components.
The polarisation fraction values at the filament location are slightly
higher than in the region on average, close to $2.5\%$ as indicated in
Fig.~\ref{fig:plot_Planck_INP}d.

\begin{figure}
\includegraphics[width=8.8cm]{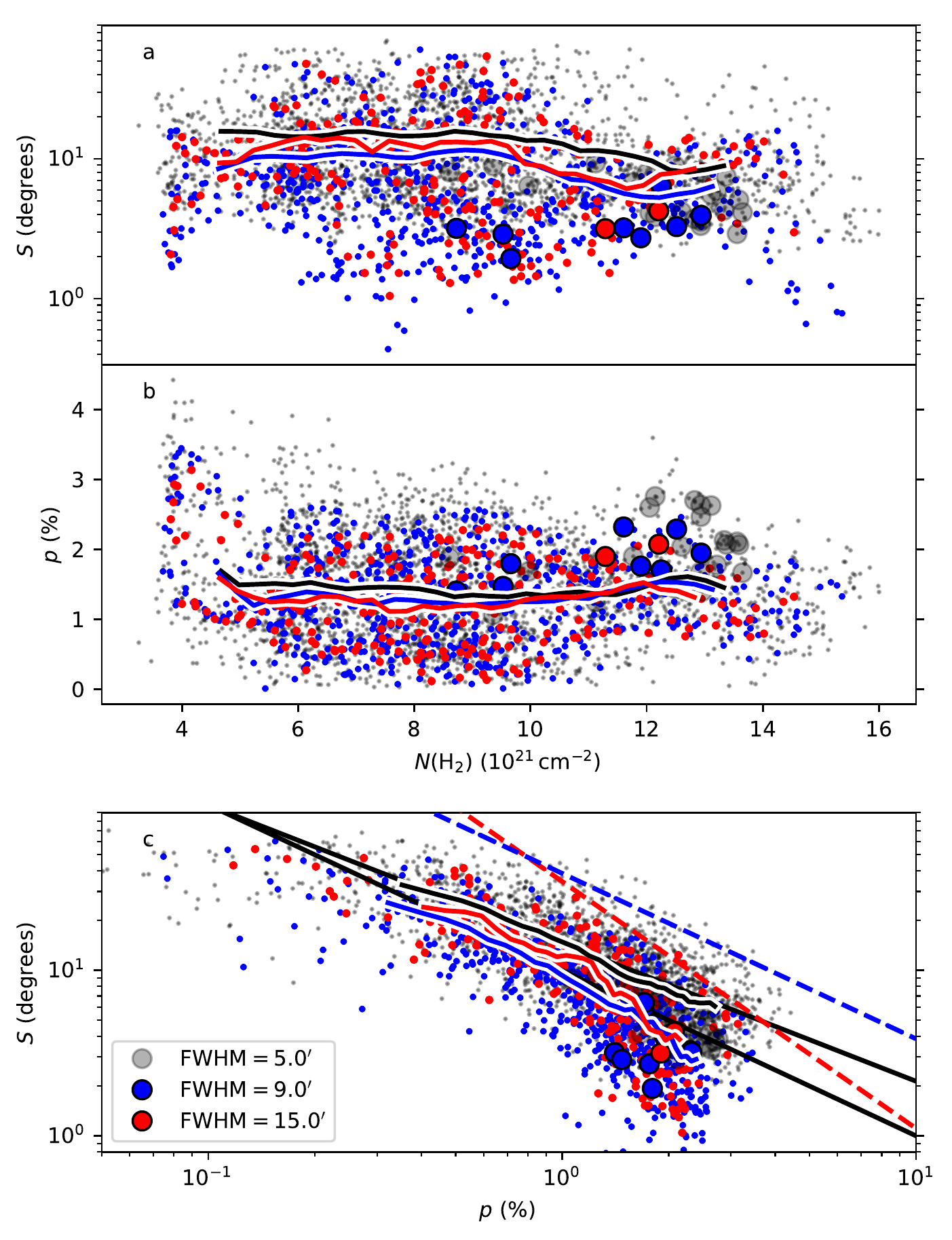}
\caption{
Polarisation angle dispersion function $S$ and polarisation fraction
$p_{\rm mas}$ from \Planck data in a $2\degr \times 2\degr$ area
centred on the POL-2 field. Frames a and b show these as a function of
dust optical depth and frame c shows their mutual correlation. The
colours of the plot symbols correspond to the data resolution, as
indicated in frame c. The solid lines show the corresponding moving
averages. The points inside the area mapped with POL-2 are plotted
with large symbols. The data are sampled at steps of FWHM/2 and $S$ is
calculated for lag $\delta=$FWHM/2. In frame c the upper solid black
line shows the relation $\log_{10}(S) = -0.834 \times \log_{10}(p_{\rm
mas}) -0.504$ from \citet{planck2014-XIX} and the lower black line the
relation $S = 0.1/p_{\rm mas}$. The red dashed line is the
relation $\log \, p_{\rm mas} = -0.670 \log \, S - 0.97$ from
\citet{Fissel2016}. The blue dashed line corresponds to the fit to
Gould Belt cloud data, $S \times p = 0.31^{\degr}
({\rm FWHM}/160\arcmin)^{0.18}$ of \citet{planck2016-l11B}, calculated with
FWHM=9$\arcmin$.
} \label{fig:SpI}
\end{figure}

We examine in Fig.~\ref{fig:SpI} how, in the case of \Planck data, the
bias-corrected polarisation fraction and the estimated polarisation
angle dispersion function depend on the column density and on the data
resolution. The changes from $5\arcmin$ to $9\arcmin$ and further to
$15\arcmin$ resolution each correspond to about a factor of three
increase of SNR. Irrespective of the resolution (and SNR), the mode of
$S$ is close to 10\% and the values in area covered by SCUBA-2 are of
similar magnitude. The polarisation fraction is mainly between 0.5\%
and 3\% and there is no significant difference between the 9$\arcmin$
and 15$\arcmin$ resolution cases. The $p$ values within the area
mapped with SCUBA-2 are higher than on average, 2-2.5\% for the
full-resolution data and $\sim 2$\% at lower resolutions. In the same
region $S$ tends to be lower than average. This anticorrelation
between $p$ and $S$ is clear in Fig.~\ref{fig:SpI}c. 
This could have its origin in either the noise (which increases the
estimates of both quantities) or in the magnetic field geometry.  
The effects of noise has been characterised in previous \Planck
studies \citep{planck2014-XIX}, and should here be small when data are
smoothed to increase the SNR. The relation are similar for
FWHM=$9\arcmin$ and FWHM=$15\arcmin$, which shows that the results are
not severely affected by noise. This is confirmed with simulations in
in Appendix~\ref{app:sim}. For given $p$, the $S$ values are lower
than previously found with BLASTPol for the Vela C molecular
\citep{Fissel2016} and with \Planck for the Gould Belt clouds
\citep{planck2016-l11B}.

\subsubsection{Polarisation fraction in SCUBA-2 observations}
\label{sect:polfrac_POL2}

\begin{figure}
\includegraphics[width=8.8cm]{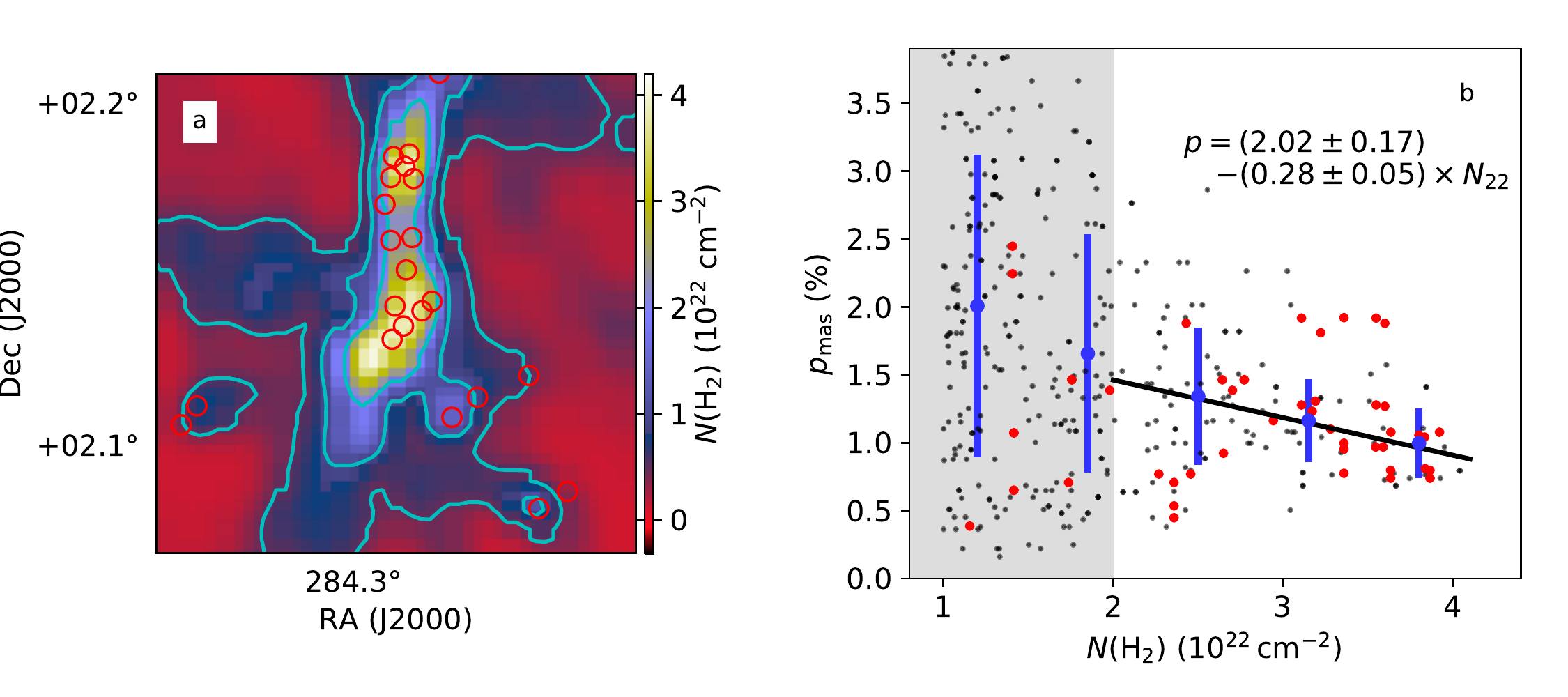}
\caption{
POL-2 polarisation fraction and column density. Frame a shows the
column density map $N_5({\rm H}_2)$ at 40$\arcsec$ resolution. The
contours are drawn at 0.5, 1, and 2 times $10^{22}$\,cm$^{-2}$ and the
red circles denote the locations of the point sources of
Table~\ref{table:PS_L}. 
In frame b, $p_{\rm mas}$ is plotted as a function of column
density for data with $N({\rm H}_2)> 10^{22}$\,cm$^{-2}$. The pixels
coinciding with point sources (red circles of Frame a) are plotted in
red. The blue bars are representative error estimates. The result of
the weighted least squares fit to data at $N({\rm H}_2)> 2 \times
10^{22}$\,cm$^{-2}$ is given in the frame and is shown as a black
solid line. The shading corresponds to $N({\rm H}_2)< 2 \times
10^{22}$\,cm$^{-2}$ where the $p_{\rm mas}$ estimates may be biased.
}
\label{fig:p_vs_N_S}
\end{figure}

We calculated the bias-corrected polarisation fraction estimates
$p_{\rm mas}$ from SCUBA-2 ($I$, $Q$, $U$) maps that were first
convolved to a resolution of $40\arcsec$ to increase their SNR.  In
Fig.~\ref{fig:p_vs_N_S} we plot $p_{\rm mas}$ as a function of column
density for with $N({\rm H}_2)>10^{22}$\,cm$^{-2}$. We avoid a
criterion based on the SNR of the polarised intensity because that
would bias the selection of the polarisation fraction values. Based on
Fig.~\ref{fig:SNR}, the plotted $p_{\rm mas}$ values should be
unbiased. The average $p_{\rm mas}$ value decreases as a
function of $N$ and, based on the formal uncertainty of the weighted
least squares fit, the decrease is significant.

The pixels associated to 70\,$\mu$m sources (Fig.~\ref{fig:p_vs_N_S}a)
do not differ from the general distribution. However, at lower column
densities (not shown), they tend to trace the lower envelope of the
($N$, $p_{\rm mas}$) distribution. This is mostly a result of them
having on average $\sim$80\% higher SNR (higher intensity for a given
column density). This makes their $p$ estimates less biased.

Appendix~\ref{app:sim} shows further how the $p_{\rm mas}$ vs.
$N$ relation changes as a function of resolution and, consequently,
as a function of the SNR. There we also present {simulations of the
$p_{\rm mas}$ vs. $N$ relation} in the presence of noise. These show
that while the noise produces significant scatter, the average
$p_{\rm mas}$ values estimated at the highest column densities are
reliable.

\begin{figure}
\includegraphics[width=8.8cm]{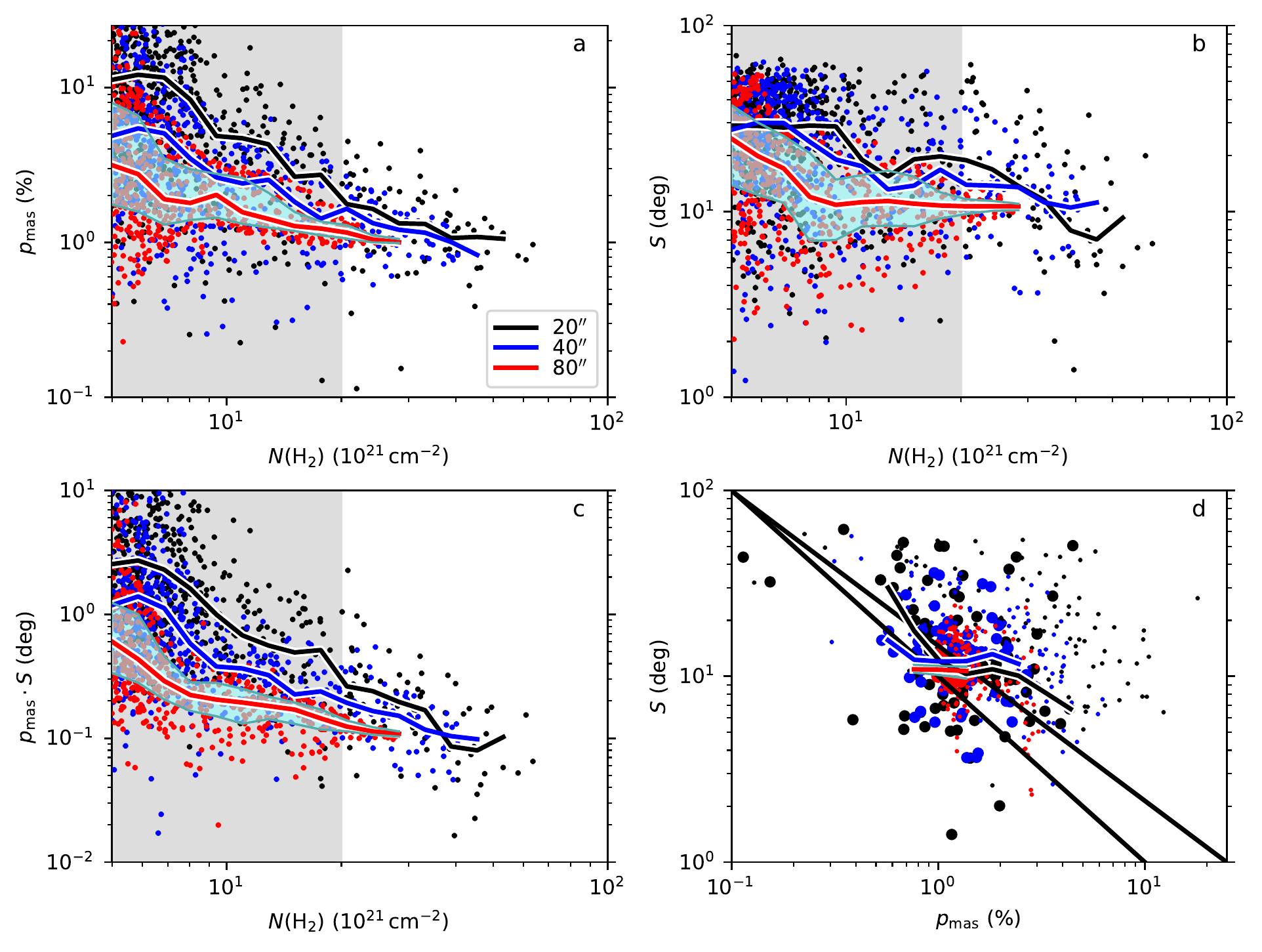}
\caption{
Correlations between POL-2 polarisation parameters $p_{\rm mas}$ and
$S$ and the column density. The quantities are calculated at four
spatial resolutions represented with the colours indicated in frame a.
The data are sampled at steps of FWHM/2 and the solid lines correspond
to running averages.
In frames a-d, the shading corresponds to $N({\rm H}_2)<2 \times
10^{22}\,{\rm cm}^{-2}$ where, according to Fig.~\ref{fig:SNR}, the 
40$\arcsec$ resolution $p_{\rm mas}$ estimates become unreliable.
In frame d, values for $10^{22}\,{\rm cm}^{-2} <
N({\rm H}_2) < 2\times 10^{22}\,{\rm cm}^{-2}$, and $N({\rm H}_2)>2
\times 10^{22}\,{\rm cm}^{-2}$, are shown with small and large
symbols, respectively, and the lines show running averages for the
higher column density interval (20 logarithmic bins over the parameter
range with a minimum of 4 points per bin). Shaded cyan region
corresponds to the interquartile range of the quantities plotted with
red lines.
In frame d the upper solid black line shows the relation $\log_{10}(S)
= -0.834 \times \log_{10}(p_{\rm mas}) -0.504 $ from
\citet{planck2014-XIX} and the lower black line the relation $S =
0.1/p_{\rm mas}$. 
}
\label{fig:p_S_N}
\end{figure}

The correlations between $p_{\rm mas}$, $S$, and $N({\rm H}_2)$
and their dependence on the data resolution are further examined in
Fig.~\ref{fig:p_S_N}. The values are independent of the resolution
only towards the highest column densities. Otherwise $p_{\rm
mas}$ and $S$ decrease with lower resolution. This is consistent with
the increasing SNR reducing the bias and data below $N({\rm H}_2)\sim
10^{22}\,{\rm cm}^{-2}$ remaining affected by noise. However, there
may be additional effects from the averaging of observations with
different polarisation angles (geometrical depolarisation). The $S$
values may reflect the fact that the \target field consists of a
single, very narrow filament. For a given column density, a larger lag
means that $S$ calculation uses data over a larger area and thus on
average with a lower SNR.

Figure~\ref{fig:p_S_N}d shows the correlation of $S$ vs. $p_{\rm
mas}$. For column densities $N({\rm H}_2)>2 \times 10^{22}\,{\rm
cm}^{-2}$, with lower resolution (higher SNR) the values converge
towards similar parameter combinations as in Fig.~\ref{fig:SpI} for
\Planck. This in spite of the fact that the \Planck result is for a
much larger area and for a data resolution lower by almost a factor of
4. Because of the small dynamical range (in part due to the
spatial filtering) and possible residual bias in $S$, no clear
anticorrelation is seen between the POL-2 estimates of $p$ and $S$.

\begin{figure}
\includegraphics[width=8.8cm]{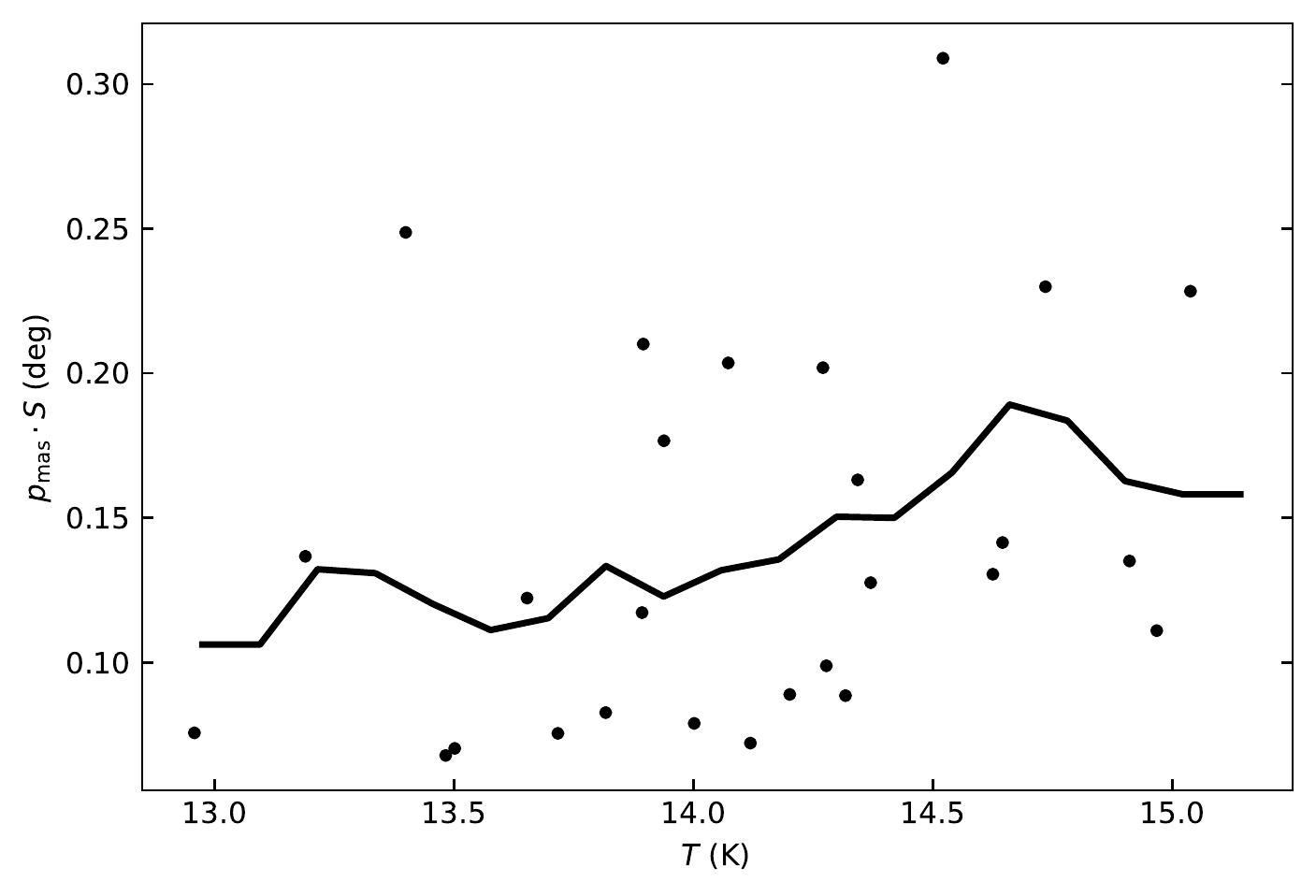}
\caption{Product $p \times S$ as a function of dust colour
temperature. The data are at 40$\arcsec$ resolution, sampled at
half-beam steps, and selected from the region with $N({\rm
H}_2)>2\times 10^{22}\,{\rm cm}^{-2}$. The solid lines shows a moving
average.} 
\label{fig:pS_vs_T}
\end{figure}

If the $p$-$N$ anticorrelation were due to a loss of grain alignment,
the product $p \times S$ should decrease as a function of increasing
column density and decreasing dust temperature.
Figure~\ref{fig:p_S_N}c shows the anticorrelation with the column
density. In Fig.~\ref{fig:pS_vs_T} we show the corresponding
correlation of $p \times S$ with the dust colour temperature. 
Although the data selection (resolution of 40$\arcsec$ and column
densities $N({\rm H}_2)>2\times 10^{22}\,{\rm cm}^{-2}$) should ensure
that $p$ values are unbiased, the polarisation angle dispersion
function $S$ may still contain some bias that contributes to increased
$p \times S$ values at higher temperatures, which mainly correspond to
lower column densities. The dispersion is calculated using data from
an area with a diameter of $1.5\times$FWHM. Therefore, high $N$ at the
central position does not fully preclude the $S$ estimate being
affected by lower SNR pixels further out. A Monte Carlo simulation
based on the $I$, $Q$, and $U$ maps and their error maps shows that
the trend in Fig.~\ref{fig:pS_vs_T} is not significant and thus
neither proves or disproves the presence of grain alignment
variations.

\section{Radiative transfer models} \label{sect:RT}

\subsection{Radiative transfer modelling of total emission}
\label{sect:RT_total}

\begin{figure}
\includegraphics[width=8.8cm]{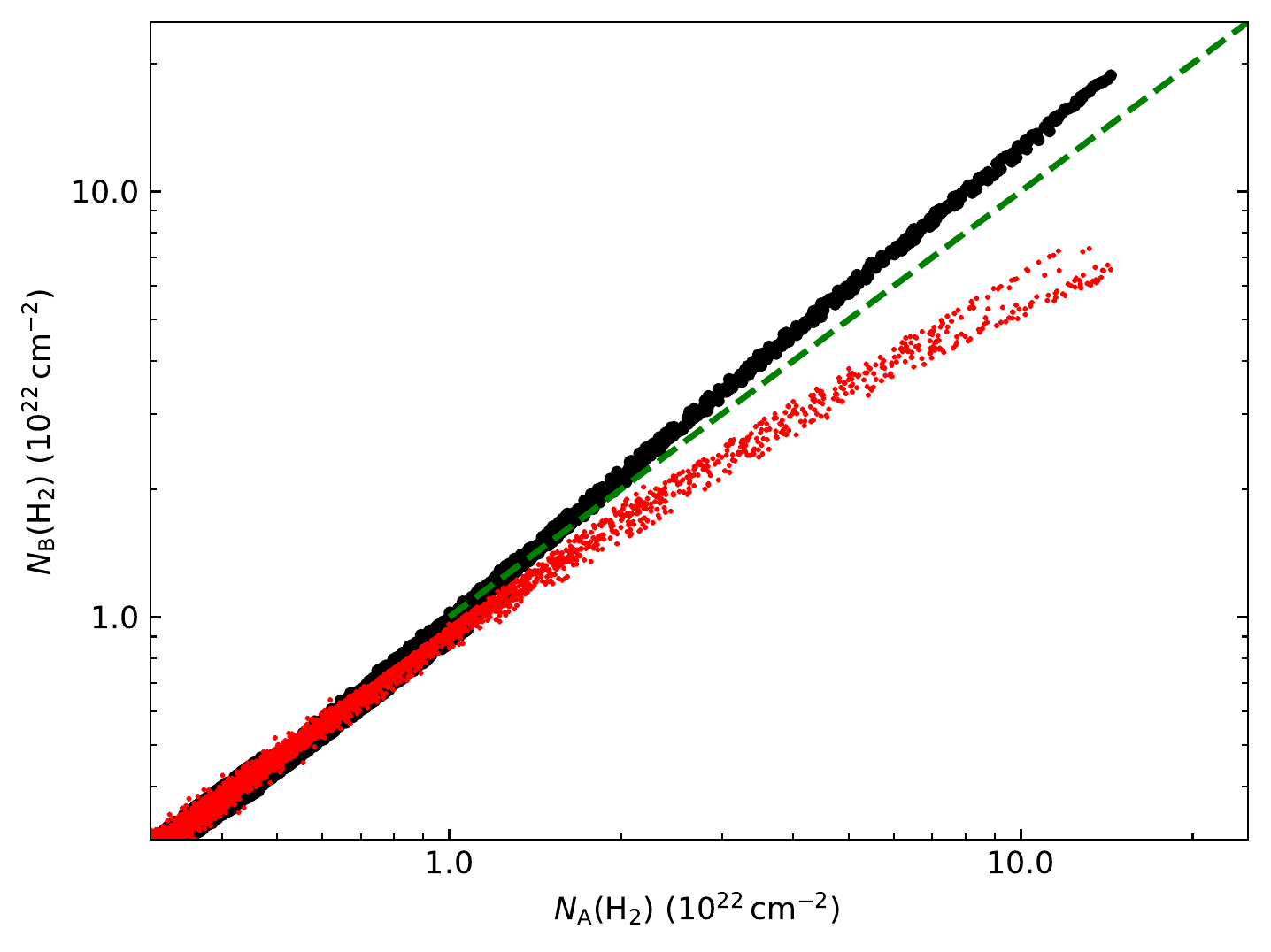}
\caption{
Correlation between the column densities $N({\rm H}_2)$ of two RT
models fitted to observations. Column densities $N_{\rm A}$ and
$N_{\rm B}$ (black dots) correspond to dust models with
$\tau(250\,\mu{\rm m})/\tau({\rm J})$ equal to $1.6\times 10^{-3}$ and
$1.0\times 10^{-3}$, respectively. For the RT model with column
densities $N_{\rm A}$, the red points show the column density
estimates derived from the synthetic surface brightness maps. The
dashed line is the one-to-one relation. All data are at $20\arcsec$
resolution. 
}
\label{fig:compare_NH2}
\end{figure}

Figure~\ref{fig:compare_NH2} compares the column densities of two RT
models fitted to SPIRE data. These differ regarding the assumed
sub-millimetre vs. NIR opacity but have identical opacity at
250\,$\mu$m. The model $A$ has an opacity ratio of $\tau(250\,\mu{\rm
m})/\tau({\rm J})=1.6 \times 10^{-3}$. This value is the average value
derived for a sample of PGCC clumps in \citet{GCC-V} and a lower limit
for the values estimated in Sect.~\ref{sect:opacity}. To test the
sensitivity to dust properties, the alternative model $B$ has
$\tau(250\,\mu{\rm m})/\tau({\rm J})=1.0 \times 10^{-3}$. Model $B$
results in 20\% higher $\chi^2$ values but both models represent the
surface brightness data of the main filament equally well. The lower
$\tau(250\,\mu{\rm m})/\tau({\rm J})$ ratio leads to higher column
densities, with a 30\% difference in the densest regions. The effect
is thus of similar magnitude as the change in the assumed opacity
ratio.

Figure~\ref{fig:compare_NH2} also shows $N({\rm H}_2)$ estimates that
were calculated using MBB fits and the simulated surface brightness
maps of the model A. As expected, the values recovered with MBB
calculations are below the true values. The difference becomes
noticeable above $N({\rm H}_2)=2 \times 10^{22}$\,cm$^{-2}$ and at
$N({\rm H}_2)=10^{23}$\,cm$^{-2}$ the error is a factor of two.

\subsection{Radiative transfer modelling of the $p$ vs. $N$ relation}
\label{sect:RT_pol}

We added to model $A$ (see Sect.~\ref{sect:RT_total}) alternative
descriptions of the magnetic field geometry to make predictions of the
polarised emission. These calculations are used to test how the field
geometry could affect the observed polarisation patterns and
especially the variations of the polarisation fraction as a function
of the column density.
A physical cloud model is needed to describe the variations of the
dust emission that depend on the temperature structure of the cloud.
In RAT grain alignment calculations, the volume density and the
variations of the radiation field (intensity and anisotropy) become
additional factors. Because the simulations are essentially free
of noise, $p$ values can be estimated directly without using the
$p_{\rm mas}$ estimator.  

\begin{figure}
\includegraphics[width=8.8cm]{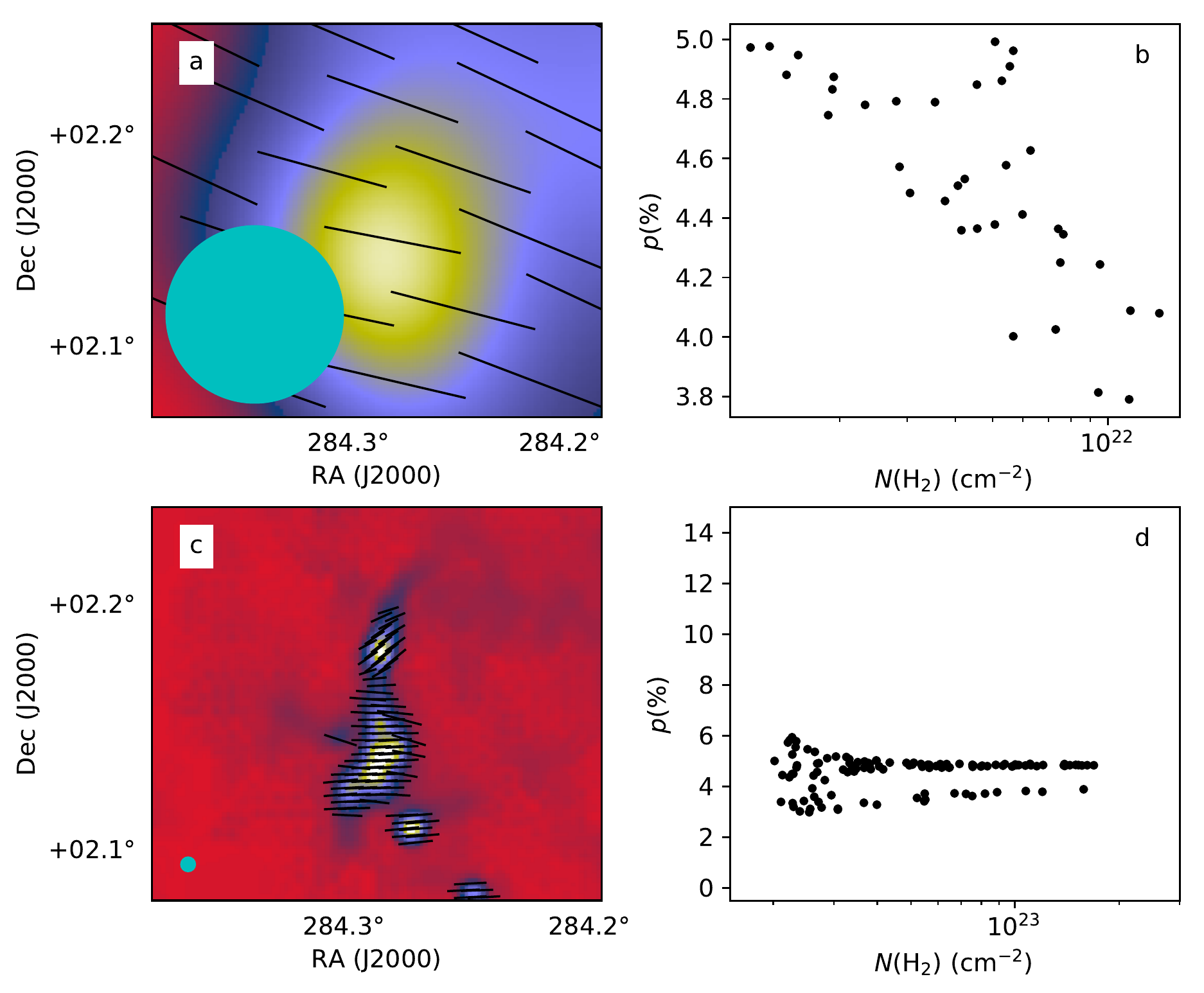}
\caption{
Predictions of the toy magnetic field model with constant grain
alignment. The polarisation vectors are shown on the column density
map in frame a (central part of the full map) and the $p$ vs. $N({\rm
H}_2)$ relation is plotted in frame b at \Planck resolution. The lower
frames are the same for synthetic POL-2 observations, at a resolution
of 20$\arcsec$, assuming high-pass filtering with $\theta=200\arcsec$.
The scaling of the absolute $p$ values is arbitrary but identical
between the frames.
\label{fig:pmod_const}
}
\end{figure}

\begin{figure}
\includegraphics[width=8.8cm]{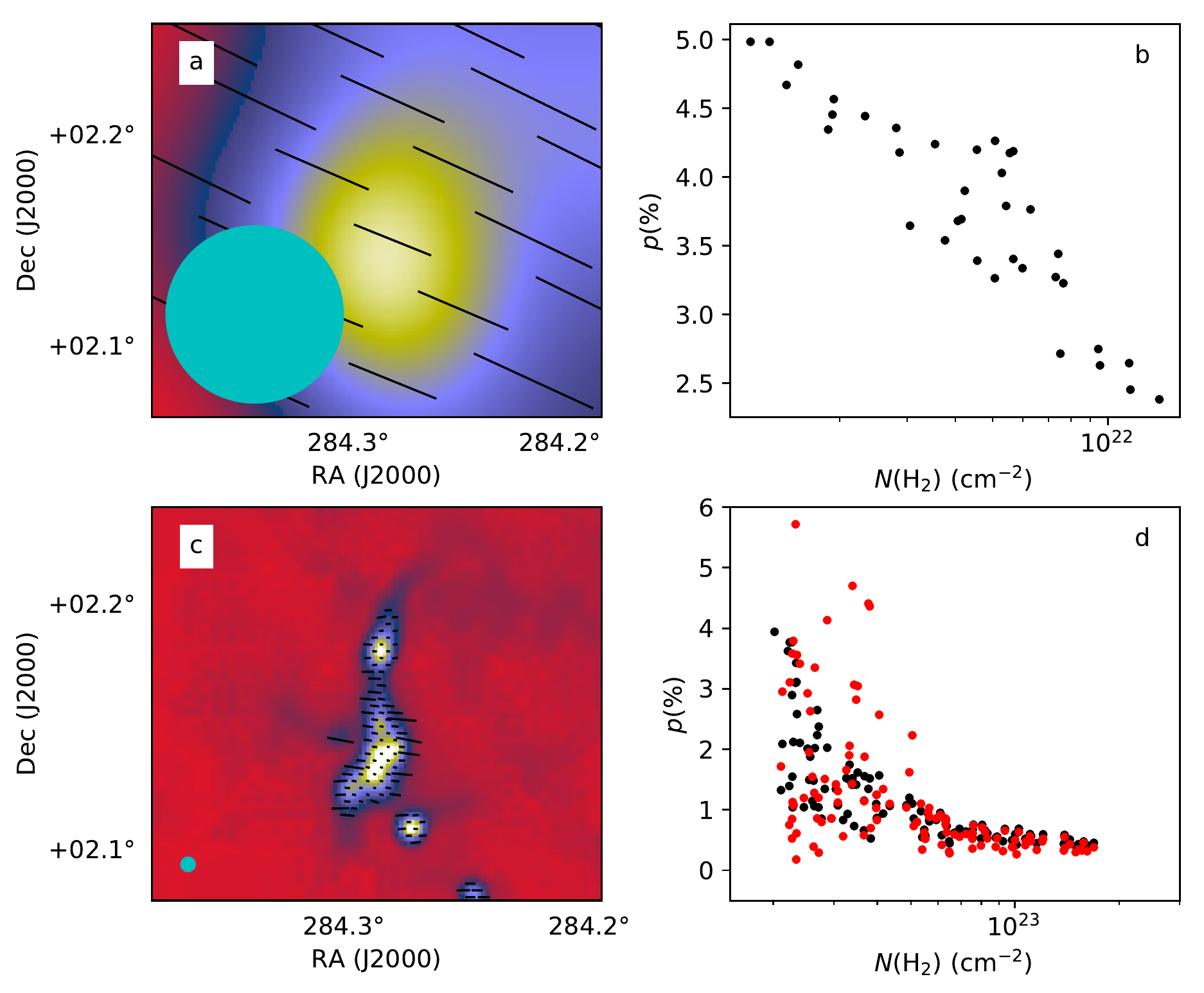}
\caption{
Same as Fig.~\ref{fig:pmod_const} but using grain alignment predicted
by RAT calculations. In frame d, the red points correspond to data
with spatial high-pass filtering with a scale of $\theta=100\arcsec$
instead of the default value of $\theta=200\arcsec$.
\label{fig:pmod_RAT}
}
\end{figure}

We used cloud models that were optimised for the $\tau(250\mu{\rm
m})/\tau({\rm J}) = 1.6 \times 10^{-3}$ dust. We started with a model
where the main volume is threaded by a uniform magnetic field in the
plane of the sky and with a position angle PA=45$\degr$, in rough
correspondence to the \Planck data in Fig.~\ref{fig:S_Planck}. At
densities above $n({\rm H}_2)=3 \times 10^3$\,cm$^{-3}$ the field is
in EW direction (PA=95$\degr$) except for the northern part DEC$>
2\degr 10\arcmin$ where it has PA=135$\degr$ and thus is perpendicular
to the large-scale field. 
The results for spatially constant grain alignment and for
calculations with RAT alignment are shown in
Figs.~\ref{fig:pmod_const}-\ref{fig:pmod_RAT}, respectively. The POL-2
simulation again assumes that the measured ($I$, $Q$, $U$) are
high-pass filtered at a scale of $\theta=200\arcsec$. The absolute
level of $p$ is scaled to give a maximum value of $5\%$ for the
synthetic \Planck observations and the same scaling is applied to the
POL-2 case.

In Fig.~\ref{fig:pmod_const} the simulated \Planck observations show
some 30\% decrease in $p$ as a function of column density. Because the
grain alignment was uniform, the drop is caused by changes in the
magnetic field orientation. In the simulated POL-2 observations the
orientation of the polarisation vectors follows the magnetic field of
the dense medium. Unlike in the actual observations, the polarisation
fraction is close to the $p=5\%$ level, the same as for \Planck. The
polarisation fraction of the northern clump is only slightly lower,
some 4\%. This is a result of the lower density (and smaller size) of
that clump and of the magnetic field orientation that is perpendicular
to the large-scale field. Appendix~\ref{app:polmap} shows results
when the change from the large-scale field takes place at a higher
density, $n({\rm H}_2)= 10^4$\,cm$^{-3}$ instead of $n({\rm H}_2)=3
\times 10^3$\,cm$^{-3}$. This has only a very small effect on the
polarisation fraction, except for the northern clump where $p$ drops
partly below 2\%.

\begin{figure}
\includegraphics[width=8.8cm]{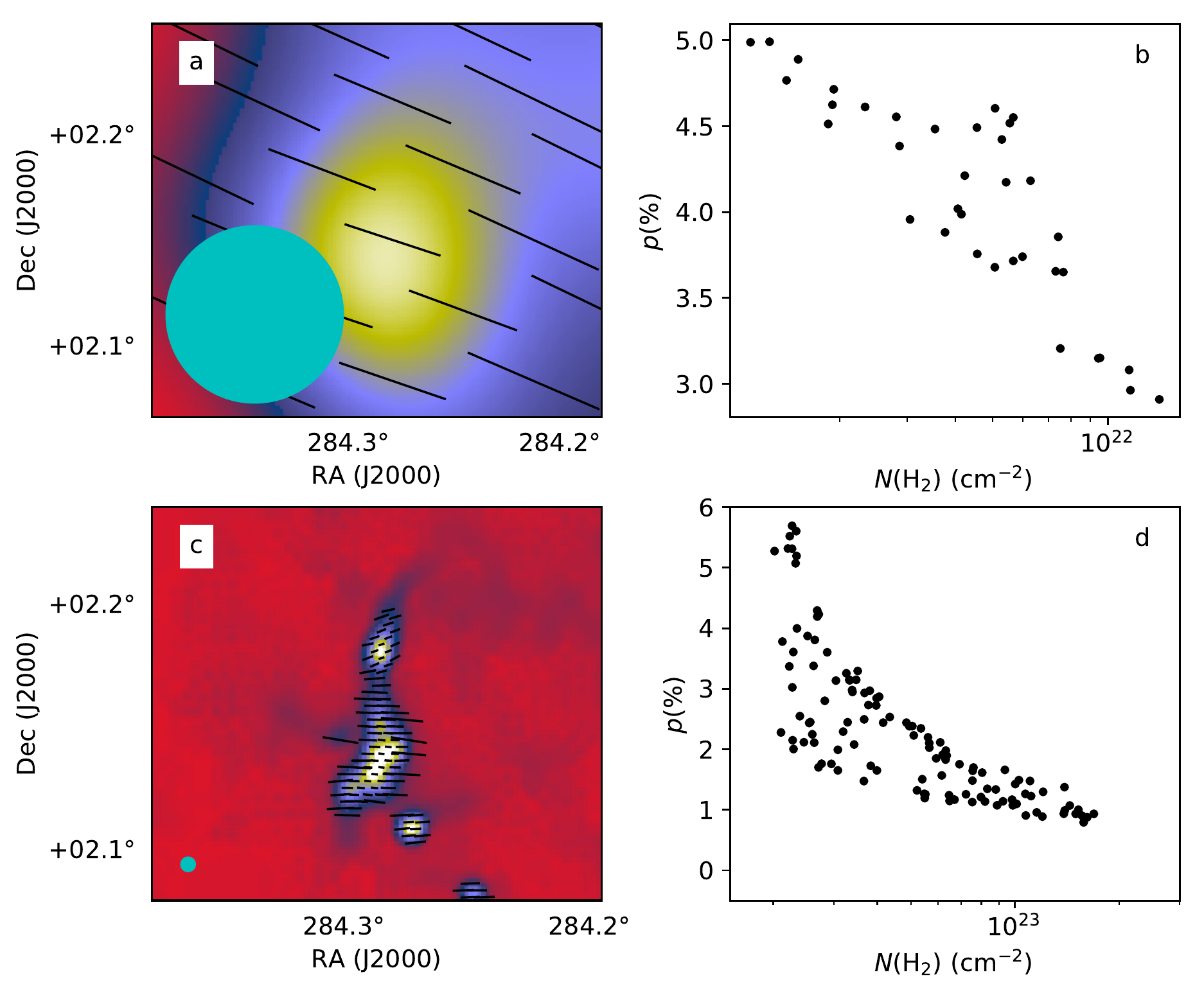}
\caption{
The same model with RAT alignment as in Fig.~\ref{fig:pmod_RAT} but
assuming a factor of two larger grain sizes.
\label{fig:pmod_RAT_DD}
}
\end{figure}

When the alignment predicted by RAT is taken into account
(Fig.~\ref{fig:pmod_RAT}), the POL-2 polarisation fractions drop below
the \Planck values but now the \Planck values show an even slightly
stronger dependence on column density, in contrast with the
observations of the \target field.
Figure~\ref{fig:pmod_RAT}d shows $p$ vs. $N({\rm H}_2)$ also for a
POL-2 simulation where the data are assumed to be high-pass filtered
at a scale of $\theta=100\arcsec$ instead of $\theta=200\arcsec$. The
different filtering does not have a strong effect but leads to some
larger values towards the edges of the filament. 

The RAT calculations of Fig.~\ref{fig:pmod_RAT} were not completely 
self-consistent because they employed the original grain size
distributions (see Sect.~\ref{methods:RT}) while assuming an increased
dust opacity at sub-millimetre wavelengths. We made an alternative
simulation where the grain alignment (and thus the polarisation
reduction factor) was calculated assuming a factor of two larger
grains. The comparison of these results in Fig.~\ref{fig:pmod_RAT_DD}
with the previous calculations of Fig.~\ref{fig:pmod_RAT} should 
partly quantify the uncertainty associated with the particle sizes. A
factor of two change in the grain size in first approximation
corresponds to a factor of two increase in the POL-2 polarisation
fractions. The effect on the simulated \Planck observations is small,
because most grains were already aligned outside the dense filament.
For RAT alignment with the larger grain sizes,
Appendix~\ref{app:polmap} shows results for an alternative model where
the POS magnetic field orientations are taken from POL-2 observations
(at 20$\arcsec$ resolution) for the model volume with $n({\rm H}_2)>3
\times 10^3$\,cm$^{-3}$. There the polarisation fractions are on
average lower only by a fraction of a percent. The difference is
larger in the northern clump, which is sensitive to changes in the
magnetic field configuration, probably because of the stronger
geometrical depolarisation that results from the orthogonality of the
local and the extended fields.

\begin{figure}
\includegraphics[width=8.8cm]{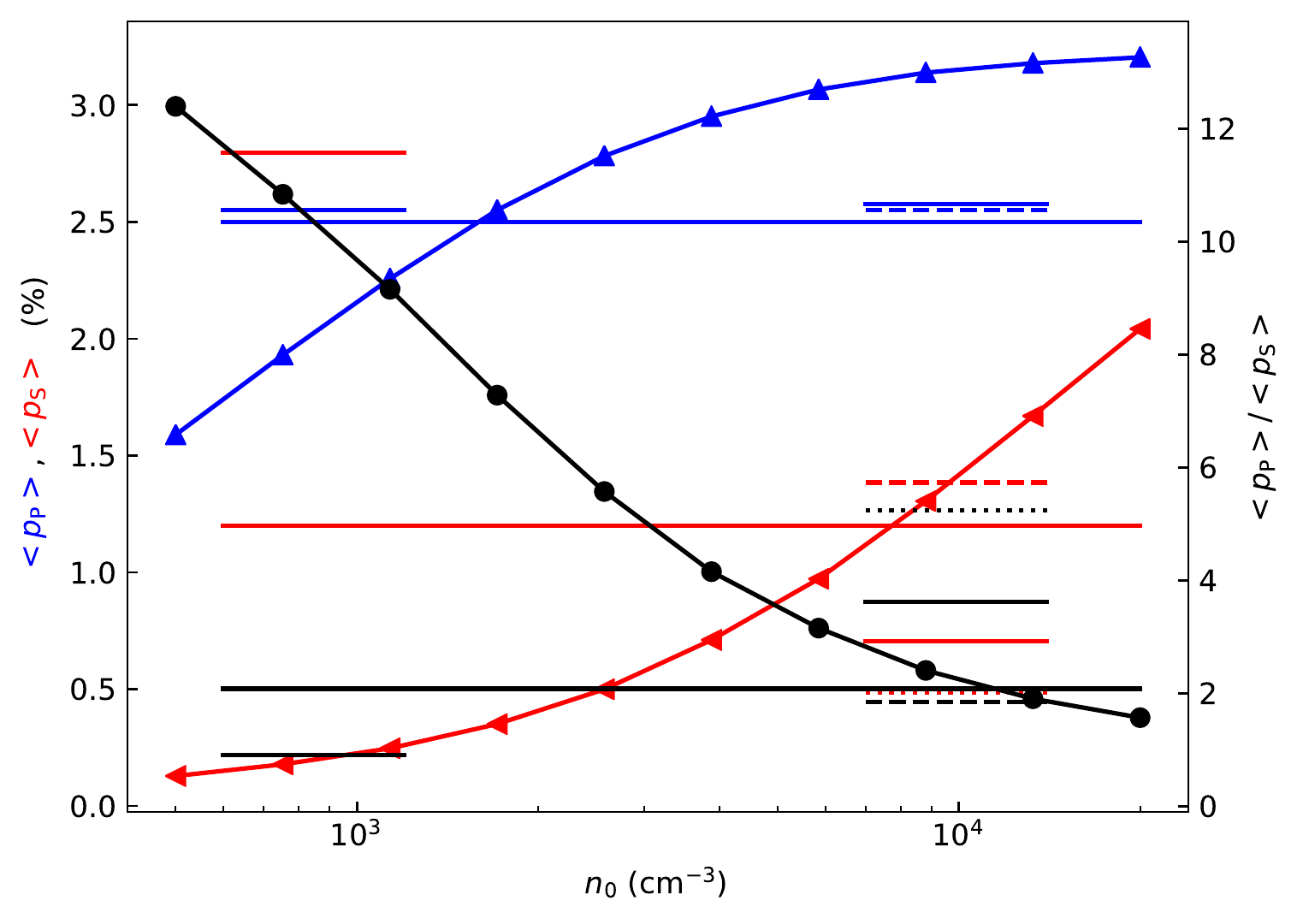}
\caption{
Polarisation fractions for different grain alignment models. The long
horizontal lines show the observed average values of $p$ (left y-axis)
for the \Planck ($p_{\rm P}$, blue line) and the POL-2 ($p_{\rm S}$,
red line) observations. The black long horizontal line stands for the
observed $\langle p_{\rm P}\rangle / \langle p_{\rm S} \rangle$ ratio
(right y-axis). 
The three curves with markers show the corresponding quantities for
models that assume a loss of grain alignment above the density
threshold $n_0$. The results for models with constant grain alignment
are indicated with short horizontal lines near $n=10^3 \,{\rm
cm}^{-3}$. The RAT models are shown as short horizontal lines near
$n=10^4 \,{\rm cm}^{-3}$, for the default model (solid lines), for the
case with larger grain sizes (dashed line), and for an alternative
model with a smaller $\tau(250\mu{\rm m})/\tau({\rm J})$ ratio
(dotted line).
\label{fig:p_vs_model}
}
\end{figure}

Figure~\ref{fig:p_vs_model} compares the constant alignment and RAT
cases to models where $R$ has an ad hoc dependence on the volume
density. The grains are perfectly aligned at low densities but $R$
decreases smoothly to zero above a density threshold $n_0$,
\begin{equation}
R = 0.5 + 0.5 \tanh \left[ \log_{\rm 10}(n_0)-\log_{\rm 10}(n) \right].
\end{equation}
In the modelling the absolute scale of $p$ is left free. In
Fig.~\ref{fig:p_vs_model} the values are scaled so that the \Planck
polarisation fraction is $2.5\%$ or have a maximum value of 3\% in the
case of the $n_0$ dependence. The only real constraint is provided by
the ratio of the \Planck and POL-2 polarisation fractions. The
observed ratio $\sim2.5$ is reached for a density threshold of $n_0 =
10^4$\,cm$^{-3}$. The constant-alignment model predicts a smaller
ratio while the initial RAT model gives a higher ratio. However, if
RAT calculation assume a factor of two larger grain sizes, the ratio
falls slightly below the observed value. 

Finally, we also examined models where the dust opacity ratio was
$\tau(250\mu{\rm m})/\tau({\rm J}) = 1.0 \times 10^{-3}$. The lower
sub-millimetre opacity means that the modelling of surface brightness
data led to larger volume densities and to a lower radiation field
intensity inside the cloud. Both factors contribute to a lower grain
alignment in RAT calculations (see Eq.~\ref{eq:RAT}). In
Fig.~\ref{fig:p_vs_model} the change from $\tau(250\mu{\rm
m})/\tau({\rm J}) = 1.6 \times 10^{-3}$ to $\tau(250\mu{\rm
m})/\tau({\rm J}) = 1.0 \times 10^{-3}$ increases the ratio of \Planck
and POL-2 polarisation fractions by over 30\%.

\begin{figure}
\includegraphics[width=8.8cm]{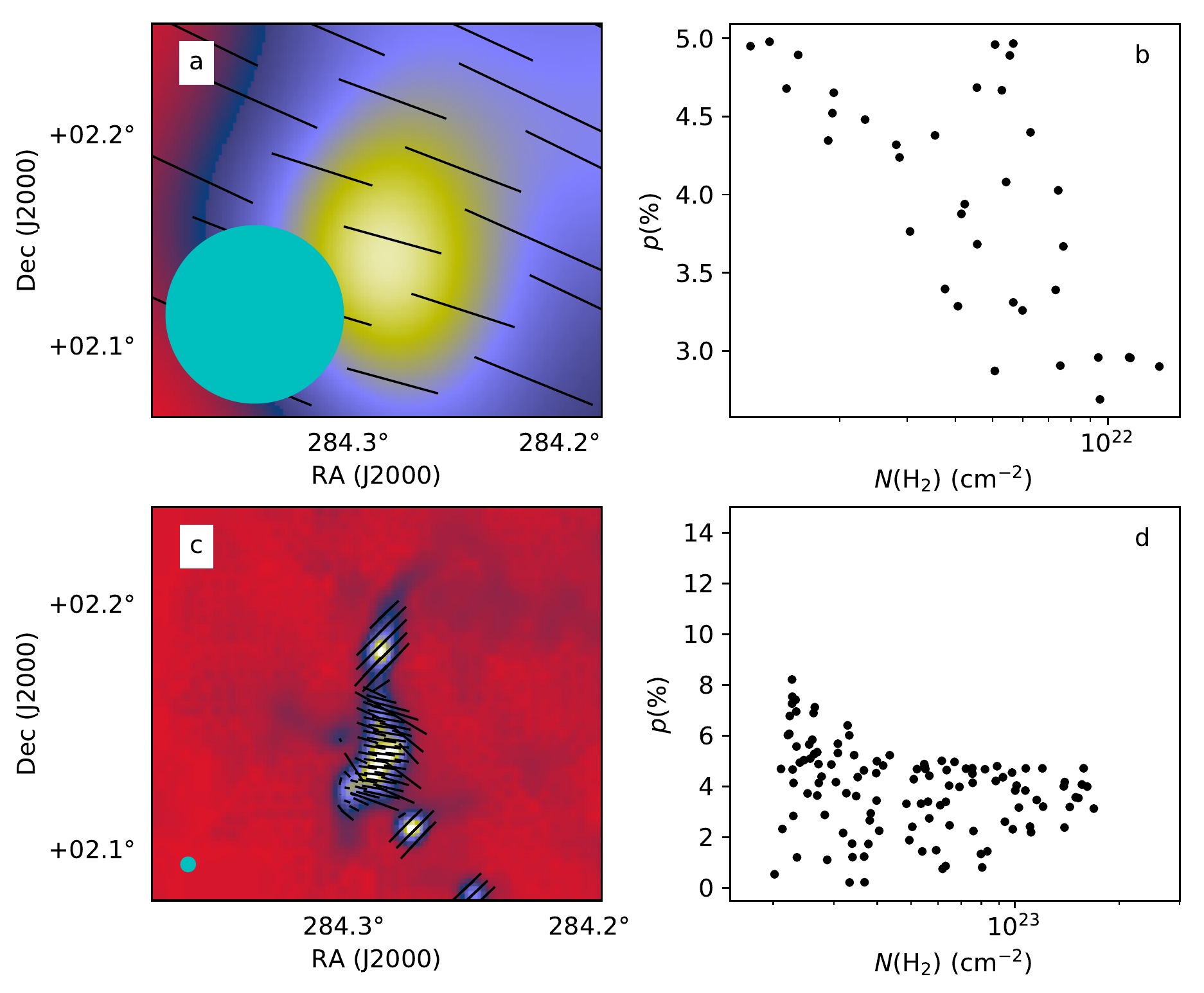}
\caption{
Same as the constant alignment case of Fig.~\ref{fig:pmod_const} but
assuming a toroidal field for the southern clump.
\label{fig:pmod_const_poltor}
}
\end{figure}

\begin{figure}
\includegraphics[width=8.8cm]{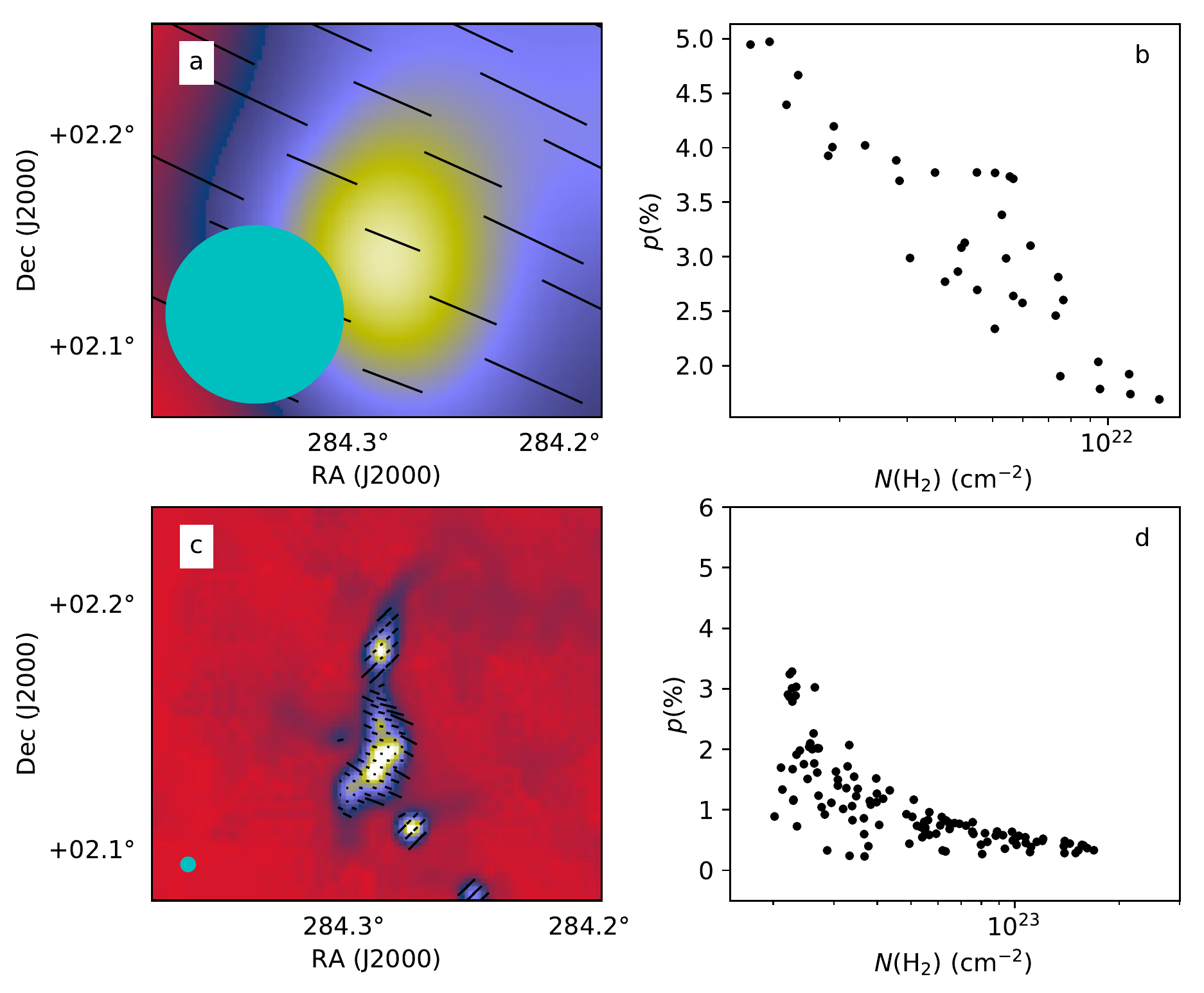}
\caption{
Same as the RAT alignment case of Fig.~\ref{fig:pmod_RAT} but assuming
a toroidal field for the southern clump.
\label{fig:pmod_RAT_poltor}
}
\end{figure}

Unfortunately the observed $p$ ratio does not provide strong
constraints on the grain alignment because the quantitative results
also depend on the assumed magnetic field geometry. As an example, we
tested a field configuration where the southern filament has a
toroidal field at densities above $n=3\times 10^3$\,cm$^{-3}$
while the field in the northern part is still uniform (poloidal). 
A toroidal field is consistent with the observed magnetic field
orientation that is perpendicular to the southern filament. Regarding
the $p$ vs. $N$ relation, it is also an interesting special case that
results in stronger geometrical depolarisation as one moves away from
the symmetry axis.
The results for the constant alignment and an RAT alignment models are
shown in Figs.~\ref{fig:pmod_const_poltor}-\ref{fig:pmod_RAT_poltor}. 
In the constant alignment case the \Planck polarisation fractions have
not changed but the POL-2 values show a larger scatter and a lower
average polarisation fraction. At the borders of the filament, where
the toroidal field of is along the LOS in the dense medium, the
polarisation vectors have turned parallel to the large-scale field.
The change is qualitatively similar for the RAT case
(Fig.~\ref{fig:pmod_RAT_poltor}). The $p$ vs. $N$ relation is
flatter than in Fig.~\ref{fig:pmod_RAT} but not significantly
different from the observations shown in Fig.~\ref{fig:p_vs_N_S}b. In
the model the toroidal configuration also causes a stronger drop in
the $Planck$-observed polarisation fraction. A more extended toroidal
component (e.g. in a test where the density threshold was reduced from
$n=3\times 10^3$\,cm$^{-3}$ to $n=5\times 10^2$\,cm$^{-3}$) would
cause clear changes also in the orientation of the $Planck$-detected
polarisation vectors. However, these effects are dependent on our
assumptions of the LOS matter distribution and would disappear if most
of the extended material was located far from the filament.

\section{Discussion}  \label{sect:discussion}

In the following we discuss the results regarding the observable dust
properties (Sect.~\ref{dis:dust}) and the polarisation
fraction(Sect.~\ref{dis:pol}).

\subsection{Dust opacity in the \target field} \label{dis:dust}

The \target field has a high-column-density background of $N({\rm
H}_2)\sim 10^{22}$\,cm$^{-2}$ (Fig.~\ref{fig:SPIRE_MBB}). Even after
the subtraction of this background, the column densities are above
$N({\rm H}_2)=2\times 10^{22}$\,cm$^{-2}$ over a filament length of
$\sim 7 \arcmin$ ($\sim$6\,pc). With a typical filament width of $\sim
40\arcsec$, the average volume density is of the order of $n({\rm
H}_2) \sim 10^4$\,cm$^{-3}$. With the high volume density and the dust
temperatures below 14\,K (with minima close to $T=12$\,K, see
Sect.~\ref{sect:MBB_Herschel}-\ref{sect:MBB_all}), the conditions are
suitable for grain evolution. The properties of dust opacity can thus
be expected to be different from those of diffuse clouds. 

The comparison of dust sub-millimetre emission and NIR/MIR
observations gave an average opacity ratio of $\tau(250\mu{\rm
m})/\tau({\rm J})=(2.55 \pm 0.03) \times 10^{-3}$, which also is close
to the behaviour of the southern clump. The relation is steeper in the
northern clump, although only in small $\sim1\arcmin$ region that is
close to some 70\,$\mu$m sources. Internal heating could reduce the
degree to which dust optical depth $\tau(250\,\mu{\rm m})$ is
underestimated. 
The fit at $\tau({\rm J})<$6 gave a lower value of $\tau(250\mu{\rm
m})/\tau({\rm J})=(1.72 \pm 0.04) \times 10^{-3}$. 
In the RT models, $\tau({\rm J})\sim6$ ($\tau(250\,\mu{\rm
m})\sim 0.012$) typically corresponds to a LOS peak volume density of
the order of $n({\rm H_2}) = 5 \times 10^{3}$\,cm$^{-3}$. 
Planck studies have found in diffuse regions values $\tau(250\,\mu{\rm
m})/N({\rm H}) \sim 0.55 \times 10^{-25}$\,cm$^2$\,H$^{-1}$
\citep{planck2013-p06b, planck2013-XVII}. With the \citet{BSD}
relation between the reddening and Hydrogen column density and with
the $R_{\rm V}=3.1$ extinction curve \citep{Cardelli1989}, this 
corresponds to $\tau(250\mu{\rm m})/\tau({\rm J})= 0.41 \times
10^{-3}$. The \target sub-millimetre opacity values relative to NIR
are thus more than four times higher than in diffuse clouds. 

The correlation between density and sub-millimetre opacity is known
from numerous studies \citep{Kramer2003, Stepnik2003,Lehtinen2004,
delBurgo2005, Ridderstad2010, Bernard2010, Suutarinen2013, Ysard2013,
Martin2012, Roy2013, Svoboda2016, Webb2017}. \citet{KAPPA} used
\Herschel observations to study sources from the Planck Catalogue
of Galactic Cold Clumps \citep[PGCC,][]{PGCC}. For a sample of 23
fields the average dust opacity was $\tau(250\mu{\rm m})/\tau({\rm
J})=1.6 \times 10^{-3}$. Given that this value corresponds to sources
with optical depths below $\tau({\rm J})\sim 3$, it is in qualitative
agreement with the results of the present study. In \citet{KAPPA}, the
maximum values derived for individual clumps were $\tau(250\mu{\rm
m})/\tau({\rm J})= 4 \times 10^{-3}$, similar to the value of the
northern clump of \target. However, in that study the NIR extinction
estimates were based on background stars only and, in the case of 
high optical depths, have a higher uncertainty. 

A large $\tau(250\,\mu{\rm m})/\tau({\rm J})$ ratio also could result
from an underestimation of the $\tau({\rm J})$ values in the \target
field. The lack of background stars should not directly affect the
$\tau({\rm J})$ estimates of the densest filament, which are based
more on MIR data but if the MIR emission had a strong foreground
component, that could also lead to low $\tau({\rm J})$ estimates.
he uncertainty of extinction depends non-linearly on the foreground
intensity but it is probably only some 10-20\% \citep[see
also][]{ButlerTan2012}.
In the analysis of \citet{KainulainenTan2013}, the MIR extinction also
was tied to the NIR extinction measurements, which makes large errors
in the extinction levels improbable. Even with an uncertainty of 30\%
\citep[see][]{KainulainenTan2013}, the 1-$\sigma$ lower limit is still
above the average value of \citet{KAPPA}.


The estimates of the sub-millimetre opacity $\tau(250\,\mu{\rm m})$
are likely to be biased because they were derived from
single-temperature MBB fits, ignoring the effects of LOS temperature
variations. Figure~\ref{fig:compare_NH2} compared the true column
densities of a model cloud to those derived from the synthetic surface
brightness maps. This also serves as an estimate for the bias of the
$\tau(250\,\mu{\rm m})$ values. At a column density of $N({\rm
H}_2)=10^{23}$\,cm$^{-2}$ the estimated bias is more than a factor of
two. This column density is higher than the values $N({\rm H}_2) \la
5\times 10^{22}$\,cm$^{-2}$ estimated for \target. However, these are
consistent if the latter are underestimated by the factor indicated by
the modelling. Quantitatively the bias predictions depend on how well
the models represent the real cloud. A stronger internal heating would
decrease the bias, at least locally. Conversely, the observations do
not give strong constraints on the maximum (column) densities and
higher optical depths would lead to a higher bias. The relative bias
is likely to be at least as large in $\tau(250\,\mu{\rm m})$ as in
$\tau({\rm J})$. Thus, the true value of $\tau(250\mu{\rm
m})/\tau({\rm J})$ may be even higher than the quoted estimate of
$2.55 \times 10^{-3}$.

The calculation of the sub-millimetre opacity assumed a fixed dust
opacity spectral index of $\beta=1.8$, which is close to the spectral
index estimated from the data (see below). An error of $\sim 0.1$ in
the spectral index would correspond only to $\sim$10\% error in
opacity. For further discussion of the effect of the spectral index
and the extinction law, see \citet{KAPPA}. 
Finally, if the actual resolution of our $\tau_{\rm 5}(250\,\mu{\rm
m})$ map were lower than the nominal 15$\arcsec$, this would lower the
$\tau(250\,\mu{\rm m})/\tau({\rm J})$ estimates. When the maps were
convolved to a lower resolution with a Gaussian beam with
FWHM=$20\arcsec$, the opacity ratio changed by less than 0.03 units.
This shows that the result is not sensitive to the resolution.

We estimated a dust opacity spectral index of $\beta \sim 1.9$ for the
main \target filament, using data at $\lambda \le 850\,\mu$m. 
For the northern part separately, the $\beta$ values were slightly
lower but there both the SCUBA-2 450\,$\mu$m and 850\,$\mu$m values
were above the relation fitted to SPIRE data
(Fig.~\ref{fig:SED_corr}). These offsets and thus the lower $\beta$
values could be caused by uncertainties in the spatial filtering or,
when using background-subtracted data, the reference region being
located at a larger distance in the south (see
Fig.~\ref{fig:overview}b).

The derived $\beta$ value is practically identical to the median value
$\beta=1.91$ that \citet{GCC-VI} reported for a sample of GCC clumps
based on \Herschel data with $\lambda\le 500\,\mu$m. In \citet{GCC-V}
the inclusion of longer wavelength \Planck data resulted in a smaller
value of $\beta=1.66$. This wavelength-dependence had been
demonstrated, for example, in \citet{planck2013-XIV}. More recently,
\citet{Juvela_2018_pilot} analysed \Herschel and SCUBA-2 observations
of cores and clumps within some 90 PGCC fields. For those objects the
median value at \Herschel wavelengths was $\beta \sim 1.8$ (although
with significant scatter) and the inclusion of the SCUBA-2 850\,$\mu$m
data point decreased the value closer to $\beta \sim 1.6$. The higher
spectral index value of the \target field is again in qualitative
agreement with \target being more dense and, in terms of dust
evolution, probably a more evolved region.  Spectral indices are
generally observed to be higher towards the end of the prestellar
phase while in the protostellar phase one may again observe lower
values \citep{Chen2016,Li2017_Class0,Bracco2017}. This may be caused
by dust evolution (e.g. further grain growth), by the temperature
variations resulting from internal heating, or directly by problems
associated with the analysis of observations at very high column
densities \citep{Shetty2009a, Shetty2009b, Juvela2012_bananasplit,
Malinen2011, Juvela2012_Tmix, Ysard2012, Juvela2013_TBmethods,
Pagani2015}. The \target field does contain a number of protostellar
objects but in this paper we only examined the average $\beta$ over
the whole filament.

\subsection{Polarisation} \label{dis:pol}

The polarisation observations trace a combination of magnetic field
morphology and grain properties. \Planck and POL-2 provided different
views into the structure of the magnetic fields. The large-scale field
was found to be uniform while the small-scale structure associated
with the dense filament was more varied, even with partly orthogonal
orientations. The \Planck and POL-2 data are not contradictory because
of the large difference in the beam sizes. POL-2 also is not sensitive
to the extended emission (Sect.~\ref{sect:geometry}). Nevertheless,
the change in the field orientations must be constrained to a narrow
region at and around the main filament. Otherwise these would be
visible also in \Planck data, as deviations from the average field
orientation and as a reduced net polarisation. If connected to the
gravitational instability of the filament and of the embedded cores,
the effects are naturally confined in space. If strong accretion flows
extend to a distance of $\sim$1\,pc, this corresponds to only $\sim 1
\arcmin$ in angular distance.

In the following we discuss in more
detail the observations and the modelling of the polarisation
fraction.

\subsubsection{Observed polarisation fractions} \label{dis:obs_p}

The polarisation fraction $p$ is generally observed to decrease
towards dense clouds and especially towards dense (prestellar) clumps
and cores \citep{Vrba1976, Gerakines1995, WardThompson2000,
Alves2014_pol, planck2014-XX, planck2014-XXXIII}. 

The Planck data were used to characterise the large-scale environment
of the \target filament. \Planck data did not show any clear column
density dependence over the examined $2\degr \times 2\degr$ area 
and the bias-corrected polarisation fraction estimate $p_{\rm mas}$
remained within a narrow range between 1\% and 3\%
(Fig.~\ref{fig:plot_Planck_INP}). The same was already evident based
on the polarisation vectors that were plotted in
Fig.~\ref{fig:S_Planck}a. These show that the large-scale magnetic
field orientation is very uniform, also in the area covered by POL-2
observations. The dynamical range of column density over this area was
only a factor of $\sim$3, which partly explains the lack of a clear
correlation. The polarised signal is largely dominated by extended
emission not directly connected to the dense filament. Because of the
low Galactic latitude, there can be contributions from many regions
along the LOS, which would tend to decrease the observed polarisation
fraction \citep{Jones1992, planck2014-XIX}. In \citet{Liu_G35_pol},
multiple velocity components were identified from $^{13}$CO and
C$^{18}$O line data.  
Only the 45 km\,s$^{-1}$ feature, the strongest of the kinematic
components, is associated with the main filament.
This is consistent with the picture shown by Fig.~\ref{fig:overview}a
and Fig.~\ref{fig:SPIRE_MBB}. At \Herschel resolution the filament
rises more than a factor of four above the extended column density
background while at the \Planck resolution it is associated with a
mere 50\% increase above the background surface brightness. With the
added effect of spatial filtering, POL-2 measurements are only
sensitive to the emission from the main filament. Conversely, \Planck
measurement could be slightly affected by the dense filament, unless
that is associated with lower polarisation fractions.

The uniformity of the magnetic field is shown quantitatively by the
polarisation angle dispersion function $S$ calculated from the
\Planck data. Figure~\ref{fig:SpI}c showed the relation 
$\log \, p = -0.670\,\log \, S - 0.97$ 
that \citet{Fissel2016} derived from BLASTPol 
observations of the Vela C molecular cloud (``ISRF-heated
sightlines''). Compared to this, the \Planck data of the \target field
indicate much lower ($p$, $S$) parameter combinations. 
Figure~\ref{fig:SpI}c also included the relation $\log_{10}(S) =
-0.834 \times \log_{10}(p_{\rm mas}) -0.504$ that
\citet{planck2014-XIX} obtained at larger scales (FWHM=$1\degr$,
$\delta=30\arcmin$), using data over a large fraction of the whole
sky. 
This relation corresponds to higher values of $S$ and $p$ than found
in the \target field. The comparison to the \Planck results
\citep{planck2014-XIX} is not straightforward because observations
probe different linear scales (FWHM=1$\degr$ compared to 
FWHM=$9-15\arcmin$ in our analysis). Figure~\ref{fig:SpI} did not
show a systematic dependence on the scale.  \citet{planck2016-l11B}
detected a shallow relation of $S \times p \propto FWHM^{0.18}$, which
would thus be detectable only by using a wider range of scales.
According to that relation, the $S \times p$ values at 15$\arcmin$
resolution should be only some 22\% smaller than at 1$\degr$
resolution.

Instead of grain alignment or the statistical averaging of LOS
emission with different polarisation angles), lower \target
polarisation fractions (in relation to $S$) could be explained by a
larger LOS component of the magnetic field, a hypothesis that we
cannot directly test.
Large angles between the magnetic field and the POS would tend to
be associated with large values of $S$ \citep{Chen2016_B}. However,
this is only a statistical correlation and cannot be used to exclude
the possibility of a large LOS field component, which also would be
consistent with the Galactic longitude and distance of the \target
field.

Compared to the low-resolution \Planck data, the POL-2 data are more
affected by observational noise. The bias-corrected polarisation
fraction estimates $p_{\rm mas}$ are reliable for the central filament
and are there $p_{\rm mas}\sim 1\%$ irrespective of the data
resolution between FWHM=$15\arcsec$ and FWHM=$60\arcsec$ (see
Fig.~\ref{fig:p_N_fwhm}). 

The POL-2 $p_{\rm mas}$ estimates should be unbiased when used at
40$\arcsec$ resolution and when the analysis is restricted to column
densities $N({\rm H}_2)>2 \times 10^{22}$\,cm$^{-2}$. In this range,
the decrease of polarisation fraction as a function of column density
is significant (Fig.~\ref{fig:p_vs_N_S}). The drop from $N({\rm
H}_2)=2 \times 10^{22}$\,cm$^{-2}$ to $N({\rm H}_2)=4 \times
10^{22}$\,cm$^{-2}$ is from 1.5\% to values below 1\%.
Because low-density regions along the LOS (in front of and behind the
filament but possibly associated to it) will produce some polarised
intensity, a non-zero polarisation fraction does not exclude the
possibility of a complete loss of grain alignment within the densest
filament. 

The simulations presented in Appendix~\ref{app:sim} confirm that the
negative {\rm correlation between $p_{\rm mas}$ and $N({\rm H}_2)$ is
larger} than expected based on the noise alone
(Fig.~\ref{fig:VI_sim_slope}). These also show that with column
densities $N({\rm H}_2)>10^{22}$\,cm$^{-2}$ and with FWHM$>20\arcsec$
most of the resolution dependence of $p$ can be attributed to
geometrical depolarisation (Fig.~\ref{fig:VI_sim_depol}). This is
related to the $S$ function. \Planck data showed some anticorrelation
{between $p_{\rm mas}$ and $S$} (Fig.~\ref{fig:VI_maps}) but this
dependence was not clear in POL-2 data. While Fig.~\ref{fig:S_SCUBA}
gave some indications, especially in the northern part, statistically
the anticorrelation remained weak (Fig.~\ref{fig:p_S_N}d). Reliable
mapping of $S$ would require higher SNR further out from the central
filament. On the other hand, $S$ also is also affected by the
filtering of the extended emission.

\begin{figure}
\includegraphics[width=8.8cm]{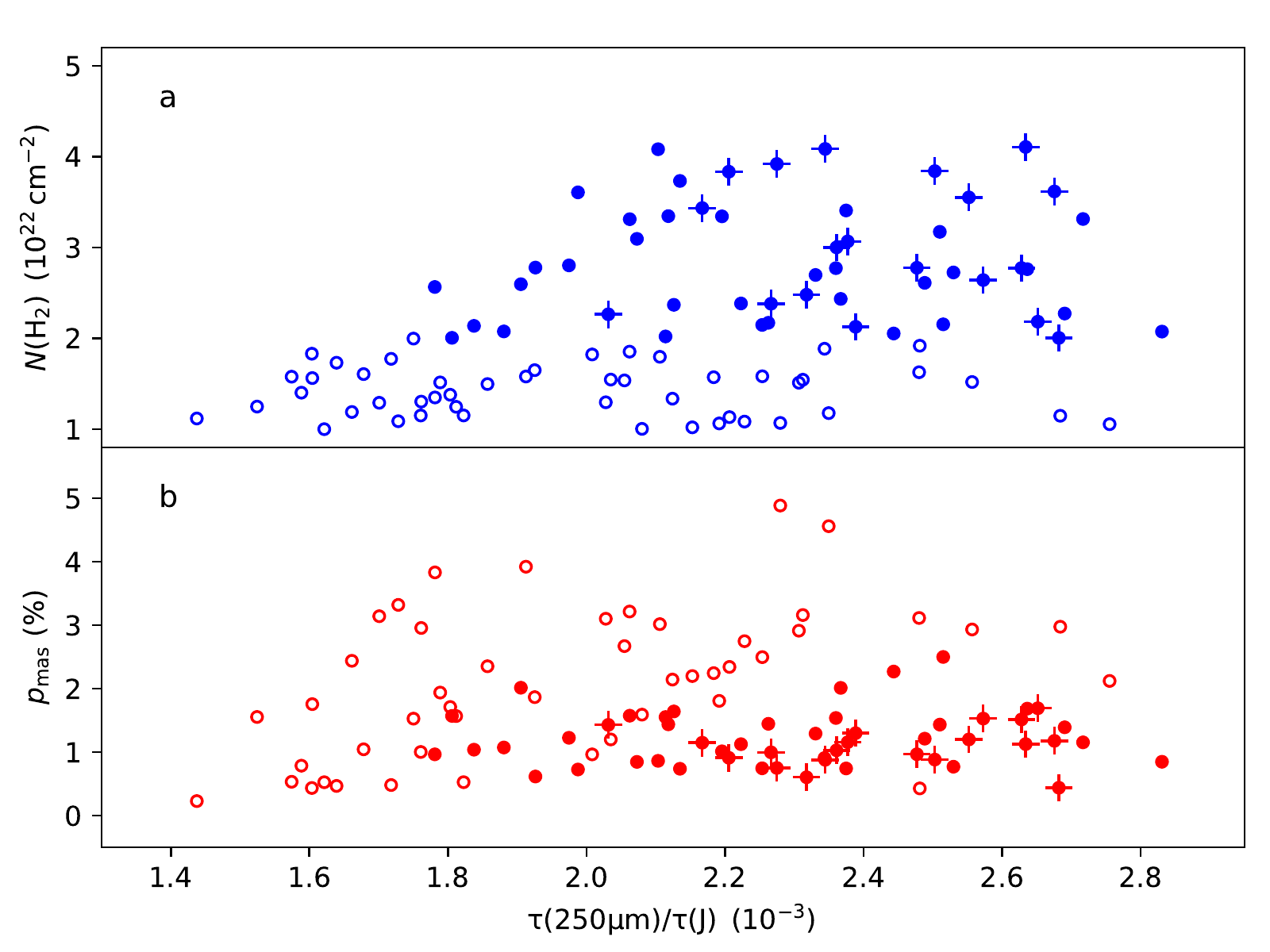}
\caption{
Correlations between opacity ratio $\tau(250\mu{\rm})/\tau({\rm J})$
and column density (frame a) or POL-2 polarisation fraction (frame b).
The data are at 40$\arcsec$ resolution and sampled at half-beam steps.
The points coinciding with the point sources marked in
Fig.~\ref{fig:p_vs_N_S} are marked with plus signs.
Data for column densities $N({\rm H}_2)<2\times 10^{22}\,{\rm
cm}^{-2}$ are plotted with open symbols and for these the $p_{\rm
mas}$ estimates may be inaccurate.
\label{fig:p_vs_tau}
}
\end{figure}

Because polarisation fraction is correlated with column density, it
could be expected to be correlated with tracers of dust evolution such
as $\tau(250\mu{\rm m})/\tau({\rm J})$. However, physically high
volume density should be associated with grain growth, which in turn
works against drop of polarisation that in the RAT scenario is caused
by the weakening of the radiation field. Figure~\ref{fig:p_vs_tau}
shows no clear correlation between $p_{\rm mas}$ and the opacity ratio
$\tau(250\mu{\rm m})/\tau({\rm J})$ and only the correlation between
$\tau(250\mu{\rm m})/\tau({\rm J})$ and $N({\rm H}_2)$ is significant
(correlation coefficient $r=0.45$, significant at $\sim 98 \%$ level
when calculated with data sampled at FWHM steps). As mentioned in
Sect.~\ref{sect:polfrac_POL2}, 70\,$\mu$m sources tend to have lower
than average polarisation fractions, in this sample 1.1\% vs. 1.4\%.
Part of this is caused by their higher SNR and thus lower bias,
although for the data in Fig.~\ref{fig:p_vs_tau} the bias should not
be very significant.

\subsubsection{Simulations of polarised emission} \label{dis:simu_p}

We used RT simulations to probe the effects that magnetic field
geometry and grain alignment variations can have on the polarisation
observations. Because the absolute values of the polarised intensity
depend on poorly known grain properties, small-scale magnetic field
structure, and the strength of the LOS magnetic field component, we
concentrated on the ratio of the simulated \Planck and POL-2
polarisation fractions.

In the case of constant grain alignment, the polarisation fractions
were even higher in the simulated POL-2 data than in the simulated
\Planck data (Fig~\ref{fig:pmod_const}), in clear contradiction with
observations.  
When grain alignment was assumed to depend on volume density, the
ratio of \Planck and POL-2 polarisation fractions could be matched
with a density threshold of $n({\rm H}_2)=10^{4}$\,cm$^{-3}$. This can
be compared to \citet{Alves2014_pol} \citep[see
also][]{Alves2014_pol_err}, who analysed optical, NIR, and
sub-millimetre observations of a starless core in the Pipe nebula.
They deduced a loss of alignment at densities $n({\rm H}_2)> 6 \times
10^{4}$\,cm$^{-3}$. The analyses are of course affected by many
uncertainties (see Sect.~\ref{dis:caveats}) and, because of the larger
distance, the linear resolution of our observations is much lower. If
the difference in the density thresholds were significant, it could be
related to the different nature of the sources (e.g. evolutionary
stage and internal heating).

The low $p$ values of the actual \target observations could be
explained by geometrical depolarisation resulting from line tangling
within the densest filament. However, this does not seem a likely
explanation given the local uniformity of the polarisation angles and
the relatively constant polarisation fraction observed over the whole
filament. The synthetic \Planck $p$ values also dropped by $\sim$30\% as a
function of column density (Fig.~\ref{fig:pmod_const}b), while in the
actual observations no clear column-density dependence was seen
(Fig.~\ref{fig:plot_Planck_INP}c). However, our models described only
a $13\arcmin \times 13\arcmin$ area and a volume of $\sim(11\,{\rm
pc})^3$. About half of the actual \Planck signal is coming from a more
extended cloud component (cf. Fig~\ref{fig:overview}a-b) that appears
to be associated with at least a factor of two higher polarisation
fractions than the main filament. If the extended component was added,
the $p$ values would not change at low column densities while the
values towards the filament would increase significantly. Therefore,
the $p$ vs. $N$ relation of the simulated \Planck observations is not
necessarily incompatible with the observations.
A more remote possibility for such $p$ vs. $N$ relations would be to
assume that the direction of the large-scale field is changing so that
(unlike in simulations) it has a larger LOS component in the
low-column-density regions \citep[cf.][]{planck2014-XXXIII}.

Figure~\ref{fig:pmod_RAT} showed the results for a model where the
grain alignment efficiency varied as predicted by RAT. In the
synthetic \Planck data the polarisation angles were again uniform,
similar to the real \target observation. 
%
The simulated POL-2 polarisation fractions were too low compared to
the \Planck values outside the filament. The discrepancy could be
corrected by using a factor of two larger grains in the alignment
calculations (Fig.~\ref{fig:pmod_RAT_DD}). The observed high
sub-millimetre opacity indicates some grain growth, which, however, is
unlikely to be as large as a factor of two. In reality, the
effects of grain growth are more complex because also the grain shapes
are probably affected. 

In the more empirical modelling we simply assumed that grain alignment
is lost above a certain volume density. The observed ratio of
polarisation fractions was recovered when the threshold was $n({\rm
H}_2) \sim 10^{4}$\,cm$^{-3}$. This would thus be consistent with no
polarised intensity being emitted from the densest filament. The
differences between the large-scale and the small-scale field
morphologies would thus only probe the envelopes of the filament and
the cores embedded within the filament (Appendix~\ref{app:polmap}).

\subsubsection{Uncertainties of polarisation simulations}
\label{dis:caveats}

There are several caveats concerning the polarisation simulations and
the comparison with the \Planck and POL-2 observations. 

In the modelling, the filtering of the POL-2 data was described using
a simple high-pass filter. Figure~\ref{fig:pmod_RAT} showed that the
difference between filtering scales $\theta=100\arcsec$ and
$\theta=200\arcsec$ was not significant. This is understandable
because any high-pass filtering with a scale larger than the filament
width will effectively remove all information of the uniform
large-scale field. However, the actual filtering in POL-2 data
reduction is not necessarily this simple. One needs simulations with
the actual POL-2 reduction pipeline, to estimate the general effect of
the filtering and to check for potential differences in the way the
different Stokes vector components get processed. 

The minimum size of the aligned grains and thus the polarisation
reduction associated to RAT was calculated for the original grain size
distributions of \citet{Compiegne2011} while the sub-millimetre
emissivity was subsequently altered (see Sect.~\ref{methods:RT}). A
subsequent factor of two increase of the grain sizes had no effect on
the simulated \Planck values but increased the POL-2 polarisation
fraction by $\sim$30\%. The total uncertainty related to the grain
properties could be higher. The results depend not only on the grain
size distribution but also on other poorly known factors that are
related to the grain composition, grain shapes, and optical
properties. Rather than an indication of specific dust opacities or
grain sizes, the model comparison in Fig.~\ref{fig:p_vs_model} should
only be taken as an indication of some of the uncertainties that
affect the modelling of polarised dust emission.

Our models did not consider the effect of internal heating sources.
While young stellar objects (YSOs) increase the radiation flux in their
environment, this does not necessarily lead to significant enhancement
of the observed polarisation. The angular momentum caused by the
radiation reaches its maximum when radiation direction is aligned with
the magnetic field \citep{HoangLazarian2009}. If the grain's angular
momentum $J$ is aligned with the magnetic field, only the RAT
component projected onto the magnetic field direction is able to spin
up the grain, leading to Eq. (38) of \citet{HoangLazarian2014}: 
$J^{\rm RAT}_{\rm max}(\psi) = J^{\rm RAT}_{\rm max}(\psi=0) \cos \psi$. 
If the aligning radiation is directed towards the observer, the
angular momentum reaches its maximum when also the magnetic field is
parallel to the observer's LOS. This leads to strong depolarisation
thanks to the $\cos^2 \gamma$ term. On the other hand, for $\cos^2
\gamma = 1$, $\cos \psi = 0$ and hence collisions are able to disrupt
the grain alignment. Since RATs are effective mostly at UV and optical
wavelengths, deeply embedded and already reddened YSOs may have an
effect only in their immediate surroundings, which may correspond to a
small part of the total dust along the LOS. Combined with the large
beams of the sub-mm telescopes, the contribution of YSOs on the
polarisation of distant clouds may thus remain negligible.

The comparison of the simulated \Planck and POL-2 observations assumed
that \Planck is able to give an upper limit for the intrinsic
polarisation. If the magnetic field had a strong random component, the
$p_{\rm P}$ values observed with the large \Planck beam would however
underestimate this maximum polarisation fraction. Conversely, if line
tangling were confined to the filament, a possibility also not probed
by our models, this would lead to lower $p$ in POL-2 observations.
However, the RAT simulations appear to leave little room for
additional geometrical depolarisation within the filament.


For the most part, our simulations assumed that the LOS magnetic
field component is constant and similar both at large scales and
within the filament. The ratio of the \Planck and POL-2 values
$p_{\rm P}/p_{\rm S}$ would be lower if the LOS field component was
systematically larger outside the filament. For example, the
large-scale field could be at 45 degree angles relative to the LOS
while a toroidal field in the filament would make the field more
perpendicular to the LOS towards the filament centre. Because the
polarised intensity is proportional to $\cos^2 \gamma$, the effect on
the $p_{\rm P}/p_{\rm S}$ ratio would be a factor of two. The effect
would be quite significant and, if it were true, would even more
strongly point to a complete loss of grain alignment inside the
filament. Of course, the opposite situation where the field
outside the filament is more closely aligned with the POS is in
principle also possible.

The dust temperatures obtained from RT calculations contain some Monte
Carlo noise but, once the temperature field is fixed, the $I$, $Q$,
and $U$ maps are mutually consistent to a very high accuracy. The
simulated $p$ values do not therefore suffer from noise bias but there
are other uncertainties that are related to the underlying RT model.
The intensity and the spectrum of the radiation field at the
boundaries of the model volume affect especially the RAT calculations.
Different assumptions of the dust sub-millimetre opacity were already
seen to lead to models with different column densities and,
consequently, different internal radiation fields. In
Fig.~\ref{fig:p_vs_model}, compared to the RT model with
$\tau(250\mu{\rm m})/\tau({\rm J})=1.6 \times 10^{-3}$, the assumption
of $\tau(250\mu{\rm m})/\tau({\rm J})=1.0 \times 10^{-3}$ lead to a
30\% higher ratio of the \Planck and POL-2 polarisation fractions. The
continuum observations also do not give strong constraints on the
volume densities. If we assumed the cloud to be more extended in the
LOS direction, this would decrease the volume density and increase the
short-wavelength radiation inside the cloud. Both factors would
enhance grain alignment (see Eq.~(\ref{eq:RAT})). An inhomogeneous
cloud structure would increase the penetration of the external
radiation but this would be partly compensated by the larger volume
densities.

The best way to combat many of the listed uncertainties would be to
study large samples of sources and use statistical arguments regarding
the field orientation and the cloud shapes.

\subsection{Field geometry vs. grain alignment} \label{sect:vs}

\Planck studies have concluded that at large scales the relation
between polarisation fraction $p$ and column density $N$ can be
explained by the structure of the magnetic field without variations in
the grain alignment.
\citet{planck2014-XX} found that the $p(N)$ relation was fairly well
reproduced by MHD simulations with constant grain alignment but their
observations only probed column densities up to $N({\rm H}_2)=5\times
10^{21}$\,cm$^{-2}$. They refrained from drawing any conclusions for
high column densities $N({\rm H}_2)>10^{22}$\,cm$^{-2}$ for which also
their MHD runs were not well suited. \citet{Soler2016} similarly
concluded that at $N({\rm H}_2) < 5\times 10^{21}$\,cm$^{-2}$ the
polarisation angle dispersion and its relation to the polarisation
fraction were mainly produced by fluctuations in the magnetic field
structure. 

\citet{planck2016-l11B} examined in more detail the relationships
between $p$, $N$, and $S$. The uniformity of the product $p \times S$
was noted as evidence that the drop in $p(N)$ is caused by the field
structure rather than a loss of grain alignment.  At a resolution of
40$\arcmin$, the $p(N)$ relation of Gould Belt clouds was followed up
to $N({\rm H}_2)=5\times 10^{21}$\,cm$^{-2}$. The probed linear scales
were thus $\sim$1.6\,pc for nearby clouds and $\sim$5\,pc for Orion.
The column densities are low because of the spatial resolution, which
means that even towards the densest structures the polarised signal
may be strongly affected by extended emission (usually with a higher
polarisation fraction). However, with data at 10$\arcmin$ resolution,
\citet{planck2016-l11B} Fig. 21 traces the product $p \times S$ up to
$N({\rm H}_2)=10^{23}$\,cm$^{-2}$. The reduction of the grain
alignment efficiency was estimated to be less than 25\% between the
diffuse ISM and the highest column densities. However, even at the
highest column densities the data do not exclusively probe the
emission from regions of high volume density. Some drop in $p \times
S$ was observed beyond $N({\rm H}_2)\sim 3 \times 10^{22}$\,cm$^{-2}$.
This was not very significant but, in principle, could hint at effects
of reduced grain alignment being visible even in \Planck data.
Alternatively, this could be related to a qualitative change in the
field morphology at scales dominated by the gravity.

These results are not directly comparable to our POL-2 study of the
massive filament G035.39-00.33. Even with the higher angular
resolution of POL-2 (0.56\,pc for the 40$\arcsec$ angular resolution
and the 2.9\,kpc distance) the filament is not fully resolved.
However, the observed column densities, $N({\rm
H}_2)>10^{22}$\,cm$^{-2}$, and especially the volume densities are
much higher than what is reached with \Planck. 
\citet{planck2014-XX} concluded that large column densities can be
associated to low polarisation fraction especially because of the
accumulation of many LOS structures. Our POL-2 observations target a
single object that dominates the LOS column density. All extended
emission also is already filtered out from the POL-2 data.  For
\targetn, the difference to the \Planck data of the same region is
particularly striking because of the low Galactic latitude (see
Fig.~\ref{fig:overview}).

\Planck studies concluded was that their observations are {\em
consistent} with constant grain alignment. Similarly, based on the
POL-2 data and radiative transfer simulations, we can conclude that
the observed $p(N)$ drop {\em can} be explained by the RAT mechanism
alone, which even tends to overestimate the drop in $p$. The two
conclusions are not necessarily contradictory. First, they apply to
different parts of the ISM, POL-2 probing a source of much higher
volume density. Second, the magnetic field geometry remains a
significant source of uncertainty in our modelling and we cannot put a
strict upper limit on its effects. For a single source, almost any
$p(N)$ relation can be explained with a suitably crafted field
geometry (small-scale structure or changes in the orientation relative
to the LOS), without a need for variations in the grain alignment
efficiency.

Another point of comparison is provided by the recent NIR polarimetry
of the starless core FeSt 1-457 presented by \citet{Kandori2018}. They
did not observe any (significant) drop in the polarisation efficiency
up to column densities of $A_{\rm V}\sim 20$\,mag. This is
qualitatively in contradiction with the RAT predictions, although the
authors noted that grain growth could provide at least a partial
explanation. Furthermore, based on the extinction data discussed in
Sect.~\ref{sect:obs_AV}, the extinction in the \target field reaches
significantly higher values.  These are some $A_{\rm V}\sim 50$\,mag
in both the northern and the southern clump, when measured at a
resolution of 15$\arcsec$. Indeed, the comparison of a set of NIR and
submm polarisation observations lead \citet{Jones_2015_alignment} to
conclude that grain alignment is lost around $A_{\rm V}\sim 20$\,mag.

It is clear that even if grain alignment is partially lost, also the
magnetic field orientation is never uniform and thus has some effect
on the polarisation observations. To determine the relative importance
of the two factors, we need more comparisons of high-resolution
observations and simulations.

\section{Conclusions}  \label{sect:conclusions}

We have used \Planckn, \Herscheln, and SCUBA-2/POL-2 data to investigate dust
emission and sub-millimetre polarisation in the \target field. 
Our main conclusions are the following:

The $\sim$6\,pc long main filament of the \target field is
characterised by dust colour temperatures $T\sim12-14$\,K and a column
density in excess of $N({\rm H}_2)=2\times 10^{22}$\,cm$^{-2}$.

The average sub-millimetre to NIR dust opacity ratio of the filament
is $\tau(250\mu{\rm m})/\tau({\rm J})=(2.55 \pm 0.03) \times 10^{-3}$.
This is more than four times higher than in diffuse clouds and
slightly higher than for previous samples of PGCC clumps of lower
column density. 

The average dust opacity spectral index of the filament is $\beta \sim
1.9$. This is similar to the values that previous studies have found
for larger samples of PGCC clumps.

At large scales, \Planck data show a relatively uniform magnetic
field orientation but a polarisation fraction of only $p\sim$2\%.  The
values of the polarisation angle dispersion function are on average
$S(\delta \sim 2.5\arcmin)\sim 9\degr$ and only slightly lower at the
location of the \target filament. The $S(p)$ relation was examined at
resolutions between FWHM=$5\arcmin$ and FWHM=$15\arcmin$ and was found
to be between the \citet{planck2014-XIX} results (obtained at much
larger scales) and a lower $S=0.1/p$ relation.

POL-2 data reveal a very low polarisation fraction of $p\sim 1\%$ for
the densest filament. The negative correlation between the
polarisation fraction and column density is significant at $N({\rm
H}_2)> 2 \times 10^{22}$\,cm$^{-2}$.

Observations are consistent with models with an almost complete loss
of grain alignment at densities above $n({\rm H}_2) \sim
10^4$\,cm$^{-3}$. RAT calculations overestimate the decrease of
polarised intensity, although this can be corrected by assuming a
strong increase in grain sizes (a factor of two). The modelling is
affected by large uncertainties in the grain properties, the magnetic
field geometry (including the small-scale structure and the average
LOS field component), and the underlying physical cloud model. 
Therefore, the relative importance of grain alignment and field
geometry for the $p(N)$ relation remains open. High-resolution
polarisation observations of a statistically significant sample of
filaments, clumps, and cores are needed.

\begin{acknowledgements}
The James Clerk Maxwell Telescope is operated by the East Asian
Observatory on behalf of The National Astronomical Observatory of
Japan, Academia Sinica Institute of Astronomy and Astrophysics, the
Korea Astronomy and Space Science Institute, the National Astronomical
Observatories of China and the Chinese Academy of Sciences (Grant No.
XDB09000000), with additional funding support from the Science and
Technology Facilities Council of the United Kingdom and participating
universities in the United Kingdom and Canada.  Additional funds for
the construction of SCUBA-2 were provided by the Canada Foundation for
Innovation.
The JCMT data were obtained under the programs M16AL003 and M17BP050.
\Planck \emph{(http://www.esa.int/Planck)} is a project of the European Space
Agency -- ESA -- with instruments provided by two scientific consortia funded by ESA
member states (in particular the lead countries: France and Italy) with contributions
from NASA (USA), and telescope reflectors provided in a collaboration between ESA and a
scientific consortium led and funded by Denmark.
{\it Herschel} is an ESA space observatory with science instruments provided
by European-led Principal Investigator consortia and with important
participation from NASA.
This research made use of Montage, funded by the National Aeronautics
and Space Administration's Earth Science Technology Office,
Computational Technologies Project, under Cooperative Agreement Number
NCC5-626 between NASA and the California Institute of Technology. The
code is maintained by the NASA/IPAC Infrared Science Archive.
This research used the facilities of the Canadian Astronomy Data 
Centre operated by the National Research Council of Canada with the
support of the Canadian Space Agency.
We thank J. Kainulainen for providing the NIR/MIR extinction map used
in the paper.
M. Juvela and V.-M. Pelkonen acknowledge the support of the Academy of
Finland Grant No. 285769.
T. Liu is supported by KASI fellowship and EACOA fellowship.
V.-M. Pelkonen acknowledges the financial support from the European
Research Council, Advanced Grant No. 320773 entitled Scattering and
Absorption of Electromagnetic Waves in Particulate Media (SAEMPL). 
L. Bronfman acknowledges support by CONICYT grant PFB-06.
Woojin Kwon was supported by Basic Science Research Program through
the National Research Foundation of Korea (NRF-2016R1C1B2013642).
C.W. Lee was supported by Basic Science Research Program through the
National Research Foundation of Korea (NRF) funded by the Ministry of
Education, Science and Technology (NRF-2016R1A2B4012593).
J. Malinen acknowledges the support of ERC-2015-STG No. 679852
RADFEEDBACK.
The french co-authors thank the support from the Programme National
Physique et Chimie du Milieu Interstellaire (PCMI) of CNRS/INSU with
INC/INP co-funded by CEA and CNES
\end{acknowledgements}

\bibliography{MJ_bib}

\begin{appendix}

\section{Modified blackbody fit to maps of different resolution}
\label{app:MBB}

Section~\ref{methods:MBB} described how column density can be
estimated using surface brightness maps that are first convolved down
to a common resolution. The resolution can be increased in several
ways.  \citet{Palmeirim2013} employed different combinations of the
observed bands. All bands were first used to derive a column density
map $N_0$ at the lowest common resolution. The lowest resolution bands
are then removed one by one and the remaining bands are used to
calculate corrections for higher spatial frequencies. On each step a
new map is obtained as
\begin{equation}
{\hat N_i}  = {\hat N}_{i-1} +  {\rm HPF}(N_i, {\rm FWHM}_{i-1}),
\end{equation}
where ${\hat N}_{i-1}$ is the previous estimate at the resolution
${\rm FWHM}_{i-1}$. $N_i$ is a resolution ${\rm FWHM}_{i}$ map that is
calculated using the maps that have a resolution better or equal to
${\rm FWHM}_{i}$, all convolved to ${\rm FWHM}_{i}$). HPF stands for
high-pass filtering for the indicated beam size. In the following, we
refer to this as method A.

In Sect.~\ref{sect:MBB_Herschel} we made high-resolution column
density maps by constructing a model that consisted of high-resolution
maps of temperature and surface brightness at a reference frequency.
The model was optimised by comparing its predictions with the observed
surface brightness maps, each at its original resolution. This
procedure (in the following method B) is time-consuming because each
optimisation step involves convolutions. The model parameters also are
not independent between pixels and must be solved simultaneously. 
After optimisation, the model maps can be converted to a column
density map. The nominal resolution could be even higher than in
method A but regularisation may be needed to avoid oscillations that
are typical for deconvolution. The pixel size of the model maps must
be small compared to the resolution of the observations. In
Sect.~\ref{sect:MBB_Herschel}, the best resolution of observations was
18$\arcsec$, the model was defined 6$\arcsec$ pixels, and the final
result was at a resolution of 20$\arcsec$. As regularisation, we
compared the model surface brightness and temperature values $X$ to
the average of its immediate neighbours ($\hat X$) and added to the
$\chi^2$ values a penalty $\xi (X-\hat X)/\hat X)^2$ with a small
weight factor $\xi=0.005$. Without regularisation $\chi^2$ values
would be better but larger small-scale temperature fluctuations could
lead to a higher average column density. This shows that analysis of
surface brightness data is not unique. The assumption of a minimal
amount of structure below the observed resolution is a special case
that leads to the lowest column density values.

In method A, the highest spatial frequencies depend on the pair of
surface brightness maps with the highest resolution. Because method B
solves a global optimisation problem, the role of different bands in
constraining structures at different scales could be different, also
because shorter wavelengths are more sensitive to temperature
variations \citep[e.g.][]{Shetty2009a,Malinen2011}.

We tested the methods using surface brightness maps obtained from
radiative transfer modelling. The density distributions of the model
clouds followed a 3D Gaussian distribution to which we had added minor
Gaussian fluctuations. The peak column densities were either $N({\rm
H}_2)$=1.3$\times 10^{22}$\,cm$^{-2}$ or a value ten times higher. The
models were heated by external radiation and, optionally, with a few
internal blackbody sources that raised the dust temperature locally to
$\sim$100\,K. We used the \citep{Mathis1983} model of the interstellar
radiation field and the dust model of \citet{Compiegne2011}. Radiative
transfer computations were used to solve the 3D temperature
distributions and to provide surface brightness maps at 160, 250, 350,
and 500\,$\mu$m. The 160\,$\mu$m band was included to increase the
sensitivity to temperature variations. We used the radiative transfer
program SOC \citep[Juvela et al. in prep.;][]{TRUST-I}, describing the
volume discretisation with hierarchical grids that were refined
according to the volume density. The effective resolution was 1024$^3$
cells and the total number of cells some 54 millions. The synthetic
observations and the subsequent analysis assume Gaussian beams that
correspond to the resolutions of the \Herschel 160-500\,$\mu$m bands.
We used observational uncertainties of 5\% for the 160\,$\mu$m band
and 3.5\% for the other bands. The column densities estimated with the
methods A and B were compared to the true column densities of the
model clouds. The analysis assumed a spectral index value of
$\beta$=1.8. The spectral index of the dust model is slightly
different and, therefore, we apply to the calculated column densities
a correction factor that in the case of an isothermal model results in
estimates equal to the true column density.

\begin{figure}
\includegraphics[width=8.8cm]{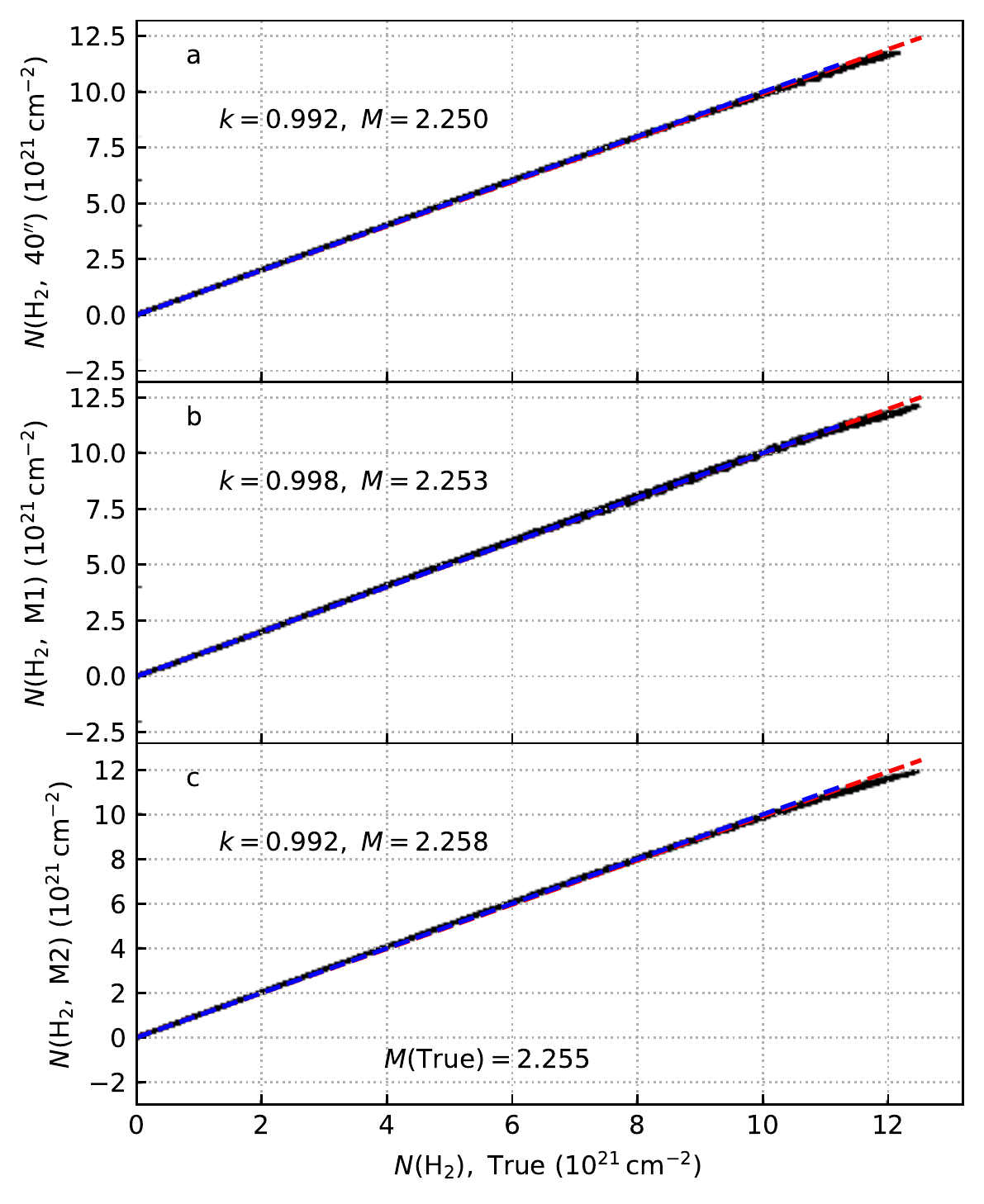}
\caption{
Correlations of column density estimates for the externally heated, 
lower column density model. The x-axis is the true column density of
the model cloud at the resolution of 40$\arcsec$ (frame a) or
18$\arcsec$ (frames b and c). On the y-axes are the results from
direct MBB fits at 40$\arcsec$ resolution (frame a) and from methods A
and B at 18$\arcsec$ resolution (frames b and c, respectively). The
dashed blue lines show the one-to-one relations and the dashed red
lines the least squares fits. The slopes $k$ of the least squares
lines as well as the estimated cloud masses (in solar masses) are
given in the frames. The true mass of the cloud is quoted in frame c.
\label{fig:cor_LBG}
}
\end{figure}

\begin{figure}
\includegraphics[width=8.8cm]{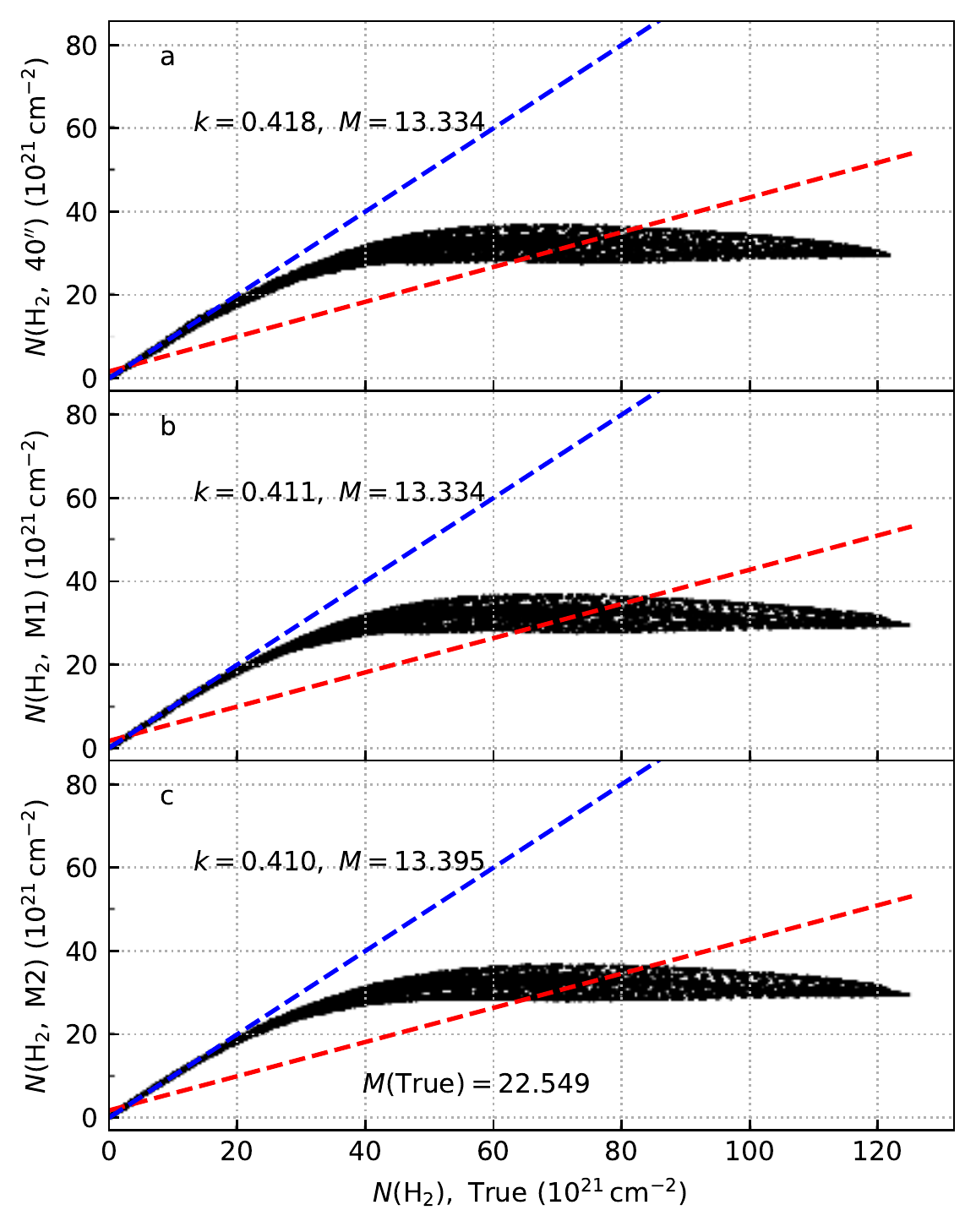}
\caption{
Same as Fig.\ref{fig:cor_LBG} but for the higher column density model.
\label{fig:cor_HBG}
}
\end{figure}

Figure~\ref{fig:cor_LBG} shows results for the lower column density
model with only external illumination and without observational noise.
The figure also includes column density estimates calculated from data
convolved to 40$\arcsec$. One pixel of the radiative transfer model
corresponded to 1$\arcsec$, which means that structures are well
resolved. All estimates are in good agreement with the true column
density values. 

\begin{figure*}
\includegraphics[width=17.7cm]{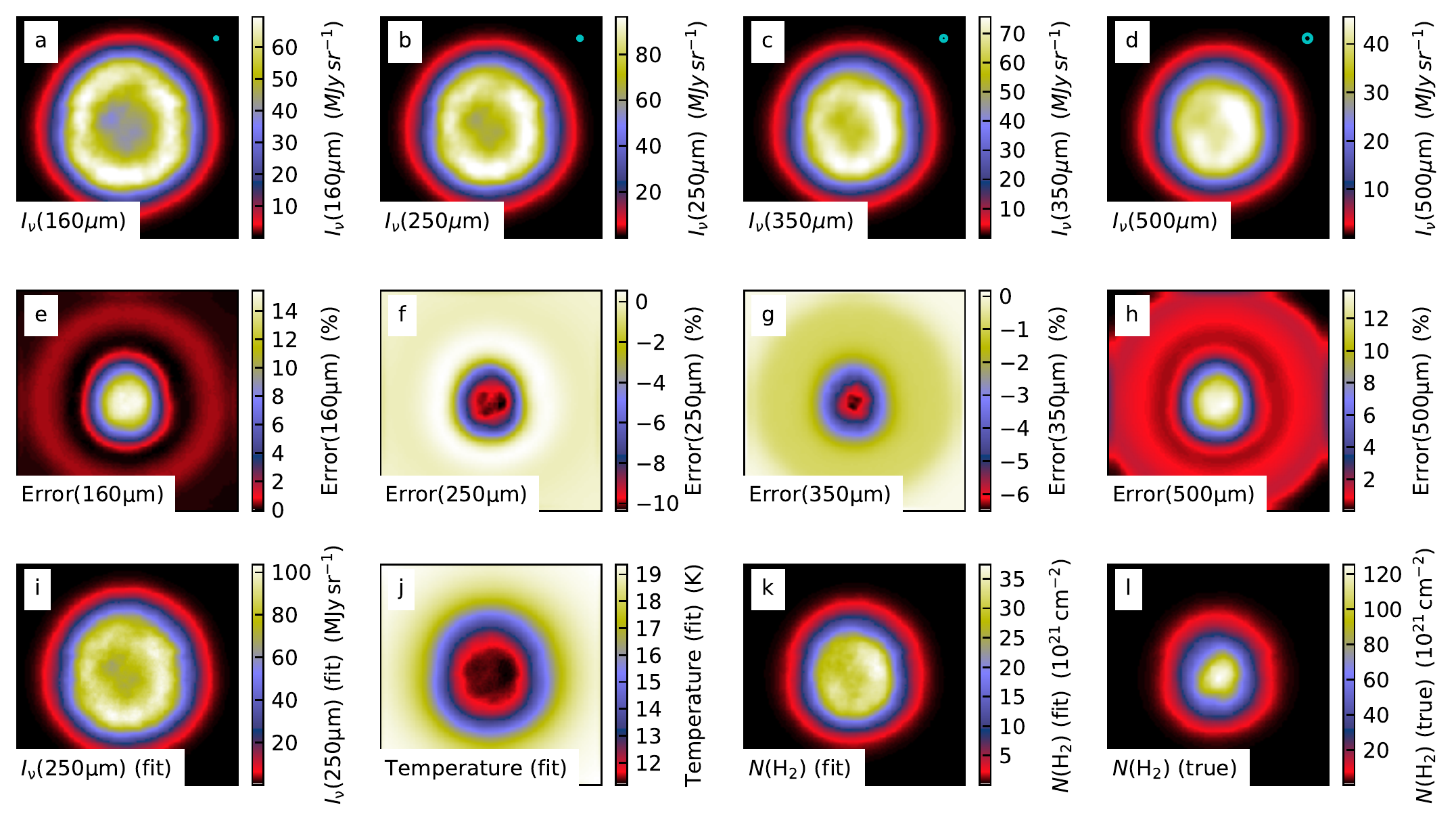}
\caption{
Fit of the simulated observations of the high-column-density model
without observational noise. The first row shows the synthetic surface
brightness maps with the beam size indicated in the upper right
corner. The second row shows the relative fit residuals. Frames i-j
show the fitted model parameters, the 250\,$\mu$m surface brightness
and the colour temperature. Frames k-l compare the final column
density map at 10$\arcsec$ resolution to the true column density.
\label{fig:sim_HBG}
}
\end{figure*}

The situation is different for the high column density model
(Fig.~\ref{fig:cor_HBG}). The column densities are severely
underestimated above $N({\rm H}_2)=2 \times 10^{22}$\,cm$^{-2}$.  The
estimates are actually decreasing towards the highest column densities
and the maximum error is close a factor of 4. The methods A and B give
practically identical results. Figure~\ref{fig:sim_HBG} shows the
method B fit in the form of maps. Because of LOS temperature
variations, the observed spectrum is much wider than the best-fit MBB
function, and the maximum residuals exceed 10\% of the measured
surface brightness.

\begin{figure}
\includegraphics[width=8.8cm]{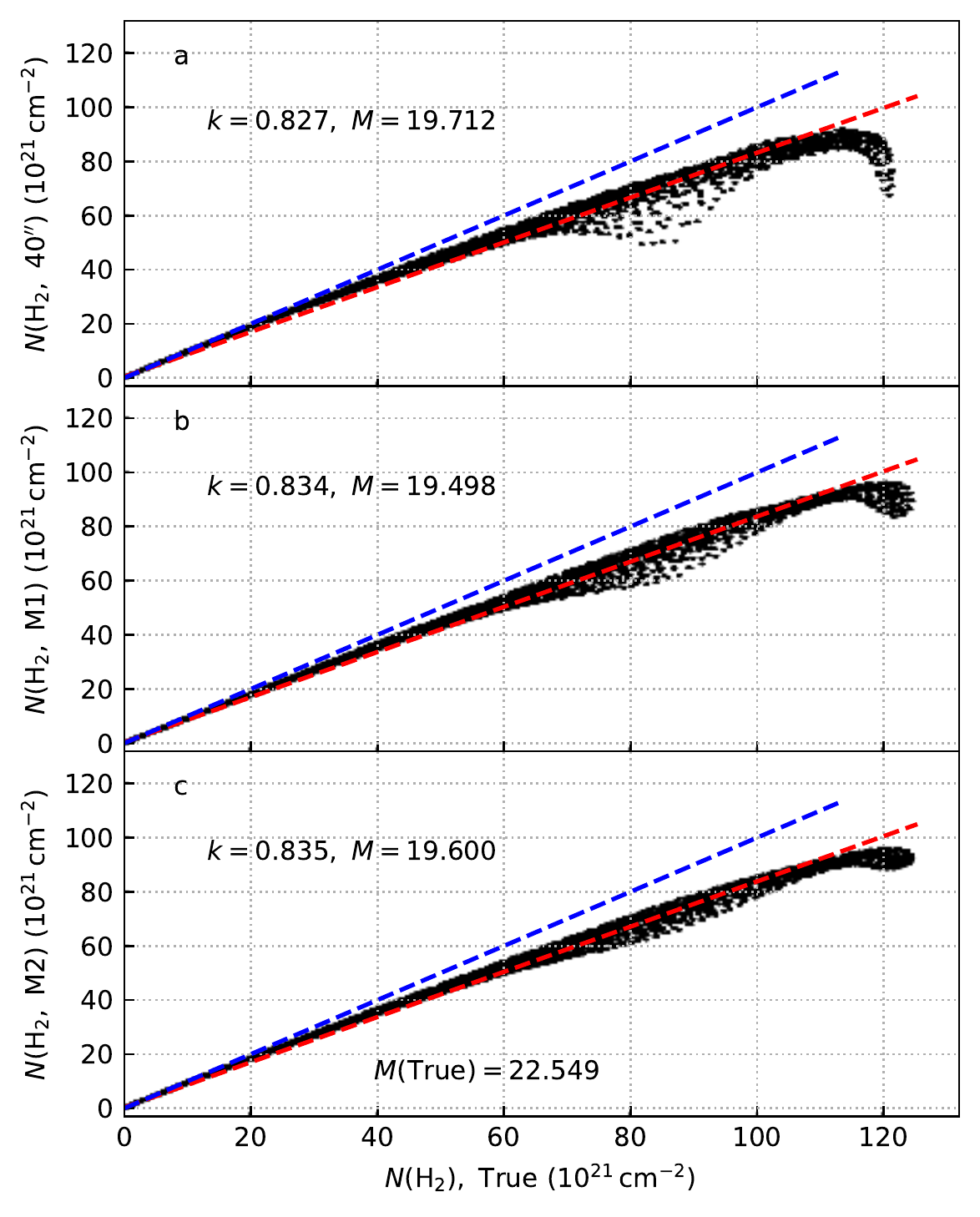}
\caption{
As Fig.\ref{fig:cor_HBG} but with added internal heating sources.
\label{fig:cor_HBG_PS}
}
\end{figure}

\begin{figure*}
\includegraphics[width=17.7cm]{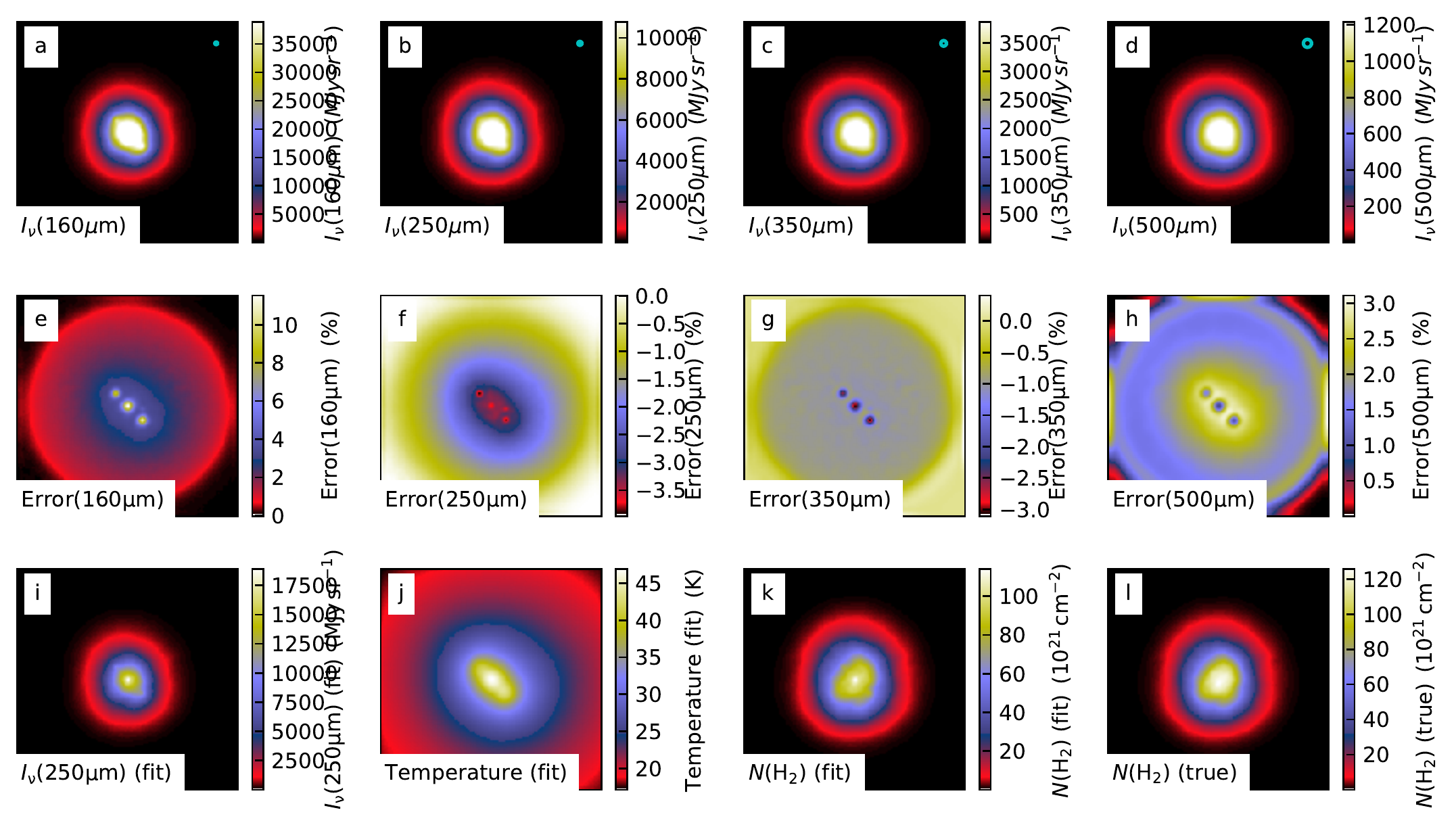}
\caption{
As Fig.~\ref{fig:sim_HBG} but with added internal heating sources.
\label{fig:sim_HBG_PS}
}
\end{figure*}

Figures~\ref{fig:cor_HBG_PS} and \ref{fig:sim_HBG_PS} show the results
for the high-column-density model after the addition of internal
sources. At the highest column densities method B now results in
marginally higher column density estimates than method A. Although the
presence of internal heating has significantly reduced the bias, both
methods underestimate the true column density by up to 25\%. The point
sources are barely resolved. If the resolution of the observations is
degraded by a factor of three, the results of the methods A and B are
identical.

\begin{figure}
\includegraphics[width=8.8cm]{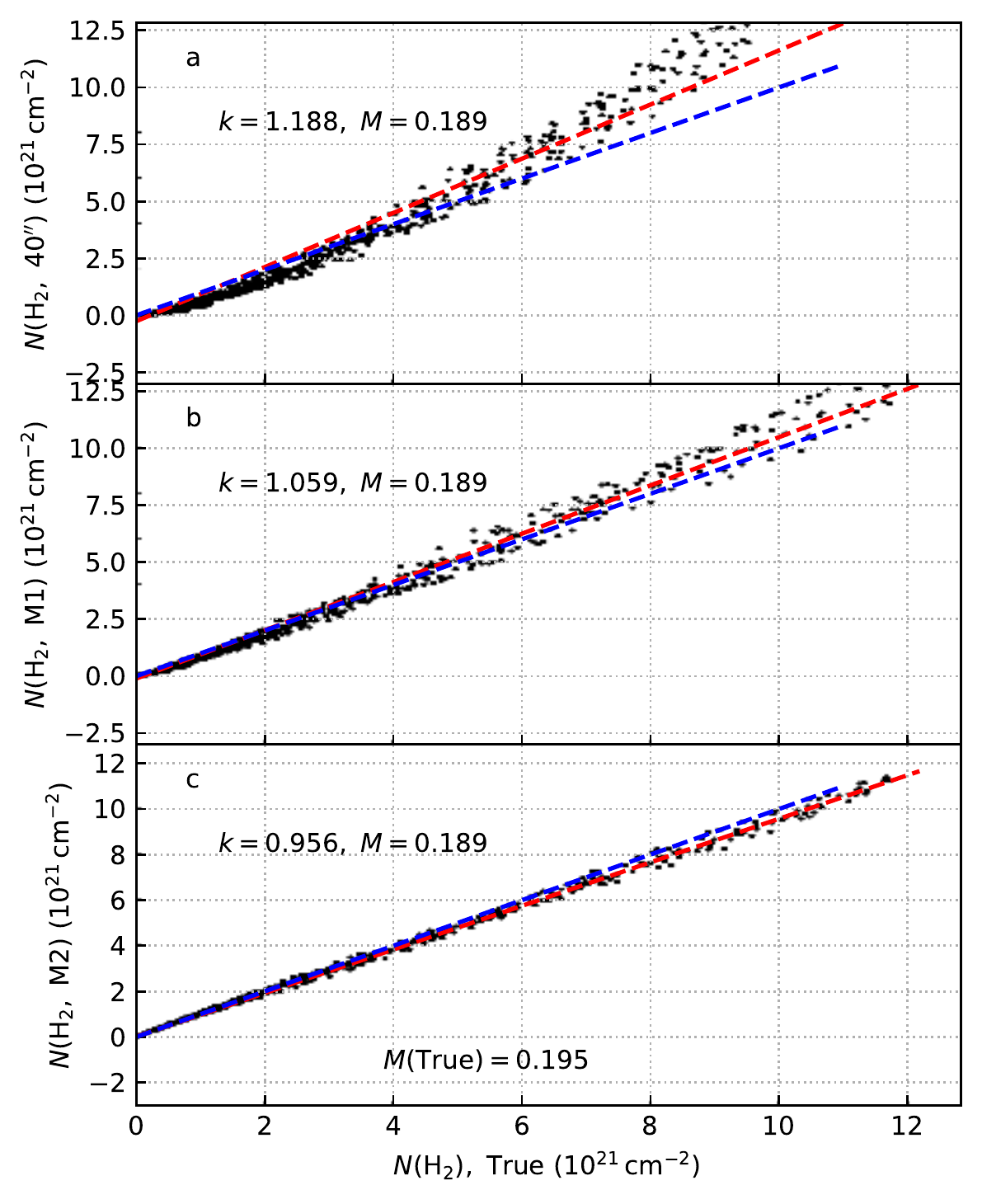}
\caption{ As Fig.~\ref{fig:cor_LBG} but with a factor of three lower
spatial resolution and with added observational noise. In each frame
the lower right corner shows the zoom-in to a column density range
from zero to 10$^{21}$\,cm$^{-2}$. The rms dispersion relative to the
linear least-squares fit is indicated in each frame.
}
\label{fig:LBG_P0p3_N0p5}
\end{figure}

\begin{figure}
\includegraphics[width=8.8cm]{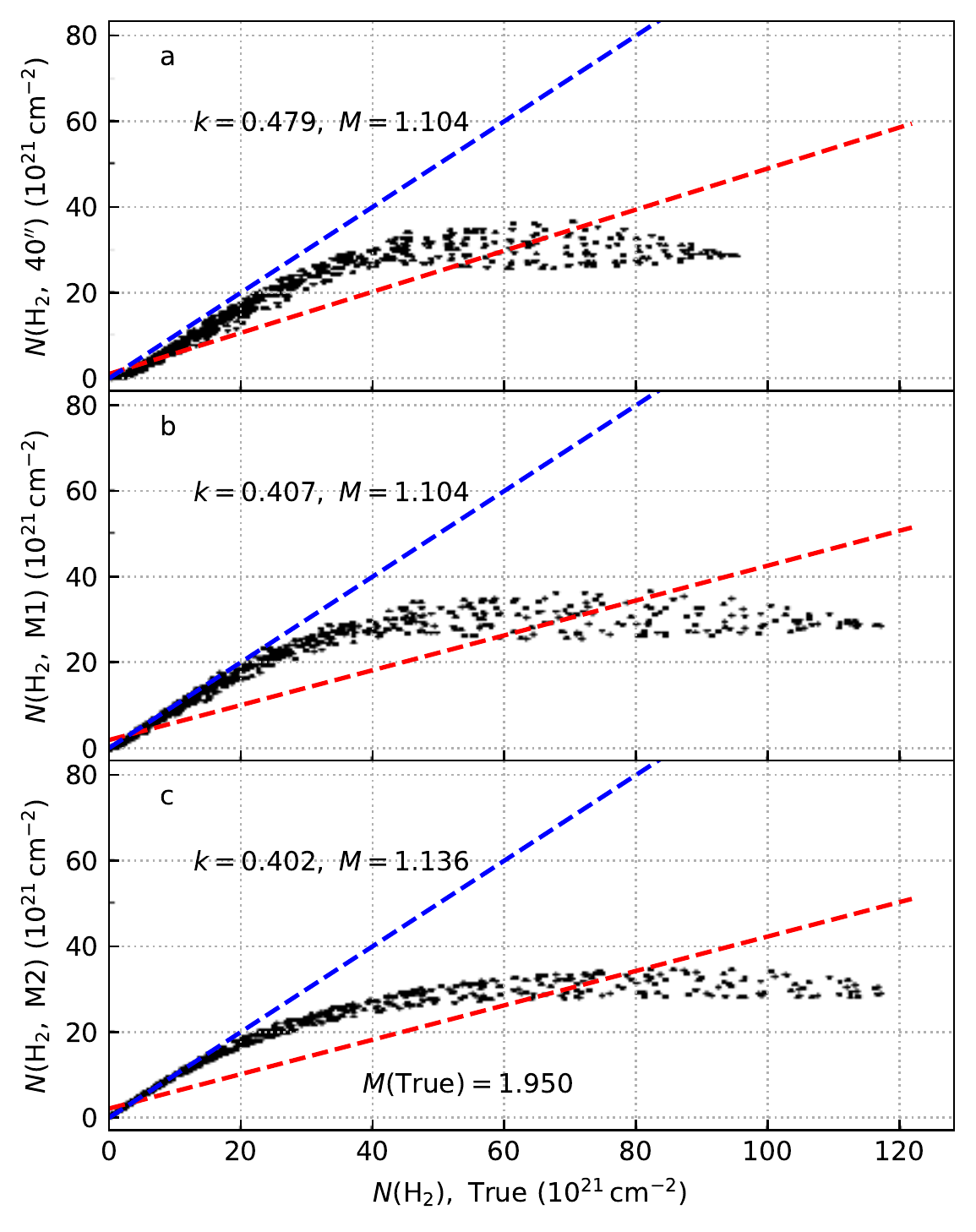}
\caption{As Fig.~\ref{fig:LBG_P0p3_N0p5} but for the models with
higher column density. Compared to Fig.~\ref{fig:cor_HBG_PS} the
spatial resolution is lower by a factor of three and the data include
added observational noise.
}
\label{fig:HBG_P0p3_N0p5}
\end{figure}

We repeated the tests after adding to the synthetic observations
Gaussian noise equal to the error estimates of the observations.
Figures~\ref{fig:LBG_P0p3_N0p5}-\ref{fig:HBG_P0p3_N0p5} show the
results for lower resolution observations where the original model
pixel corresponds to 0.3$\arcsec$. For the lower column density model
the differences are not significant, although the scatter of the
$N({\rm H}_2)$ estimates is marginally smaller for Method B
(Fig.~\ref{fig:LBG_P0p3_N0p5}).  The results at higher column
densities (Fig.~\ref{fig:HBG_P0p3_N0p5}) are qualitatively similar to
those without noise (Fig.~\ref{fig:cor_HBG}). Method B again exhibits
slightly smaller dispersion, especially at low-column-densities. This
is due to the global nature of the solution with some additional
effect from the regularisation. Apart from this, the results of the
two methods are in practice identical.

\section{Fit of the ($I$, $Q$, $U$) data of \Planck and POL-2}
\label{sect:IQU_fit}

Because of the apparent dissimilarity of the \Planck and JCMT POL-2
polarisation maps, we tested their consistency by making a model that
consisted of ($I$, $Q$, $U$) maps. These model maps were optimised through
comparison with the \Planck and POL-2 observations, both at their
original resolutions. In the case of POL-2, the filtering of low
spatial frequencies was simulated by subtracting a map obtained by
convolving the observations with a Gaussian beam with
$\theta=300\arcsec$. In the main text, the filtering was assumed to be
stronger with $\theta=200\arcsec$ and this would make it even easier
to reconcile any differences between the instruments. The model ($I$, $Q$,
$U$) maps had a pixel size of 10$\arcsec$. The model optimisation took
into account the error estimates (see Sect.~\ref{sect:obs}) and the
degree to which the model was oversampled compared to the
observations.

\begin{figure*}
\includegraphics[width=17cm]{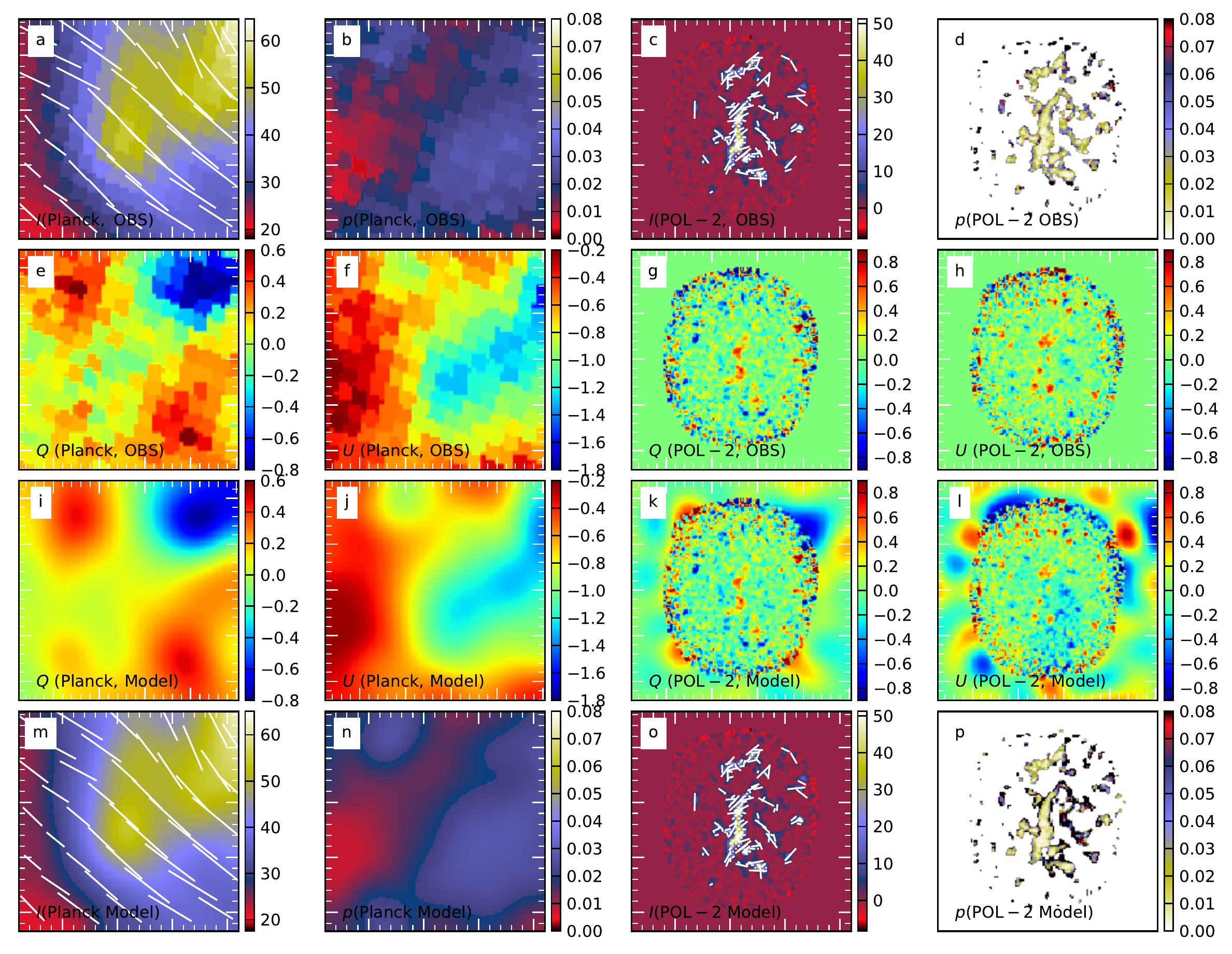}
\caption{
Results for the combined ($I$, $Q$, $U$) fir to \Planck and POL-2
data. Frames a-d show the observed \Planck and POL-2 surface
brightness data and polarisation fraction maps. Frames a and c also
show the polarisation vectors based on observations. The second row
shows the observed \Planck and POL-2 maps of $Q$ and $U$. The third
and fourth rows show the ($I$, $Q$, $U$) maps of the fitted model. In
frames c and o polarisation vectors are drawn only in regions with
$I(850\mu {\rm m})>5$\,MJy\,sr$^{-1}$.
\label{fig:IQU_fit}
}
\end{figure*}

The results are shown in Fig.~\ref{fig:IQU_fit}. The obtained ($I$, $Q$,
$U$) images fit well both the \Planck and POL-2 data. In spite of the
uncertainties (for example, the true nature of POL-2 filtering), the
exercise shows that there is no discrepancy between the data sets but
also that the spatial filtering plays a major role in the magnetic
field morphology seen in the POL-2 data.

\section{Effect of noise on the relations between $p$, $N$, and $S$}
\label{app:sim}

Noise affects the estimated values of both polarisation angles and
polarisation fraction \citep[e.g.][]{Montier2015a, Fissel2016}. In the
following we use simulations to estimate the effect of noise on the
observed relations between polarisation fraction, column density, and
polarisation angle dispersion function. {In the case of polarisation
fraction, we always refer to $p_{\rm mas}$ estimates that are already
corrected for bias, as described in Sect.~\ref{methods:pol}. The
correction is reliable at $p_{\rm mas}/\sigma_{p, \rm mas}>2$ but
values at lower SNR can be expected to be biased \citep{Montier2015b}.
However, even biased data contain some information, which can be used
when data are compared to simulated observations.}

\begin{figure}
\includegraphics[width=8.8cm]{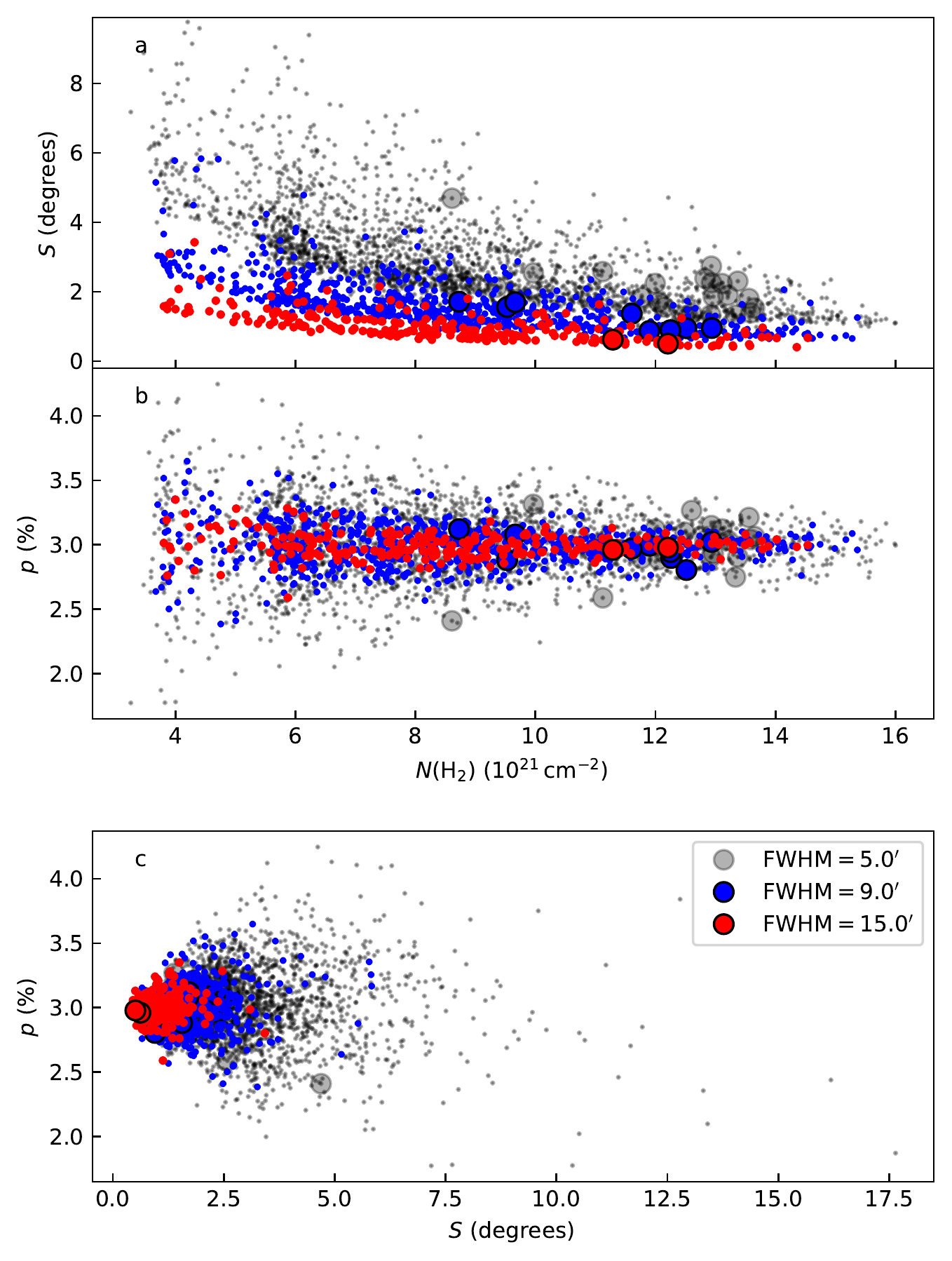}
\caption{
Polarisation angle dispersion function $S$ and polarisation fraction
$p_{\rm mas}$ based on simulated \Planck data that assume a constant
polarisation angle and a constant polarisation fraction of 3\%. Frames
a and b show these as a function of dust optical depth and frame c
shows their mutual correlation. The colours correspond to the data
resolution, as indicated in frame c. The points inside the area mapped
with POL-2 are plotted with large symbols. The data are sampled at
steps of FWHM/2 and $S$ is calculated for lag $\delta=$FWHM/2.
} \label{fig:SpI_sim}
\end{figure}

Figure~\ref{fig:SpI} showed a negative correlation between the
$p_{\rm mas}$ and $S$ values estimated from \Planck data. Such a trend
could result from noise if that causes significant bias in the
polarisation fraction estimates. For comparison with
Fig.~\ref{fig:SpI}, we made a simulation using the average error
estimates of the \Planck data (including the covariances) but assuming
constant values for both polarisation angles and polarisation
fractions. The results are shown in Fig.~\ref{fig:SpI_sim}. At full
\Planck resolution, noise increases the typical $S$ values that,
nevertheless, are only a fraction of the values in Fig.~\ref{fig:SpI}.
The polarisation fraction estimates show some dispersion but the
values are unbiased. As a result, the $p_{\rm mas}(S)$ relation shows
no significant correlations, unlike in the actual observations shown
in Fig.~\ref{fig:SpI}c.

\begin{figure}
\includegraphics[width=8.8cm]{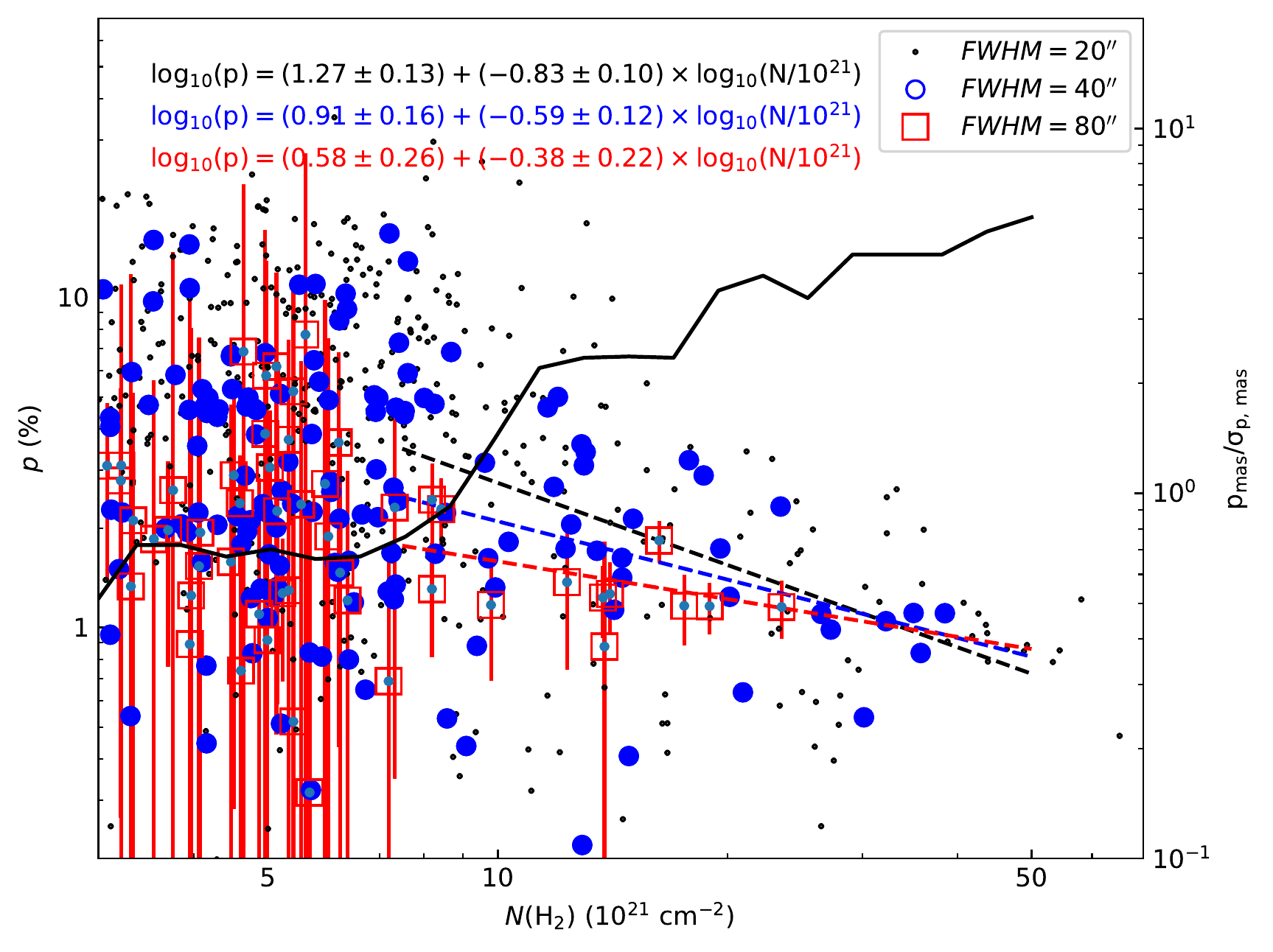}
\caption{
Polarisation fraction $p_{\rm mas}$ as a function of column
density, based on POL-2 data at 20$\arcsec$, 40$\arcsec$, and
80$\arcsec$ resolution. The data are sampled at half-beam steps. The
dashed lines correspond to least squares fits data $N({\rm H}_2)>7.5
\times 10^{21}$\,${\rm cm}^{-2}$. The error bars are shown only for
the FWHM=$80\arcsec$ data points. The solid black line and the
right-hand y-axis show the median SNR for polarisation fraction,
$p_{\rm mas}/\sigma_{p, \/ \rm mas}$, at $40\arcsec$ resolution.
}
\label{fig:p_N_fwhm}
\end{figure}

We examined the $p_{\rm mas}$ vs. $N$ relation of POL-2 data as
a function of the resolution. Lower resolution may increase
geometrical depolarisation but, more importantly, increases the SNR
and thus reduces the noise bias. In Fig.~\ref{fig:p_N_fwhm}, we plot
$p_{\rm mas}$ as a function of column density at the resolutions of
20$\arcsec$, 40$\arcsec$, and 80$\arcsec$. Only data with $N({\rm
H}_2)>7.5\times 10^{21}$\,cm$^{-2}$ are used for linear fits. As the
resolution is reduced, the slope decreases.  Above $N({\rm H}_2) \sim
20\times 10^{21}$\,cm$^{-2}$ the polarisation fraction is nevertheless
1-2\%, irrespective of the resolution.

\begin{figure}
\includegraphics[width=8.8cm]{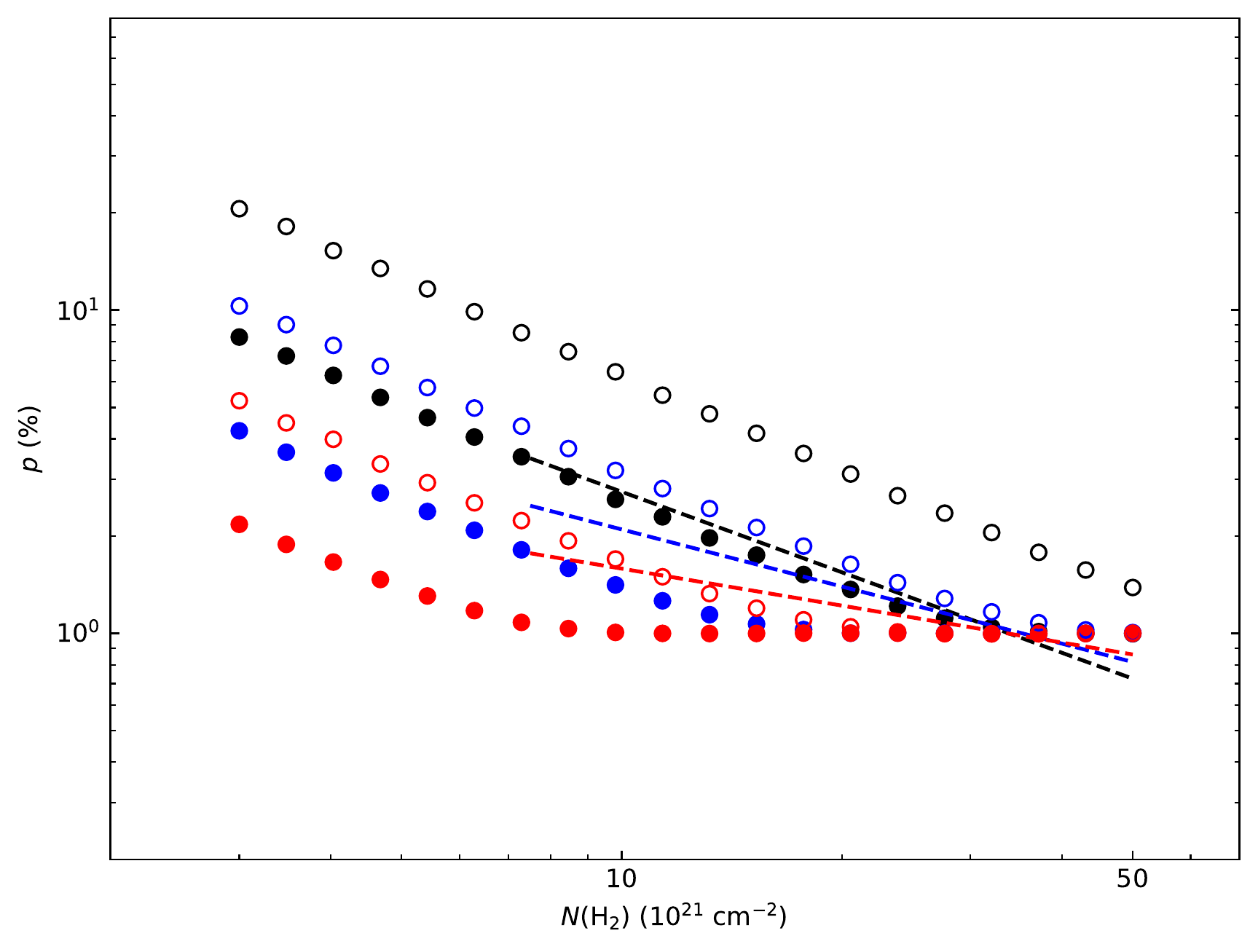}
\caption{
Simulations of the $p_{\rm mas}$ vs. $N$ relations in POL-2
observations. The calculations assumed a regular field (no geometrical
depolarisation), an intrinsic polarisation fraction of $p=1\%$, and
intensities proportional to the column density. The initial SNR of the
total intensity at $N({\rm H}_2)=10^{22}$\,cm$^{-2}$ at the full
FWHM=15$\arcsec$ resolution is either 40 (open circles) or 200 (solid
circles). The black, blue, and red colours correspond to the
convolution to a resolution of 20$\arcsec$, 40$\arcsec$, and
80$\arcsec$, respectively. For comparison, the dashed lines show the
least squares fits from Fig.~\ref{fig:p_N_fwhm}.
\label{fig:sim_p_vs_noise}
}
\end{figure}

For comparison with Fig.~\ref{fig:p_N_fwhm}, 
Fig.~\ref{fig:sim_p_vs_noise} shows simulations of the $p_{\rm
mas}$ vs. $N$ relation in the presence of noise. The simulations
assumed that the true polarisation fraction is constant 1.5\% and the
observed $p$ depends on the column density only because of the noise
bias. The observed intensity was assumed to be directly proportional
to the column density. The SNR of the total intensity at $N({\rm
H}_2)=10^{22}$\,${\rm cm}^{-2}$ is either 40 or 100.  
In the observations, the SNR is not constant for a given column
density and the values at $N({\rm H}_2)=10^{22}$\,${\rm cm}^{-2}$
range from SNR$\sim$10 to SNR$\sim$70 with a mean value of $\sim$33.
Thus the SNR=100 simulation could define a lower envelope for $p_{\rm
mas}$, the observed values being higher because of the noise bias.
Above $N({\rm H}_2)=2\times 10^{22}\,{\rm cm}^{-2}$ the average SNR is
SNR$\sim$100 or higher (see Fig.\ref{fig:p_N_fwhm}).

In Fig.~\ref{fig:sim_p_vs_noise}, when the data are convolved to a
lower resolution, the SNR increases and the $p_{\rm mas}$ values
decrease. Eventually the high-$N({\rm H}_2)$ values become independent
of the convolution while the polarisation fractions still decrease at
lower $N({\rm H}_2)$. At 40$\arcsec$ resolution $p$ values become
reliable only at $N({\rm H}_2)>10^{22}$\,${\rm cm}^{-2}$, where the
average SNR of observations is above 100 for the total intensity.

\begin{figure}
\includegraphics[width=8.8cm]{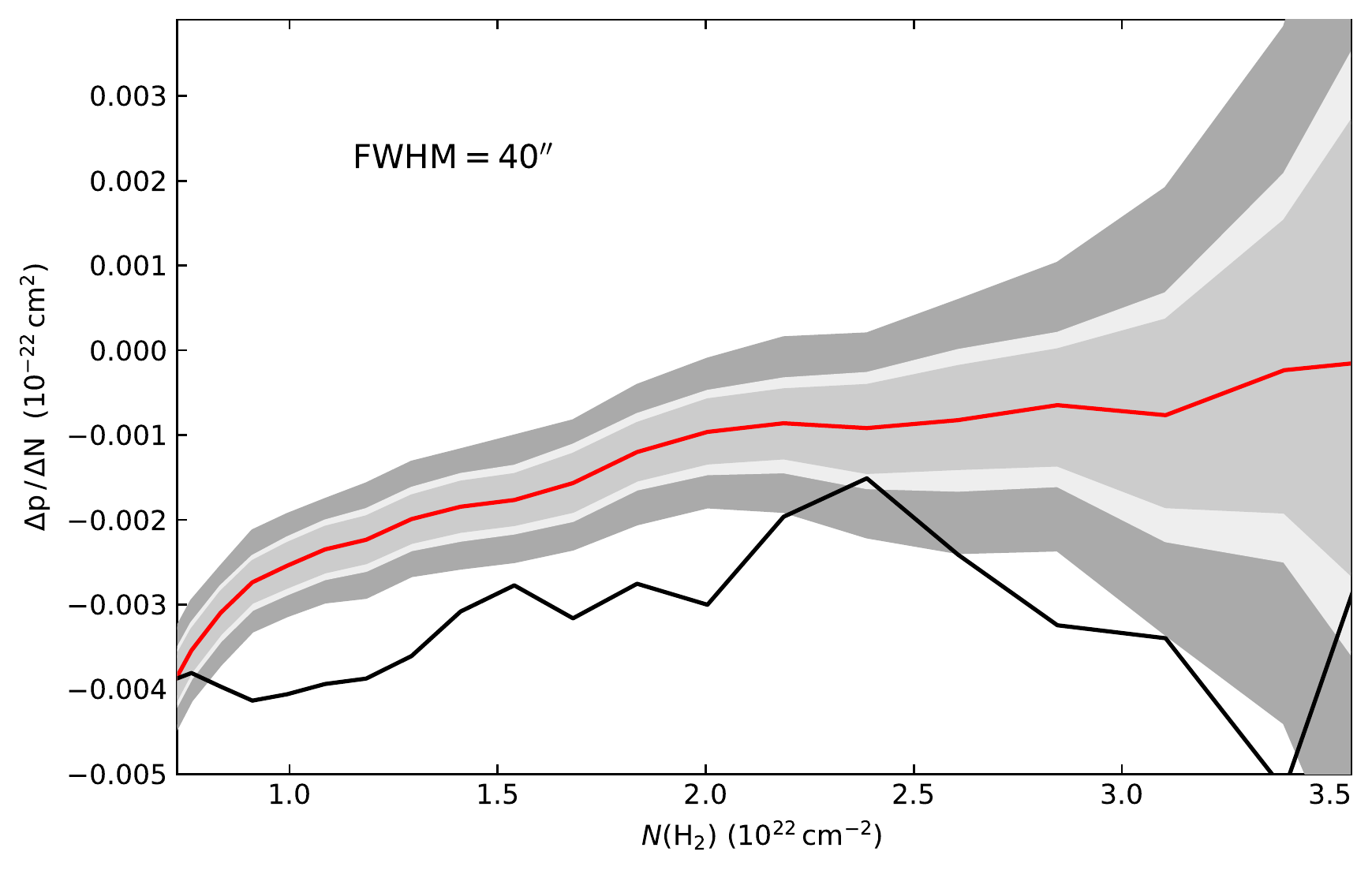}
\caption{
Simulation of $p_{\rm mas}$ vs. $N$ relation for POL-2 data. 
The figure shows the slopes of the fitted least squares lines as a
function of the column density threshold used to select data. The
black line shows the fits to the original observations. The red curve
shows the median slope for simulations that assume an intrinsically
constant polarisation fraction of $p=1.5\%$ and the shaded areas
correspond to [1\%,99\%], [10\%,90\%], and [16\%,84\%] confidence
intervals.
\label{fig:VI_sim_slope}
}
\end{figure}

In Fig.~\ref{fig:p_vs_N_S} we concluded that the anticorrelation would
be true, based on the SNR of the selected data (Fig.~\ref{fig:SNR})
and the formal errors of the fitted least squares lines. We examined
this more directly using the observed ($I$, $Q$, $U$) data and their
error estimates. We rescaled ($Q$, $U$) so that they corresponded to a
flat relation $p=1.5$\%. We constructed synthetic data sets where
noise was added according to the POL-2 error maps and the $p_{\rm
mas}(N)$ relation was fitted with a weighted least squares line.  The
simulation was repeated 2000 times for a number of column density
thresholds. Figure~\ref{fig:VI_sim_slope} indicates that there is over
99\% probability that the observed slope is steeper than what is
expected based on the constant-$p$ assumption. The slopes become
uncertain for thresholds $N({\rm H}_2)>3\times 10^{22}$\,cm$^{-2}$
because the fitted dynamical range and the number of data points
decrease. The significance also decreases below $N({\rm H}_2)\sim
10^{22}$\,cm$^{-2}$ when data become dominated by noise. In between,
the observed slope is much steeper than in the simulations.  Note
that it is not necessary for the $p_{\rm mas}$ estimates be completely
unbiased, because the bias should be the same both for the real and the
simulated observations. The significance of the result could be
decreased only if the uncertainties of $Q$ and $U$ were
underestimated, for example because of artefacts caused by the spatial
filtering or the mapping procedure.

\begin{figure}
\includegraphics[width=8.8cm]{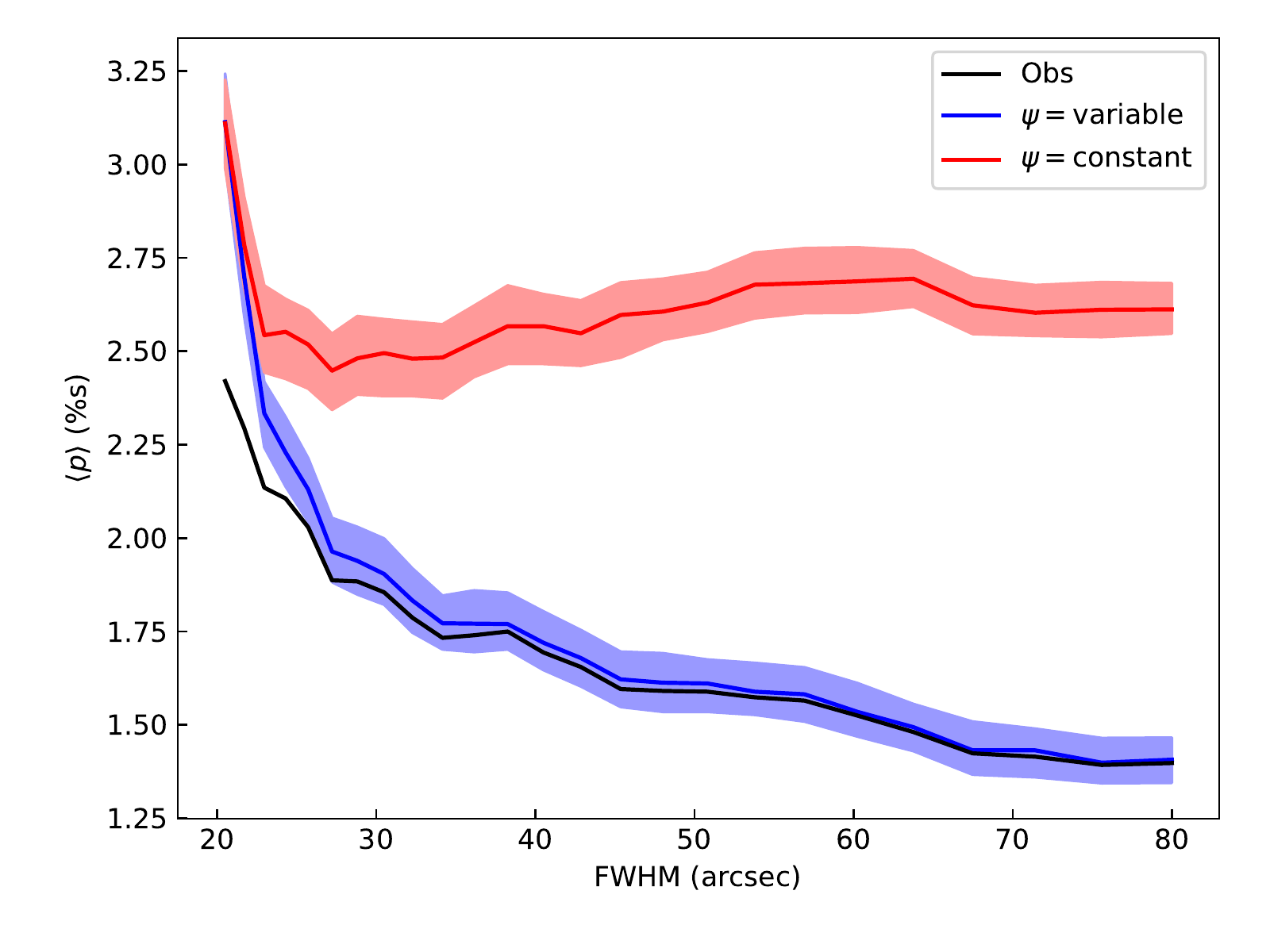}
\caption{
Simulation of $\langle p \rangle$ for POL-2 data with $N({\rm
H}_2)>10^{22}$\,cm$^{-2}$ as a function of the resolution. The black
line shows the relation for the original observations.  The blue line
corresponds to a case where ($Q$, $U$) observed at 20$\arcsec$
resolution are taken as a model of the true signal, before adding
observational noise and convolving to a lower resolution. The red
curve shows the result for data with identical $p$ and $I$ but with a
globally constant polarisation angle. The shaded regions show the
[10\%, 90\%] ranges for the random realisations.
\label{fig:VI_sim_depol}
}
\end{figure}

In Fig.~\ref{fig:p_N_fwhm} the slope was observed to change
systematically as a function of the data resolution. We carried out
separate calculation to examine the potential contribution of
geometrical depolarisation to this relation. Using data with $N({\rm
H}_2)>10^{22}$\,cm$^{-2}$, we plot in Fig.~\ref{fig:VI_sim_depol} the
observed average polarisation fraction $\langle p_{\rm mas} \rangle$
as a function of the spatial resolution. This is compared to two
simulations that start with the 20$\arcsec$ resolution observations.
First simulation assumes that these observations represent the true
(noiseless) polarisation data. We produce synthetic observations by
adding noise according to the POL-2 error estimates and  plot
again $\langle p_{\rm mas} \rangle$ as a function of the resolution.
This is thus identical to the actual observations, except for the
added noise. In the second simulation ($Q$, $U$) are first rotated so
that polarisation angle is the same in each pixel. This means that the
resolution only affects the noise level but there is no geometrical
depolarisation.

Figure~\ref{fig:VI_sim_depol} shows that without geometrical
depolarisation the $\langle p \rangle$ does not decrease beyond
FWHM$\sim 30\arcsec$, in agreement with SNR plotted in
Fig.~\ref{fig:SNR}. On the contrary, there is minor increase towards
lower resolutions. This is probably caused by the negative correlation
between $p$ and $N({\rm H}_2)$, which means that, for a given column
density threshold, the average SNR of polarised intensity decreases as
a function of the beam size. The simulation that included polarisation
angle variations naturally converges towards the observed curve. The
results indicate that between the 20$\arcsec$ and 80$\arcsec$
resolutions, more than half of the drop is caused by geometrical
depolarisation and a smaller part by the reduced noise bias.

\section{Alternative RT models for polarised emission}  \label{app:polmap}

In this section we show two variants of the polarisation models
presented in Sect.~\ref{sect:RT_pol}. 

\begin{figure}
\includegraphics[width=8.8cm]{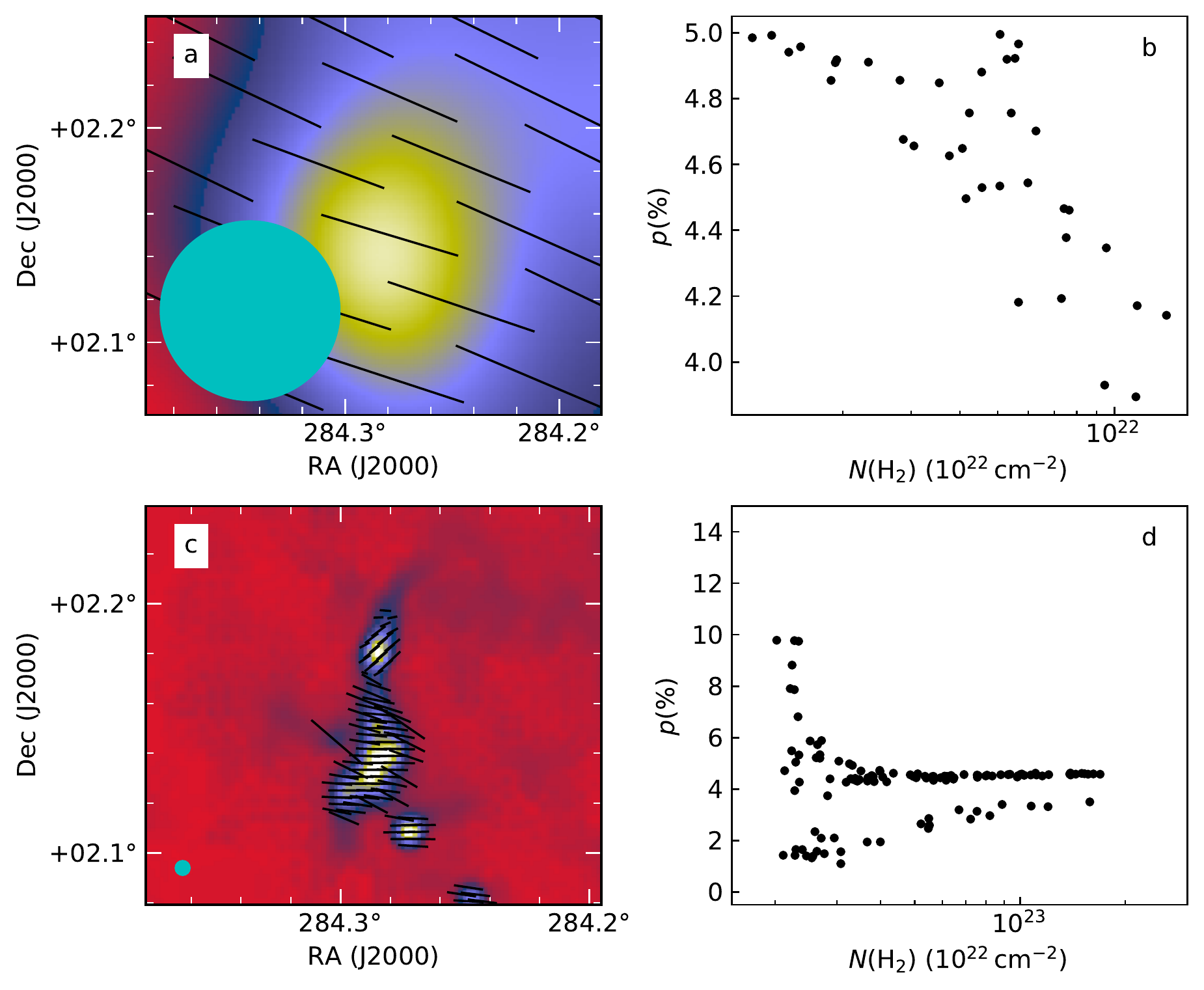}
\caption{
Predictions of polarised emission for a RT model with constant grain
alignment, using the toy magnetic field model. The figure is the same
as in Fig.~\ref{fig:pmod_const} except that the transition from the
large-scale field to different field orientations in the filament
takes place at a higher threshold of $n({\rm H}_2)= 10^4$\,cm$^{-3}$.
The polarisation vectors are shown on the column density map in frame
a and the $p$ vs. $N({\rm H}_2)$ relation is plotted in frame b at
\Planck resolution. The lower frames are the corresponding plots for
synthetic POL-2 observations.
\label{fig:pmod_const_hd}
}
\end{figure}

Figure ~\ref{fig:pmod_const_hd} shows results for constant grain
alignment. It is the same as Fig.~\ref{fig:pmod_const} except that the
transition from the uniform large-scale field to the filament field
(as defined by the magnetic field toy model) takes place at a higher
volume density threshold that is $n({\rm H}_2)= 10^4$\,cm$^{-3}$
instead of $n({\rm H}_2)= 3 \times 10^3$\,cm$^{-3}$. This leads to
decrease in the polarisation fractions, especially in the northern
clump.

\begin{figure}
\includegraphics[width=8.8cm]{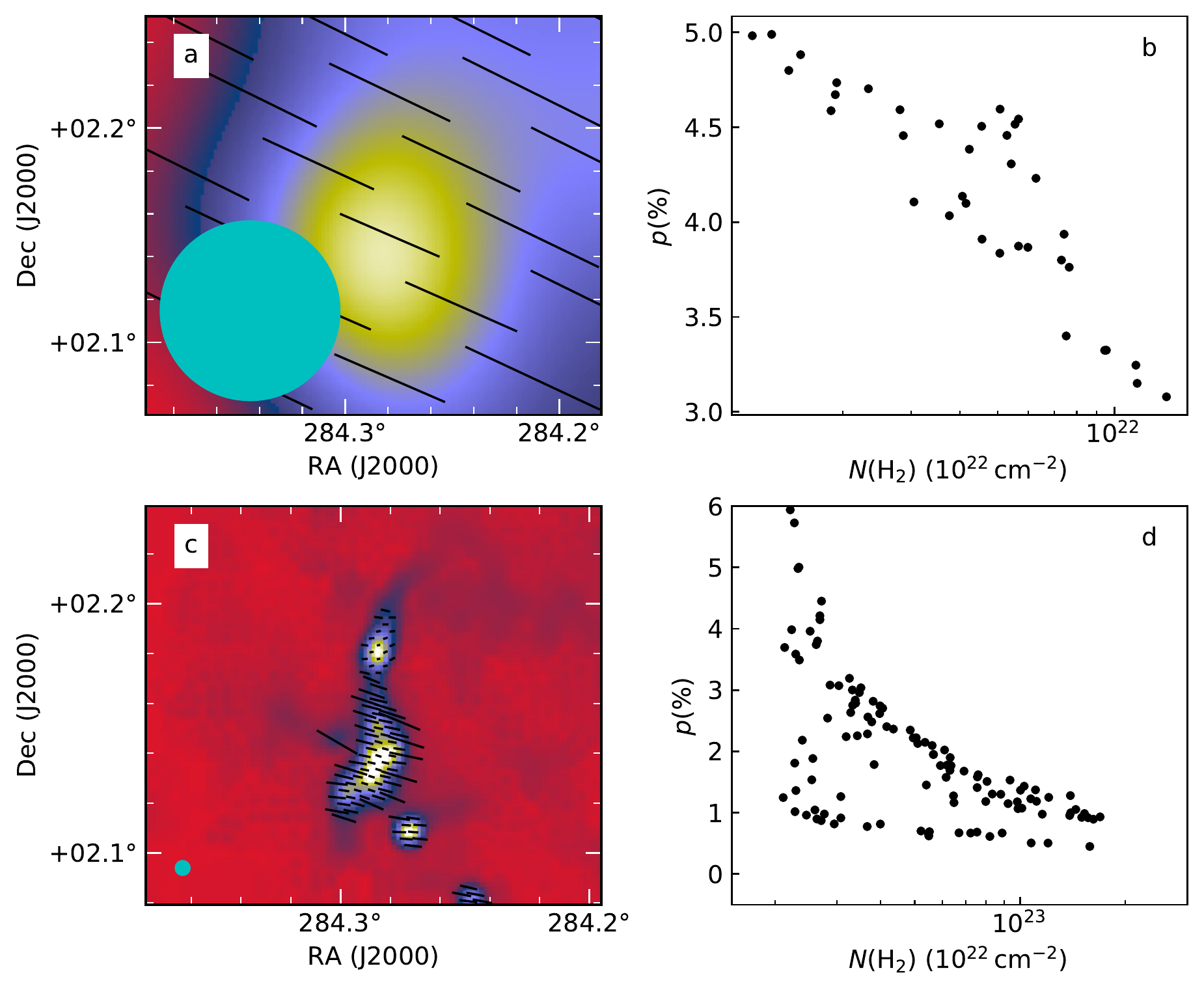}
\caption{
Model predictions for polarisation according to RAT grain alignment,
assuming larger grain sizes. The figure is the same as 
Fig.~\ref{fig:pmod_RAT_DD} but the field orientation in the filament
with $n({\rm H}_2)> 3 \times 10^3$\,cm$^{-3}$ is taken directly from
POL-2 observations at 20$\arcsec$ resolution.
\label{fig:pmod_RAT_obs}
}
\end{figure}

Figure~\ref{fig:pmod_RAT_obs} corresponds to calculations with RAT
grain alignment and grain sizes that are a factor of two larger than
in the original dust model. The figure differs from 
Fig.~\ref{fig:pmod_RAT_obs} in that the POS magnetic field
orientations are taken from POL-2 observations instead of the toy
model of Sect.~\ref{sect:RT_pol}. The polarisation fractions show a
small decrease. Part of the changes may be caused by noise in the
input polarisation angles, which here contributes to the geometrical
depolarisation (that is, angle dispersion within the beam).

\end{appendix}

\end{document}